\documentclass[review]{elsarticle}

\usepackage{lineno,hyperref}

\usepackage{hyphenat}
\modulolinenumbers[5]

\usepackage[nointegrals]{wasysym} 
\usepackage[margin=1.56cm]{geometry}

\usepackage{amsmath,xparse,mleftright}
\usepackage{gensymb}
\usepackage{amssymb}
\usepackage{subcaption}
\usepackage[export]{adjustbox}
\usepackage{float}
\usepackage{xcolor}
\usepackage{soul}

\NewDocumentCommand{\evalat}{sO{\big}mm}{%
  \IfBooleanTF{#1}
   {\mleft. #3 \mright|_{#4}}
   {#3#2|_{#4}}%
}

\journal{Icarus}

\biboptions{authoryear}

\begin{document}

\begin{frontmatter}

\title{Investigating the Feasibility of an Impact-Induced Martian Dichotomy}

\author[bernaddress]{Harry A. Ballantyne \corref{mycorrespondingauthor}}
\cortext[mycorrespondingauthor]{Corresponding author.}
\ead{harry.ballantyne@unibe.ch}

\author[bernaddress]{Martin Jutzi}

\author[bayreuthaddress]{Gregor J. Golabek}

\author[bernaddress,genevaaddress]{Lokesh Mishra}

\author[ethaddress]{Kar Wai Cheng}
\author[ethaddress]{Antoine B. Rozel}
\author[ethaddress]{Paul Tackley}

\address[bernaddress]{Space Research \& Planetary Sciences (WP), University of Bern, Bern, Switzerland}
\address[bayreuthaddress]{Bayerisches Geoinstitut, University of Bayreuth, Bayreuth, Germany}
\address[genevaaddress]{Geneva Observatory, University of Geneva, Versoix, Switzerland}
\address[ethaddress]{Institute of Geophysics, ETH Zurich, Zurich, Switzerland}

\begin{abstract}
A giant impact is commonly thought to explain the dramatic contrast in elevation and crustal thickness between the two hemispheres of Mars known as the ``Martian Dichotomy''. Initially, this scenario referred to an impact in the northern hemisphere that would lead to a huge impact basin (dubbed the ``Borealis Basin''), while more recent work has instead suggested a hybrid origin that produces the Dichotomy through impact-induced crust-production. The majority of these studies have relied upon impact scaling-laws inaccurate at such large-scales, however, and those that have included realistic impact models have utilised over-simplified geophysical models and neglected any material strength. Here we use a large suite of strength-including smoothed-particle hydrodynamics (SPH) impact simulations coupled with a more sophisticated geophysical scheme of crust production and primordial crust to simultaneously investigate the feasibility of a giant impact on either hemisphere of Mars to have produced its dichotomous crust distribution, and utilise spherical harmonic analysis to identify the best-fitting cases. We find that the canonical Borealis-forming impact is not possible without both excessive crust production and strong antipodal effects not seen on Mars' southern hemisphere today. Our results instead favour an impact and subsequent localised magma ocean in the southern hemisphere that results in a thicker crust than the north upon crystallisation. Specifically, our best-fitting cases suggest that the projectile responsible for the Dichotomy-forming event was of radius 500-750 km, and collided with Mars at an impact angle of 15-30$\degree$ with a velocity of 1.2-1.4 times mutual escape speed ($\sim$6-7 km/s).
\end{abstract}


\end{frontmatter}


\section{Introduction}

Mars is the most studied planet other than Earth, yet its most prominent feature has evaded confident explanation for nearly 50 years; the so-called ``Martian Dichotomy". This moniker predominantly refers to the stark topographical disparity between the two hemispheres of the planet, with an apparent basin encompassing around 42\% of the north, corresponding to an elevation difference of 4--8 km \citep[e.g.][]{Andrews-Hanna2008}. Additional associated features include a greater density of visible impact craters and volcanoes in the south relative to the north; however, the former may be explained via the inclusion of ancient, buried depressions that could equalise the crater counts when taken into account \citep{Frey2006,Buczkowski2007}.

Gravity measurements revealed that this feature is not limited to surface properties, with the crustal thickness distribution being highly correlated with that of elevation, resulting in an average estimated crustal thickness contrast of roughly 25 km \citep{Zuber2000,Neumann2004}. Recent seismic measurements from NASA’s InSight lander also support this claim \citep{Knapmeyer-Endrun2021}. Naturally, this has motivated many previous authors to propose geodynamic processes as the origin of such an internal structure. These studies usually support a degree-1 mantle upwelling beneath the southern hemisphere, leading to hemisphere-preferential crust production over long timescales \citep[e.g.][]{Zhong2001}. 

The length of these timescales are of great importance, however, as the Dichotomy is known to be the one of the oldest observable features on Mars. Crater distributions suggest an age of at least ${\sim}4.1$ Ga \citep{Solomon2005,Nimmo2005,Frey2006}, while recent geochemical analyses of Martian meteorites push this to ${\sim}$4.4 Ga \citep{Humayun2013,Cassata2018,Bouvier2018}. This leaves little time to form such a convection pattern; only being possible if Mars possessed a particular, layered viscosity profile that includes a large jump (by a factor of $>25$) in the mid-mantle \citep{Roberts2006,Keller2009}. Moreover, such a scenario requires vigorous convection via a high Rayleigh number which would likely lead to long-term resurfacing, erasing the initial crustal dichotomy over billion year times.

Another, more outward-looking explanation for the Dichotomy's origin is that of an impact between Mars and an object on the order of 1000 km in radius: an attractive alternative considering that such large-scale impacts were most likely during the first few 100 Myr of the Solar System \citep[e.g.][]{Morbidelli2018}. Moreover, the debris-disc formed by such an event would naturally lead to satellites with very low eccentricities and inclinations, matching the orbital properties of the Martian moons well \citep[and references therein]{Murchie2015}. Traditionally, this meant excavating the crust from the northern hemisphere while leaving the southern hemisphere intact, imprinting a classical (albeit giant) impact crater dubbed the Borealis Basin \citep{Marinova2008, Andrews-Hanna2008, Nimmo2008}. A clear caveat to this approach is the significant degree of melting that accompanies these events, as this potentially induces crust production upon re-crystallisation that must be taken into account.

This leads to the most recent group of hypotheses---a giant impact occurred that invoked significant crust production on one hemisphere. \cite{Reese2006,Reese2010a,Reese2010,Golabek2011,Leone2014} proposed this to have been in the south, whereby the heat anomaly invokes a more rapid transition to a single-plume mantle dynamic. \cite{Citron2018} suggested instead that the impact occurred in the north, stripping the hemisphere of its radiogenic-heating element enriched primordial crust and subsequently promoting a superplume in the south. In either case, the impact was modelled through highly simplified scaling laws \citep{Senshu2002,Monteux2007}. Such simplifications explicitly neglect several effects that are important on planetary-scales, such as antipodal shock-heating and re-impacting ejecta. Even when ignoring such effects, it is not clear how accurate these predictions are when extrapolated to this regime \citep{Marinova2011}.

To combat these issues, \cite{Golabek2018} successfully coupled a smoothed-particle hydrodynamics (SPH) impact model with a thermochemical model of the interior mantle dynamics for a Mars-sized object. This showed initial promise in producing a crustal dichotomy, further supporting the southern-impact hypothesis; however, only the crystallisation of the magma ocean was modelled (i.e. 0.5 Myr). Furthermore, no consideration was made for the possibility of a primordial crust, which is a highly likely pre-impact feature considering Mars' differentiated nature \citep{Stahler2021}. Finally, the study was restricted to only two impact scenarios: head-on or $45\degree$ impact angle at the mutual escape speed, with a 1000 km-radius impactor.

In this study, we explore a feasible parameter-space that could lead to such a scenario through a large set of three-dimensional SPH simulations, the details of which are described in Section \ref{model} and Section \ref{setup}, with the aim to provide much needed refinements. To estimate the post-impact crustal thickness, we develop a crust production model based on melt fraction and mantle fertility, also described in Section \ref{model}. A significant advantage of this model over previous work is its ability to include pre-impact crusts of various thicknesses, allowing us to investigate both the case of a crust-stripped northern hemisphere and the case of a magma-thickened southern hemisphere, concurrently. In Section \ref{results} and Section \ref{discussion}, we present and discuss the results of this study, respectively, {where a novel approach utilising spherical harmonic analysis is used to accurately identify the most promising regions of the explored parameter-space}. Finally, we present our conclusions and future outlook in Section \ref{conclusions}.

\section{Modelling Approach} \label{model}
\subsection{Impact Code}
To model the impacts we use SPHLATCH; an SPH code specifically designed to handle planetary-scale collisions \citep{Reufer2012,Emsenhuber2018} that has been used to study a wide range of impact regimes \citep{Emsenhuber2019,Emsenhuber2019a, Emsenhuber2020,Emsenhuber2021,Asphaug2021,Gabriel2020,Gabriel2021,Cambioni2019}.
\subsubsection{Smoothed-Particle Hydrodynamics (SPH)}
SPH is a Lagrangian method based on interpolation points (or ``particles'') of fixed mass that represent a continuous medium \citep{Lucy1977,Gingold1977}. Through the use of a Gaussian-like function known as a kernel, $W$, along with a measure of particle volume, $V$, the physical properties of the medium can be smoothed out, allowing quantities such as density and internal energy to be calculated for a given particle as a sum of the smeared-out contributions from its neighbouring particles. The degree of this smoothing is quantified by the ``smoothing length'', $h$, which is related to the density of the particle of interest, giving an approximate measure of resolution. In mathematical notation, this reads as
\begin{equation}\label{sphEq}
    B(\mathbf{x}) = \sum_{b} B_b W(\mathbf{x} - \mathbf{x}_b, h),
\end{equation}
where $B$ is the physical quantity of interest, $\mathbf{x}$ is the associated spatial location and the subscript $b$ indicates a specific neighbouring particle. The ``standard'' kernel function, also used in SPHLATCH, is the cubic B-spline given explicitly in \ref{bspline}.
For further details of this method, we refer the reader to an in-depth review such as those by \cite{Rosswog2009a} or \cite{Monaghan2005}.
\subsubsection{Solid Strength}\label{strength}
One advantage of SPHLATCH over other similar codes is its ability to include the effects of shear strength and plasticity. Typically, it is assumed that the force of gravity will dominate over any material strength in planetary-scale impacts, leading most previous works to neglect such effects entirely \citep[e.g.][]{Marinova2008}. This approach conveniently allows for greatly reduced computational cost; however, \cite{Golabek2018} (and its sister study \cite{Emsenhuber2018}) showed that the inclusion of material strength in this size-regime still plays a non-negligible role, indicating important physics that cannot be ignored.

The implementation of these effects in the SPHLATCH code is described in \cite{Emsenhuber2018}. To avoid unnecessary repetition, therefore, we shall only give a brief description of such aspects of the code, emphasising the key adaptations that have been made since.

At its core, adding solid mechanics to the standard, inviscid fluid SPH requires the generalisation of pressure into a stress tensor, $\sigma^{\alpha\beta}$, made up of a hydrostatic pressure component $p$ and a shear stress component $S^{\alpha\beta}$:

\begin{equation}
    \sigma^{\alpha\beta} = -p\delta^{\alpha\beta} + S^{\alpha\beta},
\end{equation}

where $\delta^{\alpha\beta}$ is the Kronecker delta. We follow \cite{Benz1994,Benz1995}, who (following \cite{Wingate1993,Libersky1991}) applied Hooke's law to expand $S^{\alpha\beta}$ as follows:
\begin{equation}
    S^{\alpha\beta} = 2\mu ( \epsilon^{\alpha\beta} - \frac{1}{3} \epsilon^{\gamma\gamma}),
\end{equation}
where $\mu$ is the material's shear modulus, $\epsilon^{\alpha\beta}$ is the strain tensor and the superscript $\gamma$ follows the Einstein summation rule. The time derivative of this equation is
\begin{equation}\label{dshearEq}
    \dot{S}^{\alpha\beta} = 2\mu ( \dot{\epsilon}^{\alpha\beta} - \frac{1}{3} \dot{\epsilon}^{\gamma\gamma}) + S^{\alpha\gamma}R^{\gamma\beta} + S^{\beta\gamma}R^{\gamma\alpha},
\end{equation}
where $R^{\alpha\beta}$ is the rotation rate tensor. The rotation terms are necessary for reference frame independence; without them the calculations would correspond to the material reference frame, unlike all other calculations that are in the laboratory frame. 

This approach has been used in a great number of subsequent papers, however there appears to be a small, but very important disparity between different works regarding the exact form of the rotation terms in Equation~\ref{dshearEq}. In the original work of \cite{Benz1994,Benz1995}, the rotation terms were of the form
\begin{equation}
    S^{\alpha\gamma}R^{\beta\gamma} + S^{\beta\gamma}R^{\alpha\gamma},
\end{equation}
which is the variant used for all previous work with SPHLATCH other than \cite{Emsenhuber2021} and \cite{Asphaug2021}. Other studies such as \cite{Jutzi2008} replaced $S^{\alpha\gamma}$ with $S^{\alpha\beta}$. In contrast to this, \cite{Schafer2007} (and related subsequent work such as \cite{Schafer2016}) use the same form as Equation \ref{dshearEq}.
We find that it is important to use this variation to avoid significant angular momentum conservation issues in certain impact-regimes such as those detailed in this study (see Figure~\ref{fig:conservation}).\footnote{We note that the angular momentum conservation issues related to the rotation terms only occur at large, planetary scales. Previous studies at smaller scales would experience only a negligible improvement if they were to be repeated with the correct form.}

Plastic deformation effects are included through a Drucker-Prager-like yield strength model, with a temperature dependence such that strength decreases as the material approaches its melting temperature \citep{Collins2004}. For this purpose, we use the Mars-specific solidus of \cite{Duncan2018}.

\subsubsection{Other details}
SPHLATCH uses the Barnes-Hut hierarchical tree method to find the SPH nearest neighbours and calculate self-gravity through the associated multipole approximation \citep{Barnes1986,Hernquist1987,Hernquist1989}, with an improved neighbour-list method used to ensure force symmetry that is described in \ref{neighlist}. To calculate thermodynamic quantities such as pressure and temperature, we use the sophisticated equation of state ANEOS \citep{Thompson1972,Thompson1990}.

\subsection{Crust Production}\label{crustproduction}
In this study, we are investigating the immediate distribution of crust due to melt, and in the case of a pre-impact crust, displaced primordial material. Due to limitations in direct simulation output quantities and resolution, this required the development of a new post-processing scheme to convert our results into a relevant format. 

{First, the SPH data is smoothed onto a uniform spherical grid, described in Section~\mbox{\ref{sphericalgrid}}. At each grid point, a melt fraction is then calculated via the simulation temperature and pressure values, as described in Section~\mbox{\ref{magma2crust}}. This therefore gives a mass of molten material for each cell. Through summation of these molten grid cell masses across all depths for each longitude and latitude coordinate on the grid, the total mass of melt can be calculated. By assuming that a fraction of this melt will crystallise to become crust, and by utilising reasonable density estimates for this crustal material, a crustal thickness is then calculated across the entire spherical surface. The precise crust fraction used for this scheme also takes primordial crust (and its associated mantle depletion) into account, which is calculated by tracking the displacement of the SPH particles that initially composed the surface of the Mars-like body prior to the impact. This aspect of the scheme is described in Section~\mbox{\ref{primcrust}}.}

\subsubsection{Spherical Grid}\label{sphericalgrid}
The mesh-free nature of SPH means that the coordinates of our simulation results follow those of the SPH particles and are thus distributed in an arbitrarily uneven fashion, which can be a hindrance for in-depth analysis. We therefore use Equation~\ref{sphEq} to smooth our results onto a spherical grid, uniformly-spaced in radius, latitude and longitude. Specifically, we use a resolution of $13.\dot{3}$ km in radius and $\frac{\pi}{100}$ in latitude and longitude. 

Simply smoothing all of our thermodynamic quantities in this manner would be inconsistent with our equation of state, however, as it is only internal energy and density that are calculated with SPH alone. We therefore mirror the simulation method; only smoothing these two values directly onto the grid, then using these as input to the ANEOS equation of state to calculate all other thermodynamic values such as pressure and temperature. One caveat of this method is ANEOS' inability to determine the physical properties of a mixture of two (or more) materials, only using one integer material number for each calculation. As a practical solution, we assign each SPH particle the mass fraction for each material (either a 1 or 0 as each particle is made up of only 1 material) and then use Equation~\ref{sphEq} again to calculate a continuous mass fraction at every point on our grid for each material. The material number corresponding to the largest mass fraction is then passed to ANEOS. {For further details on the results of the spherical grid compared to those of the SPH particles see Section~\mbox{\ref{grid_accuracy}}.}

\subsubsection{From Magma to Crust}\label{magma2crust}

To give a measure of melt fraction, $\xi$, at each grid point, we use the following expression \citep{Burg2005,Golabek2011}:
\begin{equation}
    \xi = 
    \begin{cases} 
          0 & T \leq T_{\textrm{sol}} \\
          \evalat{\frac{ T - T_{\textrm{sol}} }{ T_{\textrm{liq}} - T_{\textrm{sol}}}}{P} & T_{\textrm{sol}} < T < T_{\textrm{liq}} \\
          1 & T \geq T_{\textrm{liq}} 
    \end{cases}
    ,
\end{equation}
where $T$ is temperature and $T_{\textrm{sol}}$ and $T_{\textrm{liq}}$ are the solidus and liquidus temperatures corresponding to the pressure, $P$. $T_{\textrm{sol}}$ is calculated using the previously mentioned function in \cite{Duncan2018}, and $T_{\textrm{liq}}$ is found through the following linear relation for peridotite \citep{Wade2005}:
\begin{equation}
    T_{\textrm{liq}} = 1973 + 28.57 P,
\end{equation}
with $T_{\textrm{liq}}$ in K and $P$ in GPa. To account for the additional energy necessary for melting due to latent heat, a temperature offset of 400 K is also added to $T_{\textrm{liq}}$ \citep{Turcotte2002}.

If we assume a bulk mass fraction, $\beta$, of this molten material that will rise to the surface to crystallise and form basaltic to andesitic crust, we can calculate a crustal mass associated with each grid cell as follows:
\begin{equation}\label{mcrust}
    m_{\textrm{crust},ijk} = \min\left(\beta, \xi_{ijk} \right) \left(\rho V \right)_{ijk},
\end{equation}
where the $\min\left(\xi_{ijk},\beta \right)$ term is due to the crust-bearing material preferentially partitioning into the melt, $\rho$ is the density and $V$ is the volume of the grid cell given by:
\begin{equation}
    V_{ijk} = \int_{\phi_{k-\frac{1}{2}}}^{\phi_{k+\frac{1}{2}}} \int_{\theta_{j-\frac{1}{2}}}^{\theta_{j+\frac{1}{2}}} \int_{r_{i-\frac{1}{2}}}^{r_{i+\frac{1}{2}}} r^2 \sin{\theta}~\mathrm{d}r \, \mathrm{d}\theta \, \mathrm{d}\phi,
\end{equation}
which when evaluated becomes:
\begin{equation}\label{Vi}
    V_{ijk} = \frac{\Delta\phi}{3} \left[ r_{i+\frac{1}{2}}^3 - r_{i-\frac{1}{2}}^3 \right] \left[ \cos{\theta_{j-\frac{1}{2}}} - \cos{\theta_{j+\frac{1}{2}}} \right] 
    ,
\end{equation}
where the indices $(i,j,k) - \frac{1}{2}$ and $(i,j,k) + \frac{1}{2}$ indicate the values at the lower and upper boundaries of the grid cell, respectively ($(i,j,k)$ represent the grid cell centres), and $\Delta\phi = \phi_{k+\frac{1}{2}} - \phi_{k-\frac{1}{2}}$, which is constant due to our grids having equal spacing.

{As magma is more compressible than solid rock, it experiences negative buoyancy at high pressures. For Mars, laboratory experiments indicate that this occurs at pressures greater than 7.4 GPa, or roughly 600 km depth} \citep{Ohtani1998,Suzuki1998}. We therefore neglect any melt at such pressures in our crust production model. In addition, we assume that melt cannot be extracted at melt fractions below 4\%. Significantly lower thresholds have been assumed in various earlier studies \citep[e.g.][]{Fraeman2010,Ruedas2013,Ruedas2021}; however, we follow the \cite{Citron2018} train of thought in that Mars should have a higher melt extraction threshold than Earth due to its lower gravity (and thus weakened buoyancy forces), where values predicted for Earth lie in the range 1-4\% \citep[][and references therein]{Wang2021}.

{For melt above these thresholds, we make the assumption that the material will rise radially upward (i.e. at a constant latitude and longitude) meaning that the associated crustal masses calculated from Equation~\mbox{\ref{mcrust}} can be summed up to give a total crustal mass, $M_{\textrm{crust},jk}$, for each latitude and longitude. Buoyant melt is known to propagate to the surface on much faster timescales than rock deforms \mbox{\citep{Condomines1988}}, and thus mantle convection should not affect this assumption. The magma pond that forms at the surface would most likely crystallise rapidly, within 0.1--1 Myr \mbox{\citep{Solomatov2007}}, again on a faster timescale than mantle convection. This simple geophysical scheme should therefore give a reasonable estimate for the distribution of crust directly after the crystallisation of the impact-induced melt, albeit with some shortcomings discussed in Section~\mbox{\ref{discussion}} and Section~\mbox{\ref{outlook}}.}

{To convert these total crustal masses to a crustal thickness}, we assume a crustal density, $\rho_{\textrm{crust}}$, which we can divide $M_{\textrm{crust},jk}$ by to give us a crustal volume, $V_{\textrm{crust},jk}$, corresponding to a crustal thickness given by:
\begin{equation}\label{zcrust}
    z_{\textrm{crust},jk} = \left[ R_{\mars}^3 - \frac{3V_{\textrm{crust},jk}}{\Delta\phi\left(\cos{\theta_{j-\frac{1}{2}}} - \cos{\theta_{j+\frac{1}{2}}}\right)} \right]^\frac{1}{3}
    ,
\end{equation}
where $R_{\mars}$ is the radius of Mars. Note that this is simply a rearrangement of Equation \ref{Vi}, with $r_{i+\frac{1}{2}}$ and $r_{i-\frac{1}{2}}$ becoming $z_{\textrm{crust},jk}$ and $R_{\mars}$, respectively, and $V_{ijk}$ becoming $V_{\textrm{crust},jk}$.

\subsubsection{Primordial Crust}\label{primcrust}
Unfortunately, current computational limitations prohibit the inclusion of a fully-resolved crustal layer in our simulations, so we must implement further post-processing approaches to quantitatively estimate its influence. Firstly, we identify the uppermost SPH particles in the original, pre-impact target body that represent the surface. As the primordial crustal thicknesses of interest are thinner than an SPH smoothing length, we approximate the mass fraction of crust, $\beta_{\textrm{prim}}$, for a given SPH particle based on its radial position relative to the mantle-crust boundary, $R_{\textrm{MCB}}$. SPH particles represent a mass that has been smeared-out radially to follow the smoothing kernel, and thus the precise 3-dimensional calculation of this crustal mass fraction is highly non-trivial. To avoid this, while still retaining good accuracy, we use the integrated 1-dimensional kernel, $W'_{\textrm{1D}}(r,h)$, given in \ref{bsplineprime}, as follows:
\begin{equation} \label{betaprim}
    \beta_{\textrm{prim}} = 
    \begin{cases} 
          \frac{1}{2} - W'_{\textrm{1D}}(|r - R_{\textrm{MCB}}|, h) & r \leq R_{\textrm{MCB}} \\
          \frac{1}{2} + W'_{\textrm{1D}}(|r - R_{\textrm{MCB}}|, h) & r > R_{\textrm{MCB}}
    \end{cases}
    ,
\end{equation}
where the $\frac{1}{2}$ term comes from the symmetric nature of $W'_{\textrm{1D}}$.

The presence of a primordial crust infers a depleted mantle that will no longer produce as much crust if melted again. To include this effect, we calculate the global crustal mass $M_{\textrm{crust,global}}$ through the summation of all previously calculated $M_{\textrm{crust},jk}$ values, and then remove this mass from the initially fertile mantle in one of two end-member scenarios. 

In the first case, we assume that the entire mantle has had time to fully mix, meaning that our $\beta$ in Equation~\ref{mcrust}, which from here onward we will call the ``fertility'', is still constant across the mantle, but is decreased to match the mass of fertile mantle required to produce $M_{\textrm{crust,global}}$, i.e. :
\begin{equation}
    \beta_{\textrm{dep}} = 
    \left(1-\frac{M_{\textrm{crust,global}}}{\beta_0 M_{mantle}}\right)\beta_0
    ,
\end{equation}
where $M_{mantle}$ is the mass of the mantle and $\beta_0$ is the initial mantle fertility prior to primordial crust formation. If we assume that all primordial crust mass will form new crust upon remelting, we therefore redefine $\beta$ for each particle as:

\begin{equation}\label{betamixed}
    \beta = 
    \begin{cases} 
          \beta_{\textrm{dep}} & r \leq R_{\textrm{MCB}} - 2h \\
          \beta_{\textrm{dep}} (1 - \beta_{\textrm{prim}}) + \beta_{\textrm{prim}} & R_{\textrm{MCB}} - 2h < r < R_{\textrm{MCB}} + 2h \\
          1 & r \geq R_{\textrm{MCB}} + 2h
    \end{cases}
    .
\end{equation}

The second case corresponds to a fully stratified mantle, whereby the mantle directly below the crust is entirely depleted ($\beta=0$), eventually reaching untouched, fertile mantle at depth ($\beta=\beta_0$). The exact radial position of this fertile-depleted boundary is determined again using $M_{\textrm{crust,global}}$, instead converting this into a volume of depleted mantle, $V_{\textrm{dep}}$, that is used in a modified form of Equation~\ref{zcrust} given by:
\begin{equation}
    R_{\textrm{FDB}} = \left[ R_{\textrm{MCB}}^3 - \frac{3V_{\textrm{dep}}}{4\pi} \right]^\frac{1}{3}
    ,
\end{equation}
with $V_{\textrm{dep}}$ given by:
\begin{equation}
    V_{\textrm{dep}} = \frac{M_{\textrm{crust,global}}}{\beta_0 \overline{\rho}},
\end{equation}
where $\overline{\rho}$ is the mean density of the mantle. We introduce this boundary using a similar method to Equation~\ref{betaprim}, now including both the fertile-depleted and mantle-crust boundaries as follows:
\begin{equation}\label{betastrat}
    \beta = 
    \begin{cases} 
          \beta_0 \left(\frac{1}{2} + W'_{\textrm{1D}}(|r - R_{\textrm{FDB}}|, h) \right) + \beta_{\textrm{prim}} & r < R_{\textrm{FDB}} \\
          \beta_0 \left(\frac{1}{2} - W'_{\textrm{1D}}(|r - R_{\textrm{FDB}}|, h) \right) + \beta_{\textrm{prim}} & R_{\textrm{FDB}} \leq r \leq R_{\textrm{FDB}} + 2h \\
          \beta_{\textrm{prim}} & r > R_{\textrm{FDB}} + 2h
    \end{cases}
    .
\end{equation}

Another important aspect of a pre-impact crust is the contribution to crustal thickness by primordial crustal material that has remained solid throughout the giant impact event (or has re-solidified due to a change in pressure). This material is included in our crustal calculations by simply adding the solid crust contribution to our grid cell crustal masses as follows:
\begin{equation}
  m_{\textrm{crust},ijk} = \Bigg\{ \left[\min\left(\beta, \xi \right)
  + \beta_{\textrm{prim}} \left(1 - \frac{\min(\beta,\xi)}{\beta}\right)\right]  \rho V \Bigg\}_{ijk}
  ,
\end{equation}
where the first term represents the molten material that will solidify into crust and the second term represents the intact primordial crust that is in the solid phase at the end of the simulation.

Clearly, this primordial crust scheme relies upon a homogeneous SPH particle distribution, as small fluctuations in position (relative to a given particle's smoothing length) can have a dramatic effect on a particle's primordial crust fraction, as is clear from equation~\ref{betaprim}. Our pre-impact bodies are initially composed of a Hexagonal Close-Packed (HCP) lattice particle structure that undergoes a relaxation step before being used in the full impact simulation. Unfortunately, this leaves some remnant heterogeneities in the particle distribution of purely numerical origin, which in turn influences the initial primordial crust distribution.\footnote{More recently proposed initial particle distributions such as those described in \cite{Reinhardt2017} or \cite{Diehl2015} may lead to smaller numerical artefacts in the particle distribution.} To counteract these artefacts, we normalise the particles' radial positions used for the primordial crust calculations such that the maximum value in a given latitude-longitude range is equal to the global maximum value. The size of these longitude-latitude ranges is given by the equal-area iso-latitudinal grid of \cite{Malkin2019} as to keep the resolution of this normalisation step constant across the entire body.

\section{Parameter Study} \label{setup}
The chosen parameters for our impact simulations are as follows: impact angles of 0-90\degree~in steps of 15\degree~(with 0\degree~being the head-on case); impact velocities of 1.0-1.4 $v_{\textrm{esc}}$ ($v_{\textrm{esc}}$ is the mutual escape speed of the two bodies given by $v_{\textrm{esc}} = \sqrt{\frac{2G(m_{\textrm{imp}} + m_{\textrm{imp}})}{r_{\textrm{imp}} + r_{\textrm{imp}}}}$, where $G$ is the gravitational constant, $m$ is mass, $r$ is radius and the subscripts $_{\textrm{imp}}$ and $_{\textrm{tar}}$ signify the impactor and target, respectively); and impactor radii of 500 km, 750 km, 1000 km, 1500 km and 2000 km. Each simulation consists of 200,000 SPH particles in total, with near head-on impacts (angles $<$ 45\degree) simulated for 50 hours post-impact and oblique impacts (angles $\geq$ 45\degree) simulated for 200 hours as to allow for any potential secondary impacts. Material strength is included in all cases; however, select cases were repeated without this feature for comparison.

Both impactor and target are modelled as differentiated bodies with an iron core and dunite mantle, using the input parameters of \cite{Emsenhuber2018} and \cite{Benz1989}, respectively. As the impactor's internal structure is unknown, we investigate a core mass fraction of approximately 25\% and 50\% for each set of parameters. For our Mars-like body we use a core mass fraction of $\sim$25\%, consistent with geodetic and seismic measurements \citep{Rivoldini2011,Stahler2021}. The internal temperature profiles of the bodies approximately match the \cite{Duncan2018} solidus at the surface, as this is the most likely scenario at early times after formation, with a $\sim$150 K temperature jump at the core-mantle boundary of our Mars-like body as proposed by \cite{Williams2004}. For the crustal density of Mars, $\rho_{\textrm{crust}}$, used in the crust production scheme, we take a value of 2582 kg m$^{-3}$ from \cite{Goossens2017} which matches with recent observational constraints \citep{Knapmeyer-Endrun2021}. {For the sake of simplicity, neither the target nor the impactor possess any pre-impact rotation.}

{For the primordial crust calculations of our Mars-like body,} we investigate initial fertilities ($\beta_0$) of 0.1-0.25 in steps of 0.025, and crust thicknesses of 10-70 km in steps of 15 km. {For each of these cases we investigate both the fully-stratified and fully-mixed depletion schemes.} For the sake of simplicity, the impactor fertility was taken to be $0.2$ in all cases, and any impactor crust was neglected.

{Considering all of these parameters, the total number of SPH impact simulations (other than repeated cases without material strength) is therefore 210 (i.e. 5 impactor sizes $\times$ 7 impact angles $\times$ 3 impact velocities $\times$ 2 impactor core sizes), and the total number of cases after applying the post-production primordial crust scheme is 14,700 (i.e. 210 impact simulations $\times$ 7 initial fertilities $\times$ 5 primordial crust thicknesses $\times$ 2 depletion schemes).}

\section{Results}\label{results}
\subsection{``Classical'' Cases}\label{classical}
For the sake of comparison to previous studies we will first focus upon the ``classical'' cases: a 1000 km-radius impactor with a 25\% core fraction impacting at the mutual escape velocity at either 0\degree or 45\degree.

\begin{figure}[htbp]
    \centering
    \begin{subfigure}{.49\textwidth}
        \centering
        \textbf{\large Fluid}
        \adjincludegraphics[trim={{.15\width} 0 {.06\width} 0},clip,width=\linewidth]{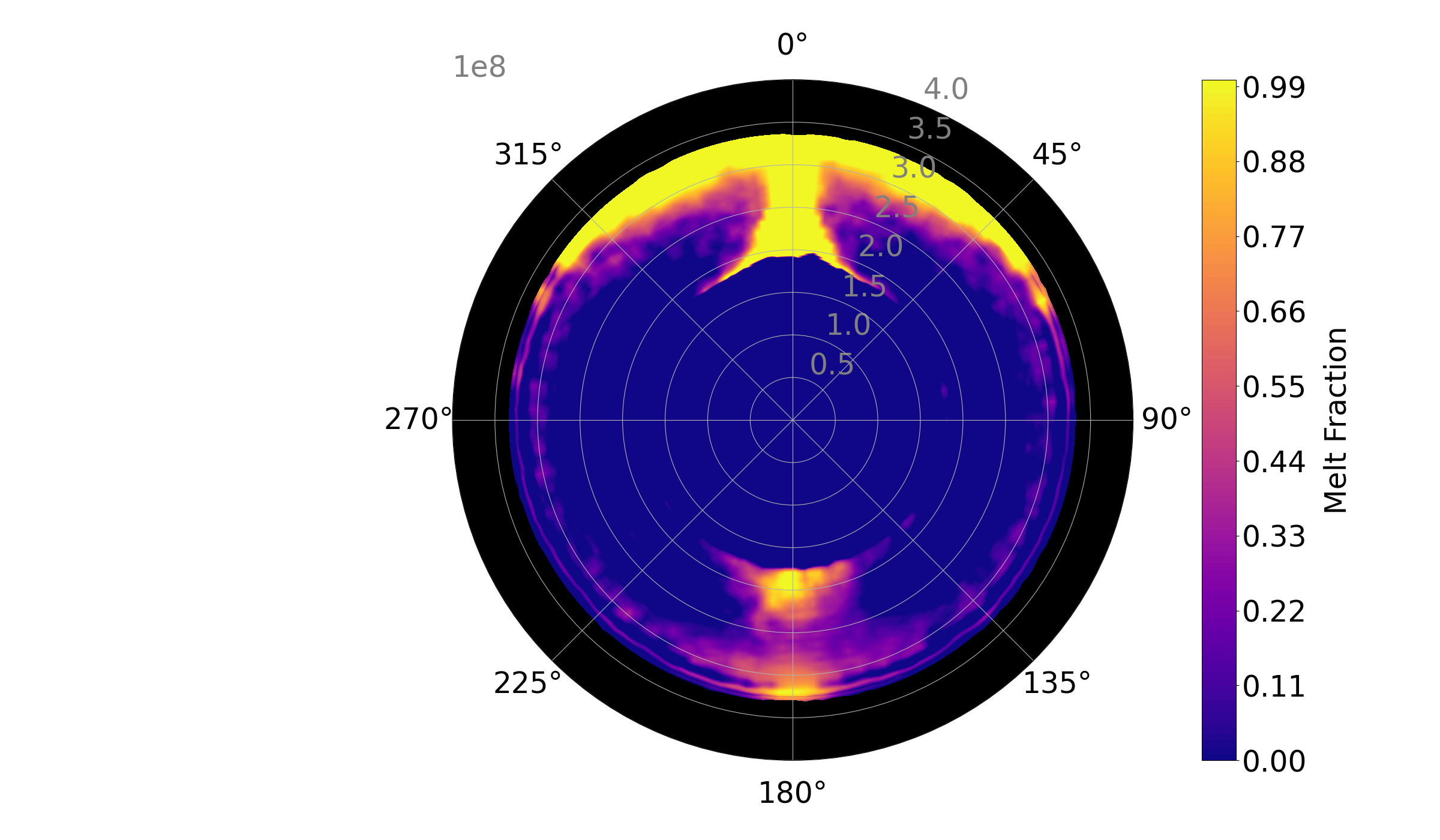}
    \end{subfigure}
    \begin{subfigure}{.49\textwidth}
        \centering
        \textbf{\large Solid}
        \adjincludegraphics[trim={{.15\width} 0 {.06\width} 0},clip,width=\linewidth]{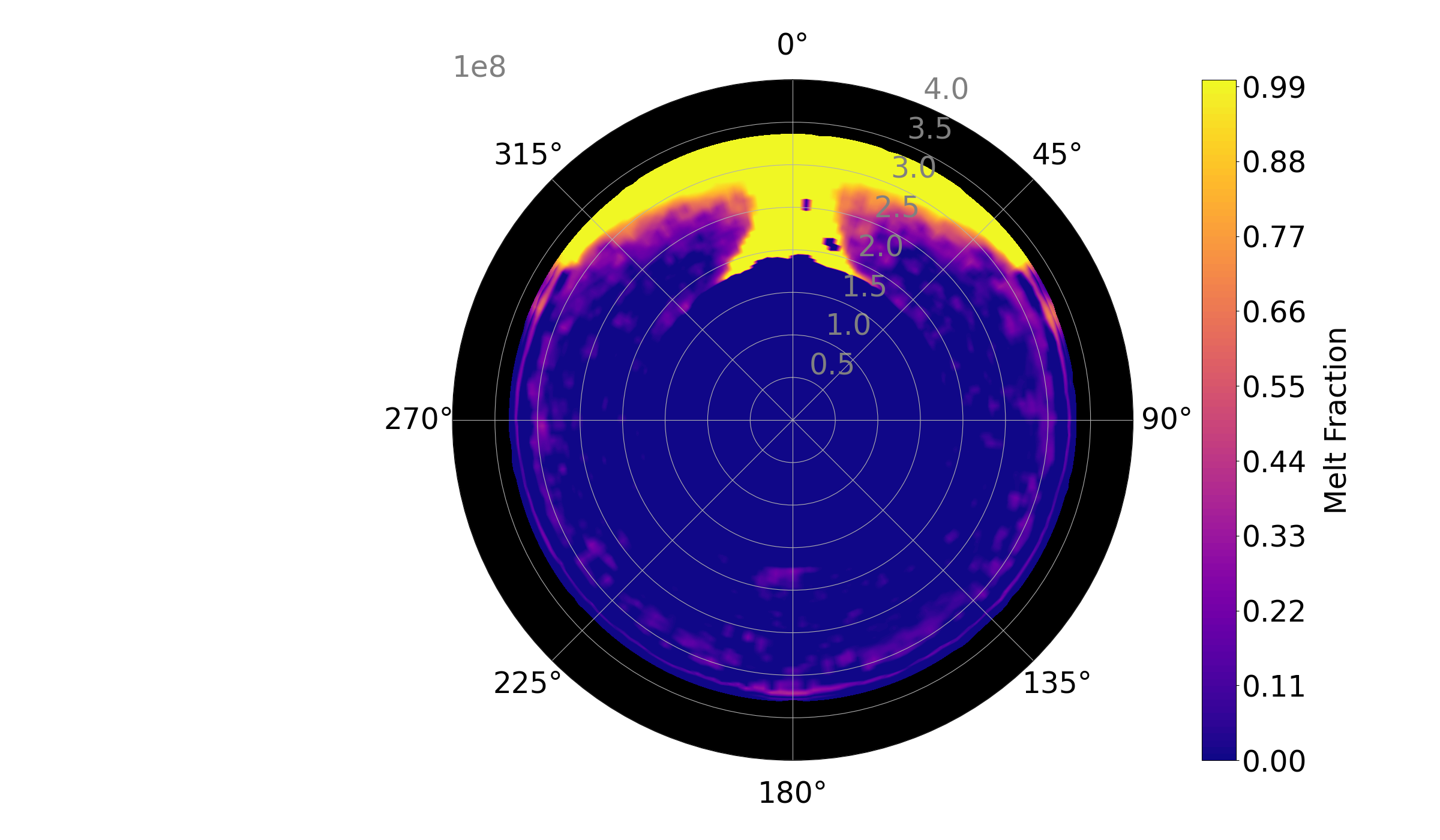}
    \end{subfigure}
    \begin{subfigure}{.49\textwidth}
        \adjincludegraphics[trim={{.15\width} 0 {.06\width} 0},clip,width=\linewidth]{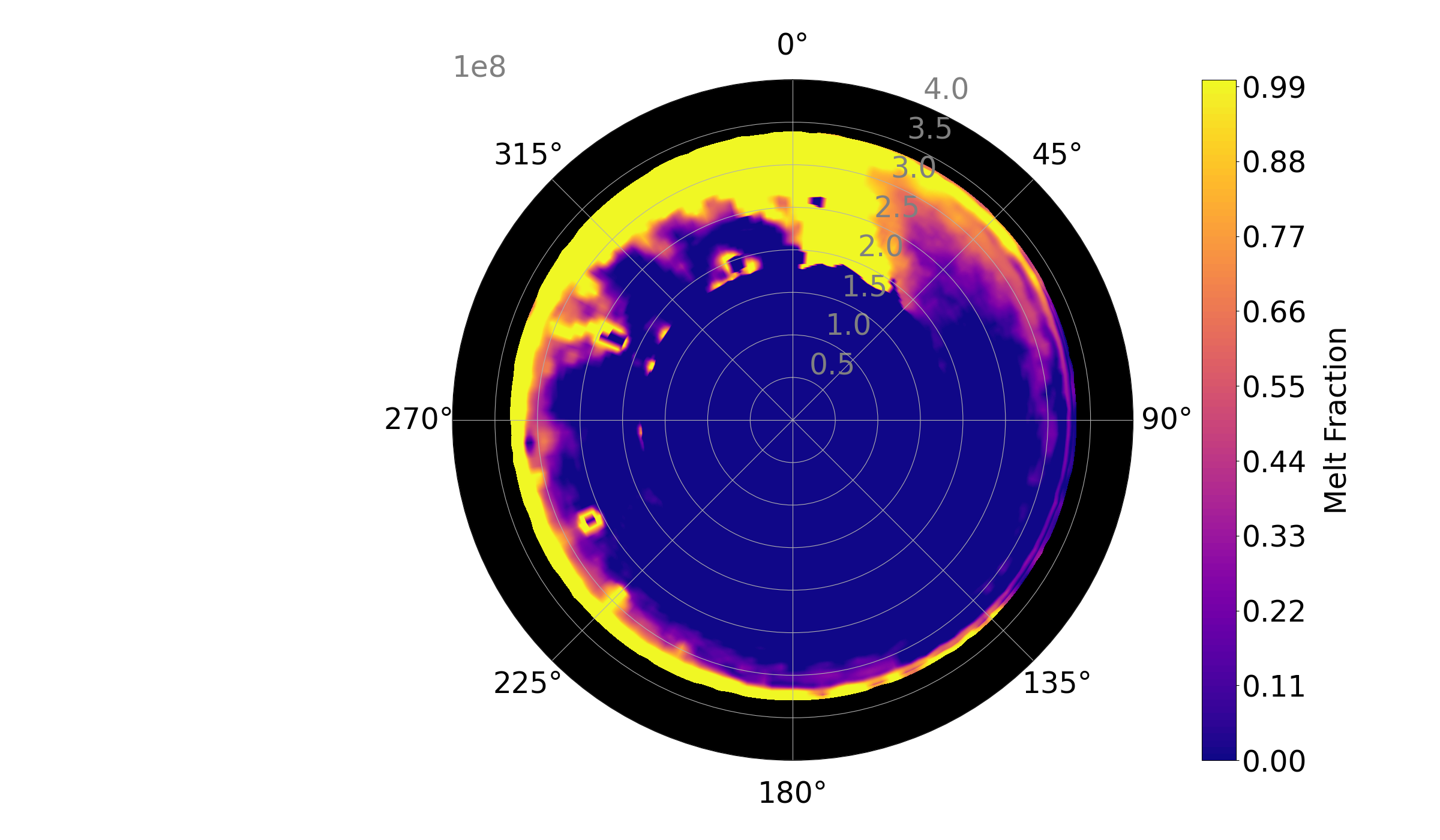}
    \end{subfigure}
    \begin{subfigure}{.49\textwidth}
        \adjincludegraphics[trim={{.15\width} 0 {.06\width} 0},clip,width=\linewidth]{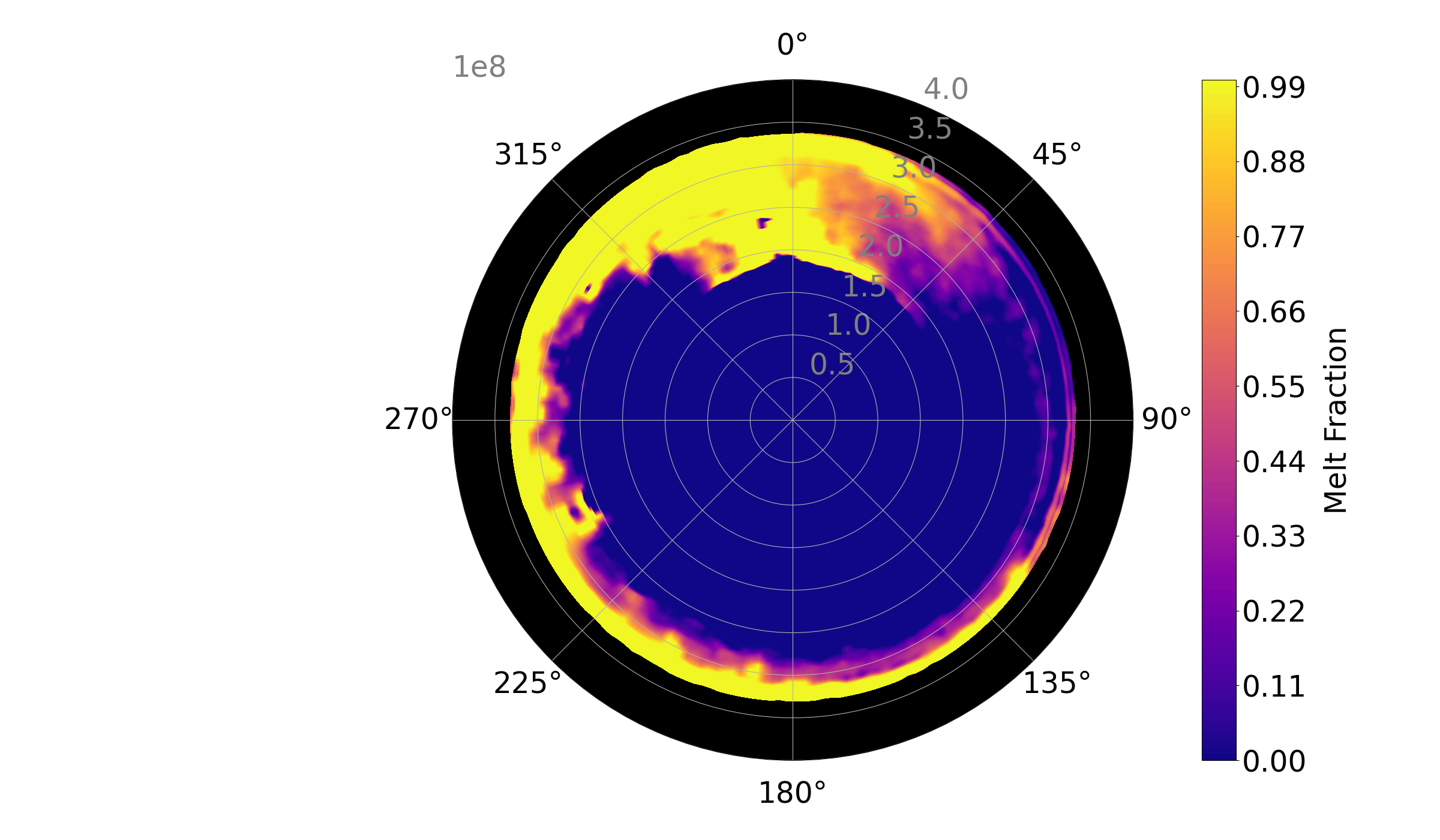}
    \end{subfigure}
    \caption{Melt fraction distribution for each of the classical cases, with (right) and without (left) solid strength, where a slice has been made in the impact plane to display the interior. The top row shows the head-on cases, whereas the bottom row shows the $45\degree$ cases. Note that the oblique impact induces a rotation in the anticlockwise direction.}
    \label{fig:fluidcases}
\end{figure}

These cases were performed both with and without solid strength. As can be seen in Figure~\ref{fig:fluidcases}, strength clearly plays an important role in this impact regime, effectively acting to dissipate the impact-generated shockwave through increased frictional heating. This effect is most apparent in the head-on case, where the strength-less model shows a significant antipodal feature that is not present at all when strength is included.

\begin{figure}[htbp]
    \centering
    \begin{subfigure}{.49\textwidth}
        \adjincludegraphics[width=\linewidth]{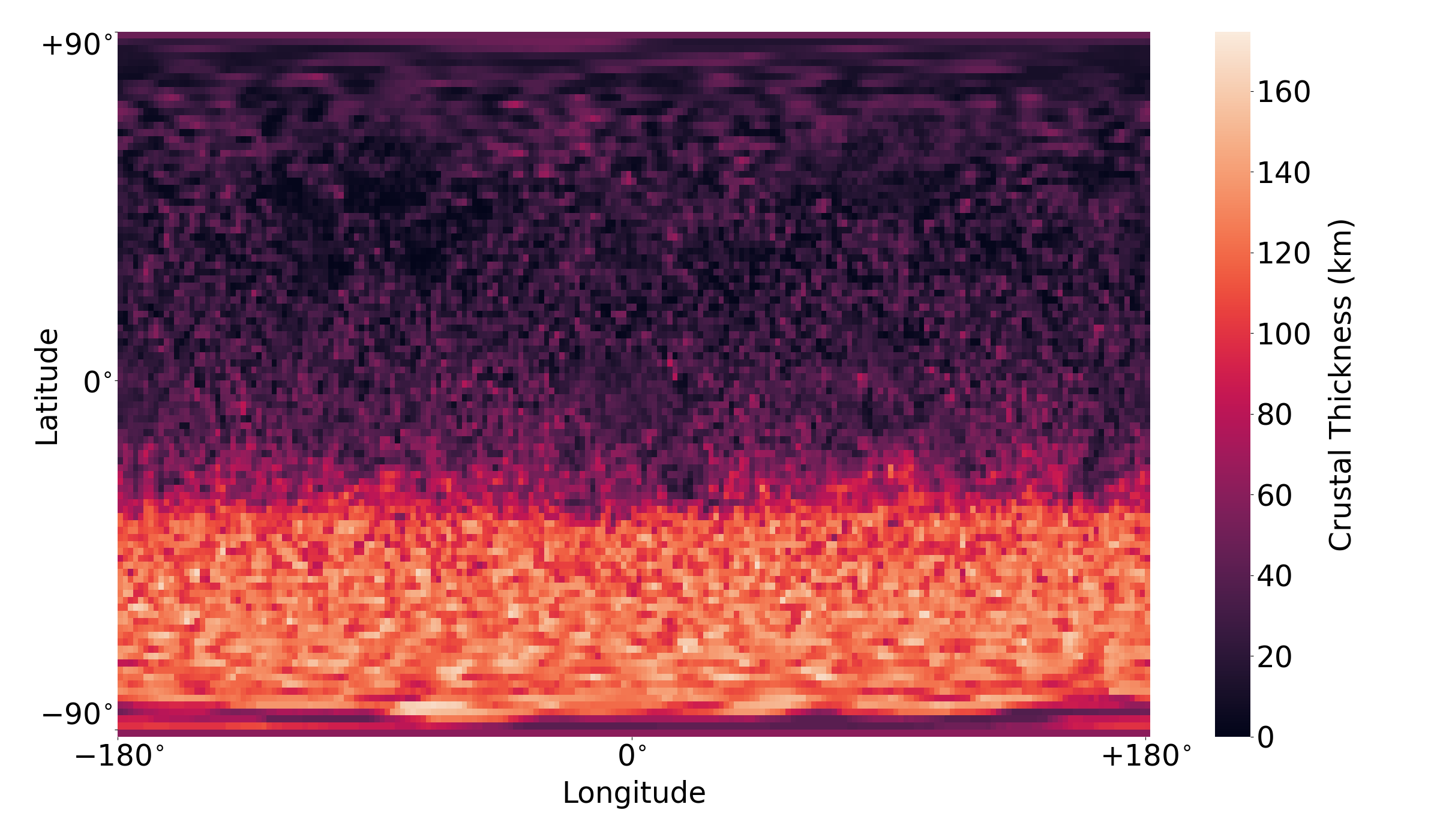}
    \end{subfigure}
    \begin{subfigure}{.49\textwidth}
        \adjincludegraphics[width=\linewidth]{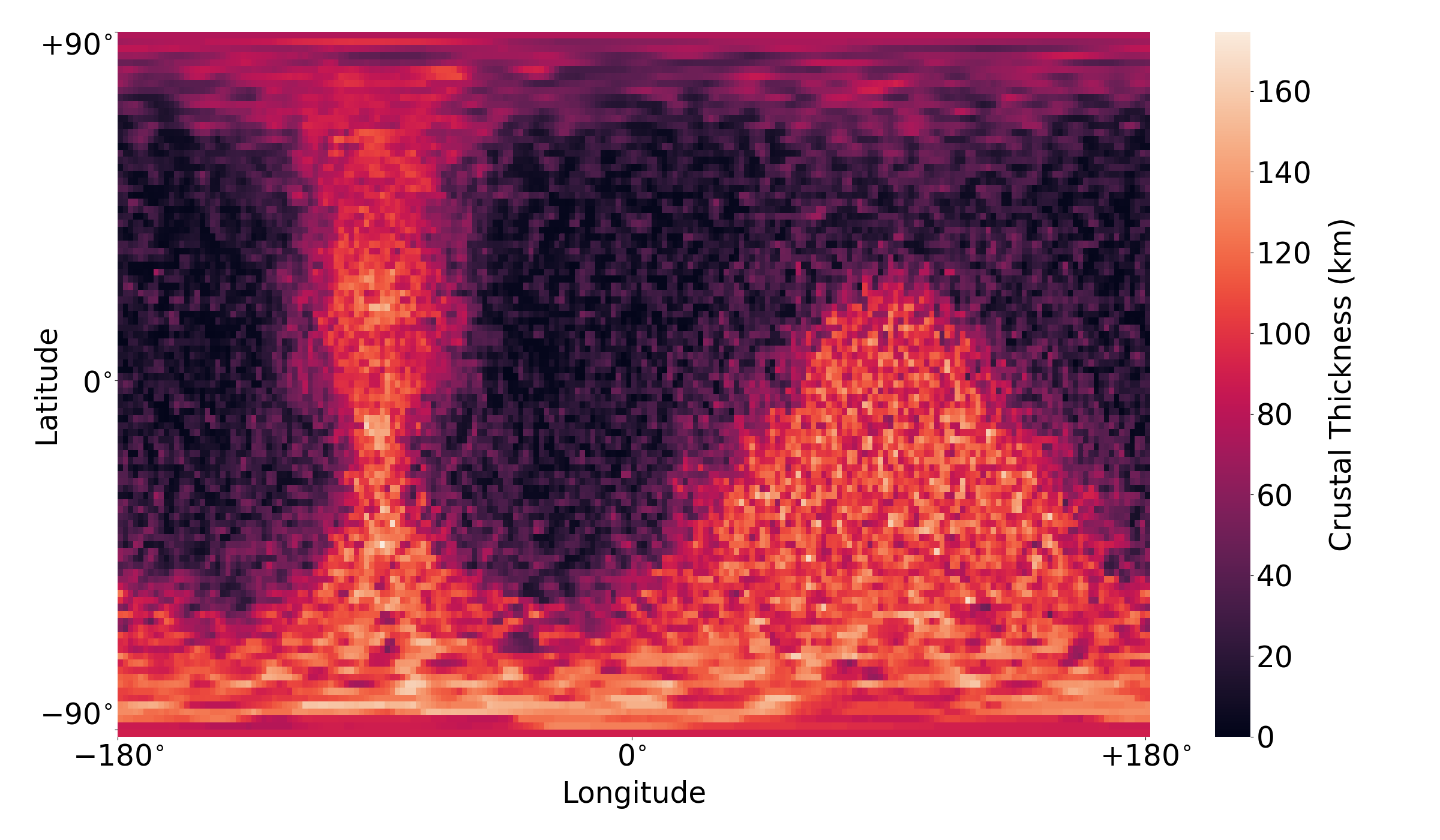}
    \end{subfigure}
    \caption{Crustal thickness distributions without any primordial crust calculation for the head-on (left) and 45\degree (right) classical cases, with an initial mantle fertility of 0.2. The simple equirectangular projection has been used for this plot, as well as all other similar plots in this paper. The 45\degree case displays an iso-longitudinal ring of crust-production at roughly -90\degree longitude due to re-impacting ejecta from the initial impact.}
    \label{fig:classicalmaps_nocrust}
\end{figure}

\begin{figure}[htbp]
\centering
\includegraphics[width=\textwidth]{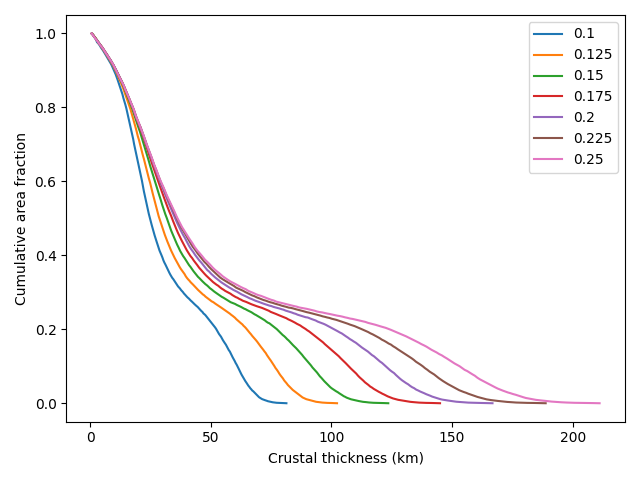}
\caption{Cumulative fractional area as a function of crustal thickness for the classical head-on case without any primordial crust, where each colour denotes a different initial mantle fertility.}
\label{fig:cumsum}
\end{figure}

The corresponding crustal thickness maps for the cases including material strength (calculated using the method of Section~\ref{magma2crust}) can be seen in Figure~\ref{fig:classicalmaps_nocrust}, where no primordial crust has been included and an initial mantle fertility of 0.2 has been used. While the head-on case may have a much clearer hemispherical distribution than the oblique impact, the most overwhelming feature of both scenarios is an average crustal thickening that is significantly too large when compared to Mars' true hemispherical contrast of $\sim$25 km. This effect is still apparent even when the initial mantle fertility is reduced to 0.1, where the majority of the impact-induced crust has a thickness of approximately 60 km (see Figure~\ref{fig:cumsum}).

\begin{figure}[htbp]
    \centering
    \begin{subfigure}{.49\textwidth}
        \adjincludegraphics[width=\linewidth]{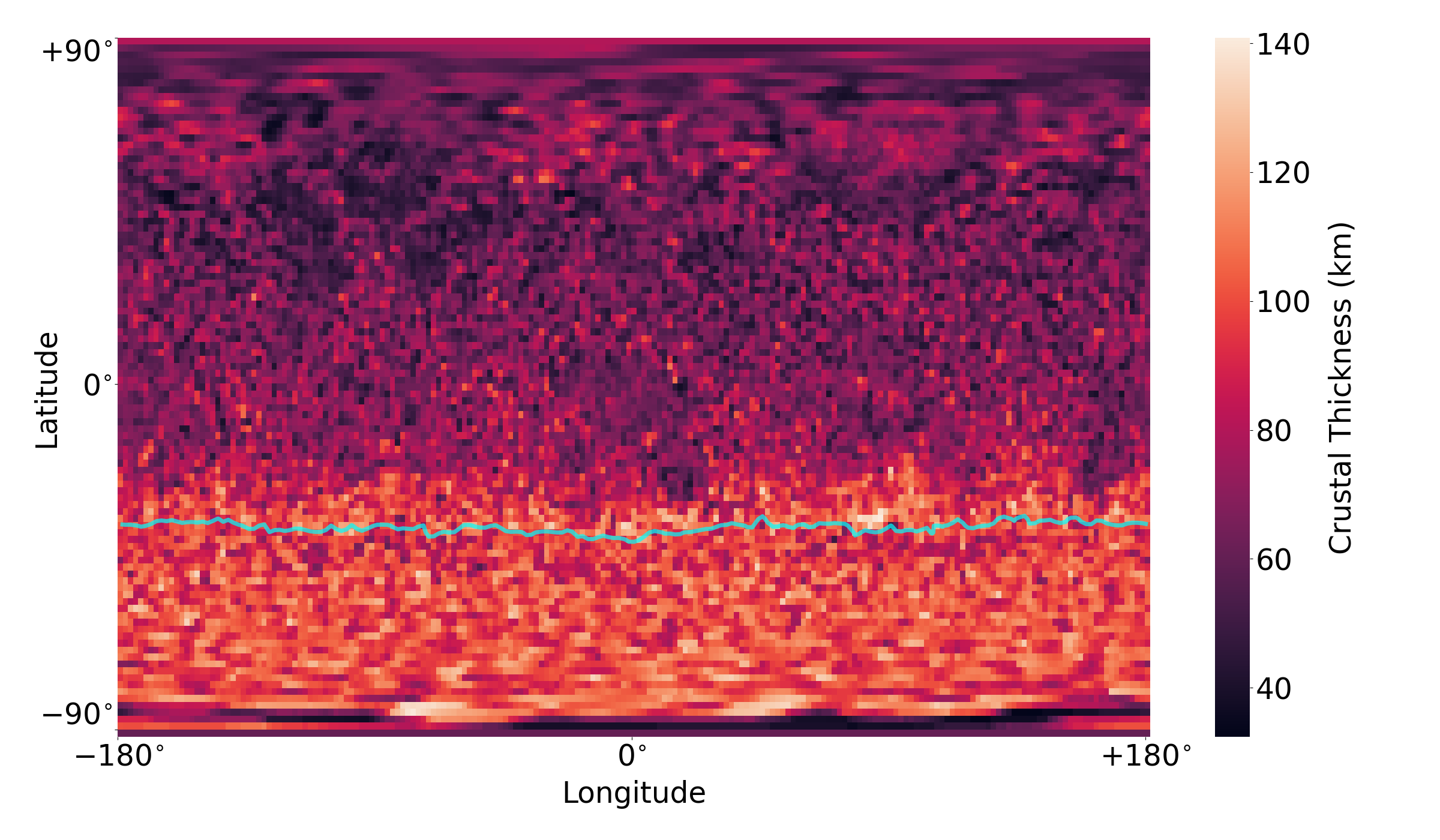}
    \end{subfigure}
    \begin{subfigure}{.49\textwidth}
        \adjincludegraphics[width=\linewidth]{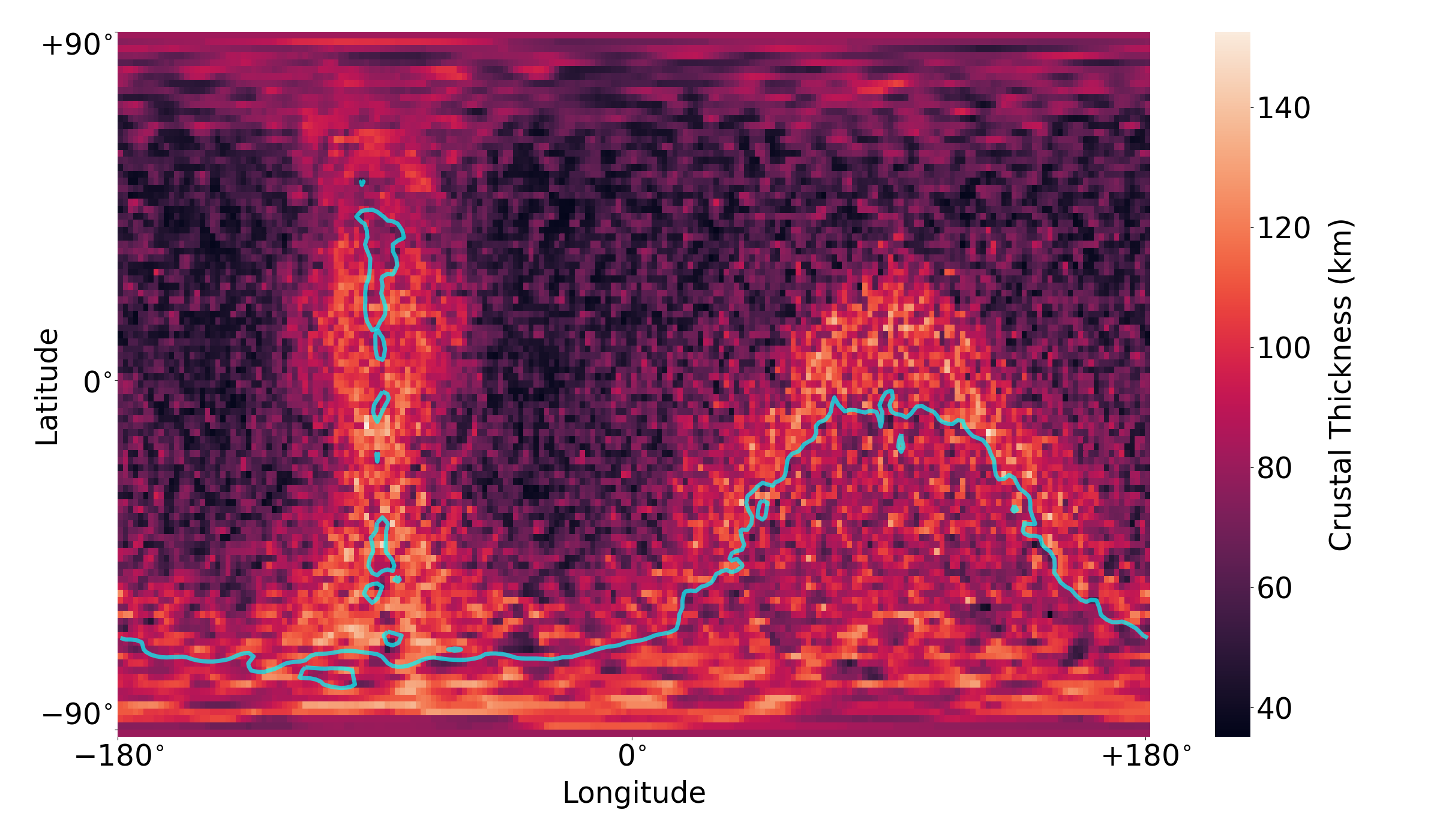}
    \end{subfigure}
    \begin{subfigure}{.49\textwidth}
        \adjincludegraphics[width=\linewidth]{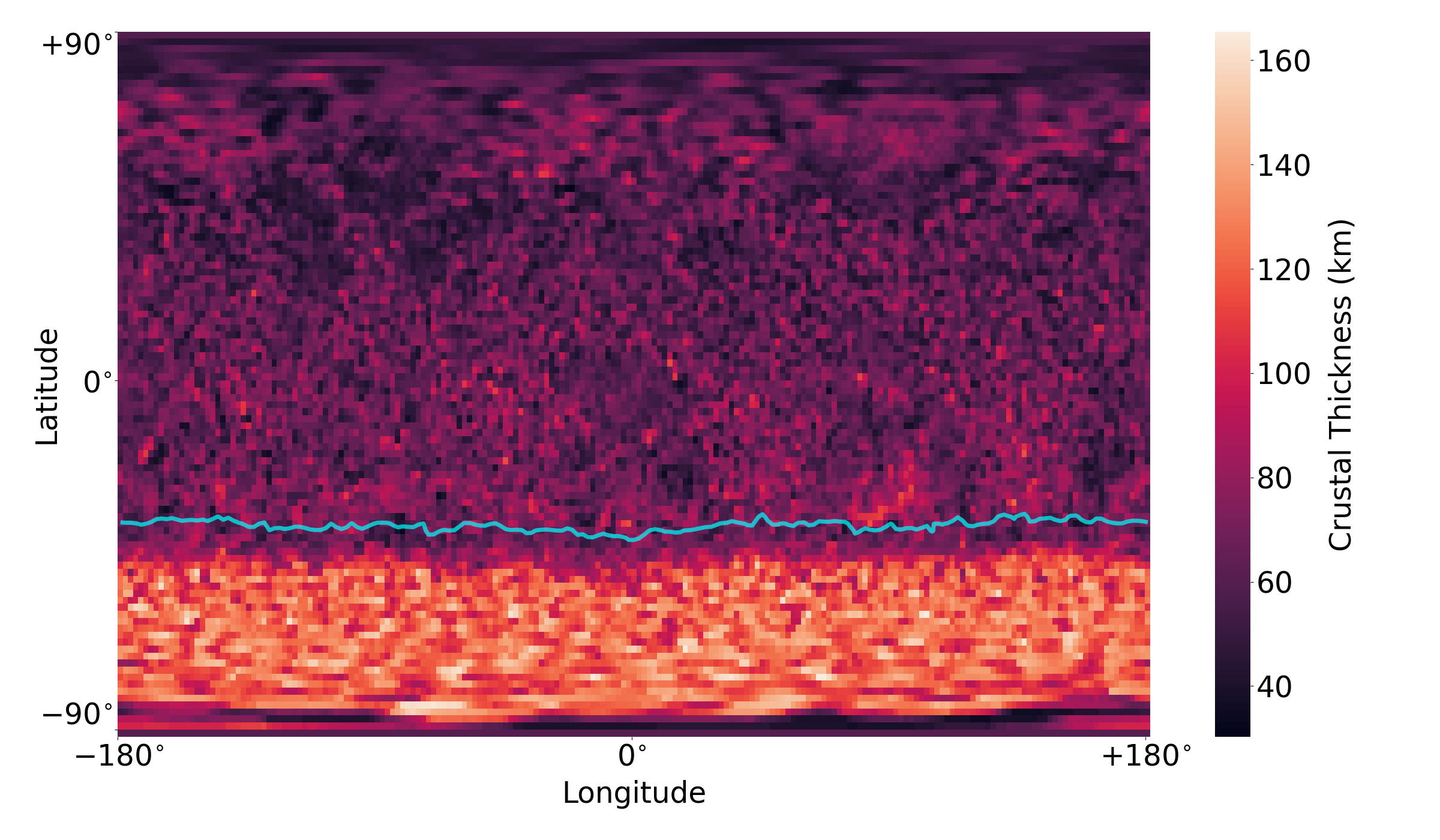}
    \end{subfigure}
    \begin{subfigure}{.49\textwidth}
        \adjincludegraphics[width=\linewidth]{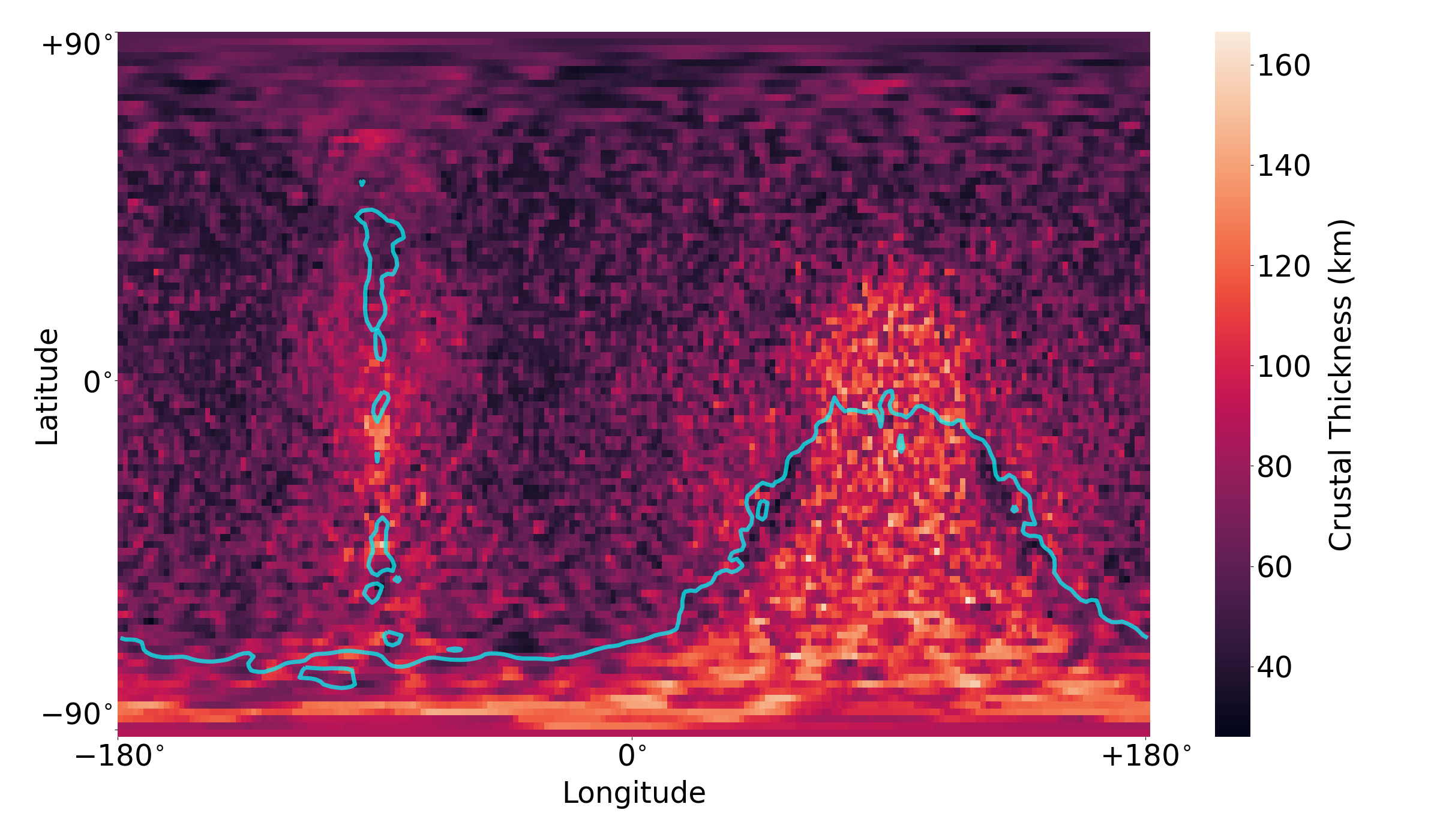}
    \end{subfigure}
    \caption{Crustal thickness distributions for the head-on (left) and 45\degree (right) classical cases, with a 40 km primordial crust and initial mantle fertility of 0.2. The upper and lower panels correspond to the mixed and stratified mantle models, respectively. The cyan contour line indicates $\sim$25\% of primordial crust remaining, and displays the approximate transition at which primordial crust begins to contribute significantly to the post-impact crust distribution. The crust-production outside (north) of this contour is a result of a crossover region where significant melting occurs but primordial crust still remains, leading to a ridge of particularly thickened crust at the perimeter of the impact site. This effect is particularly visible in the 45\degree cases.}
    \label{fig:classicalmaps_withcrust}
\end{figure}

As described in Section~\ref{primcrust}, any primordial crust acts to deplete the Martian mantle of incompatible elements, making any remelting less efficient in producing crust. In addition, the primordial crust originally located at the impact site is excavated, with the majority being redistributed around the planet (see Appendix~\ref{fig:classicalmaps_onlycrust}). Both of these effects lead to a reduced final thickness of crust at the impact site. Figure~\ref{fig:classicalmaps_withcrust} illustrates the result of these effects for the same impact cases as earlier through the inclusion of a 40 km pre-impact, primordial crust, for each of the mantle depletion models. 

\begin{figure}[htbp]
    \centering
    \begin{subfigure}{.49\textwidth}
        \centering
        \adjincludegraphics[trim={{.15\width} 0 {.05\width} 0},clip,width=\linewidth]{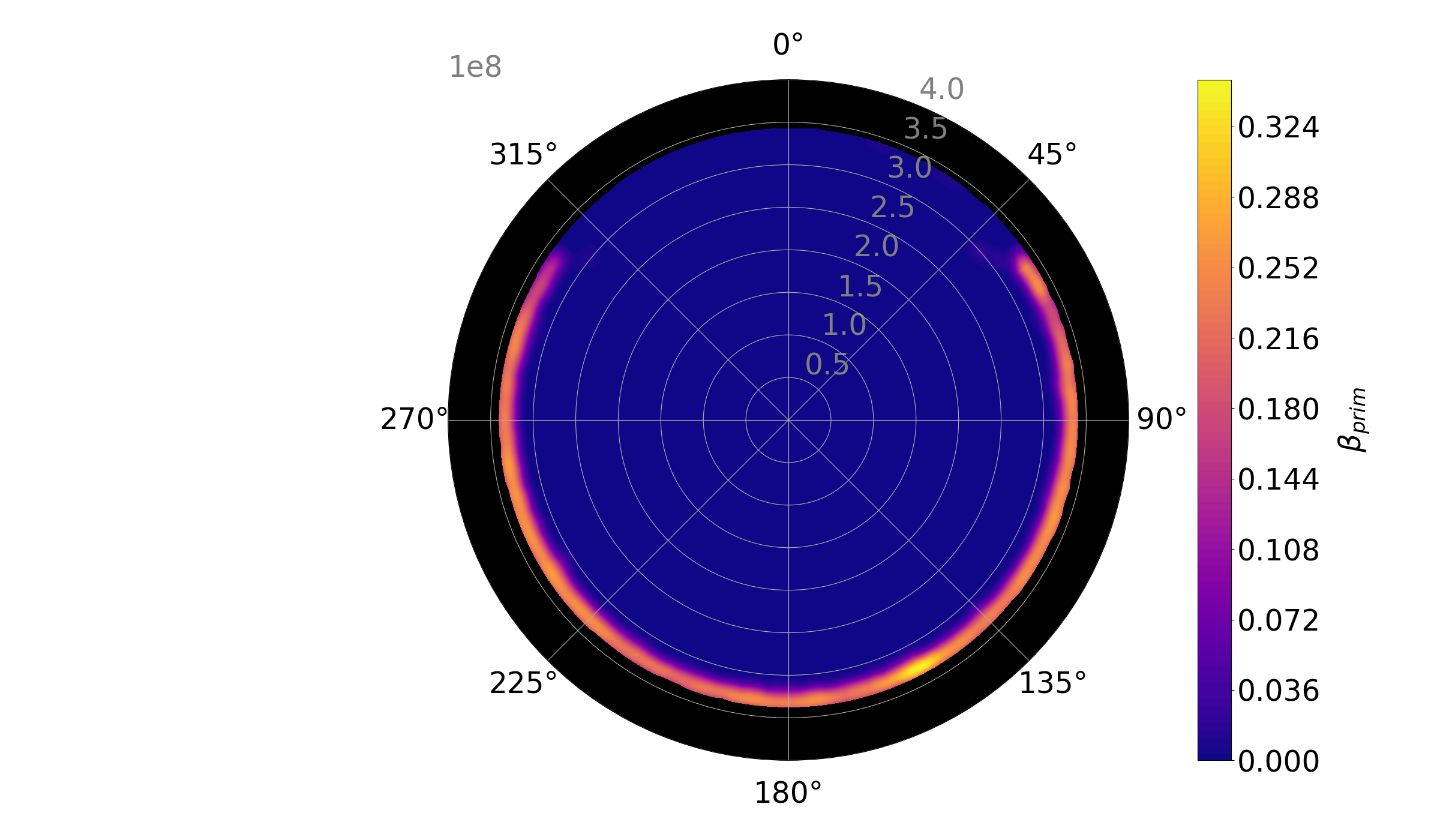}
    \end{subfigure}
    \begin{subfigure}{.49\textwidth}
        \centering
        \adjincludegraphics[trim={{.15\width} 0 {.05\width} 0},clip,width=\linewidth]{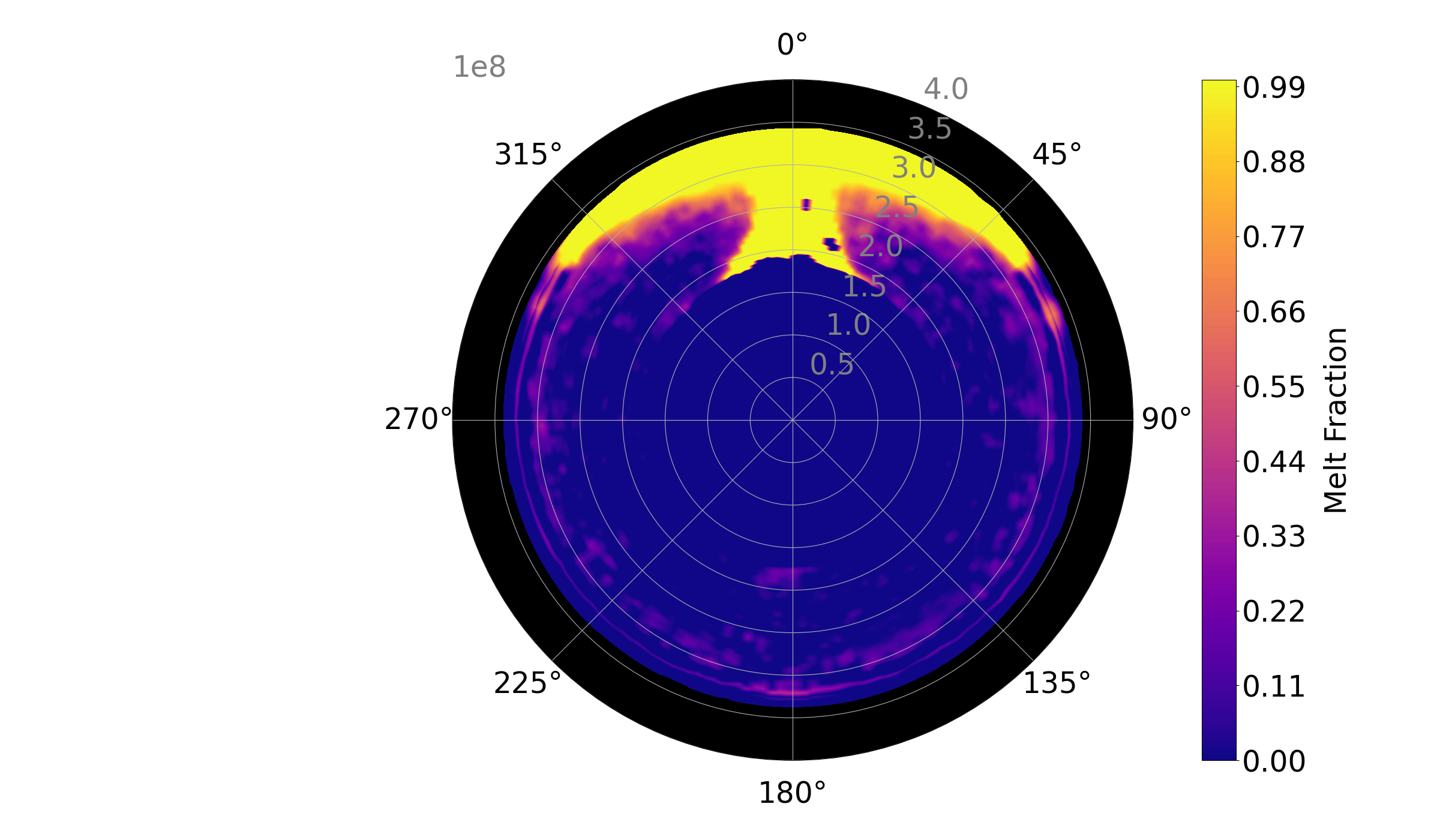}
    \end{subfigure}
    \begin{subfigure}{.49\textwidth}
        \adjincludegraphics[trim={{.15\width} 0 {.05\width} 0},clip,width=\linewidth]{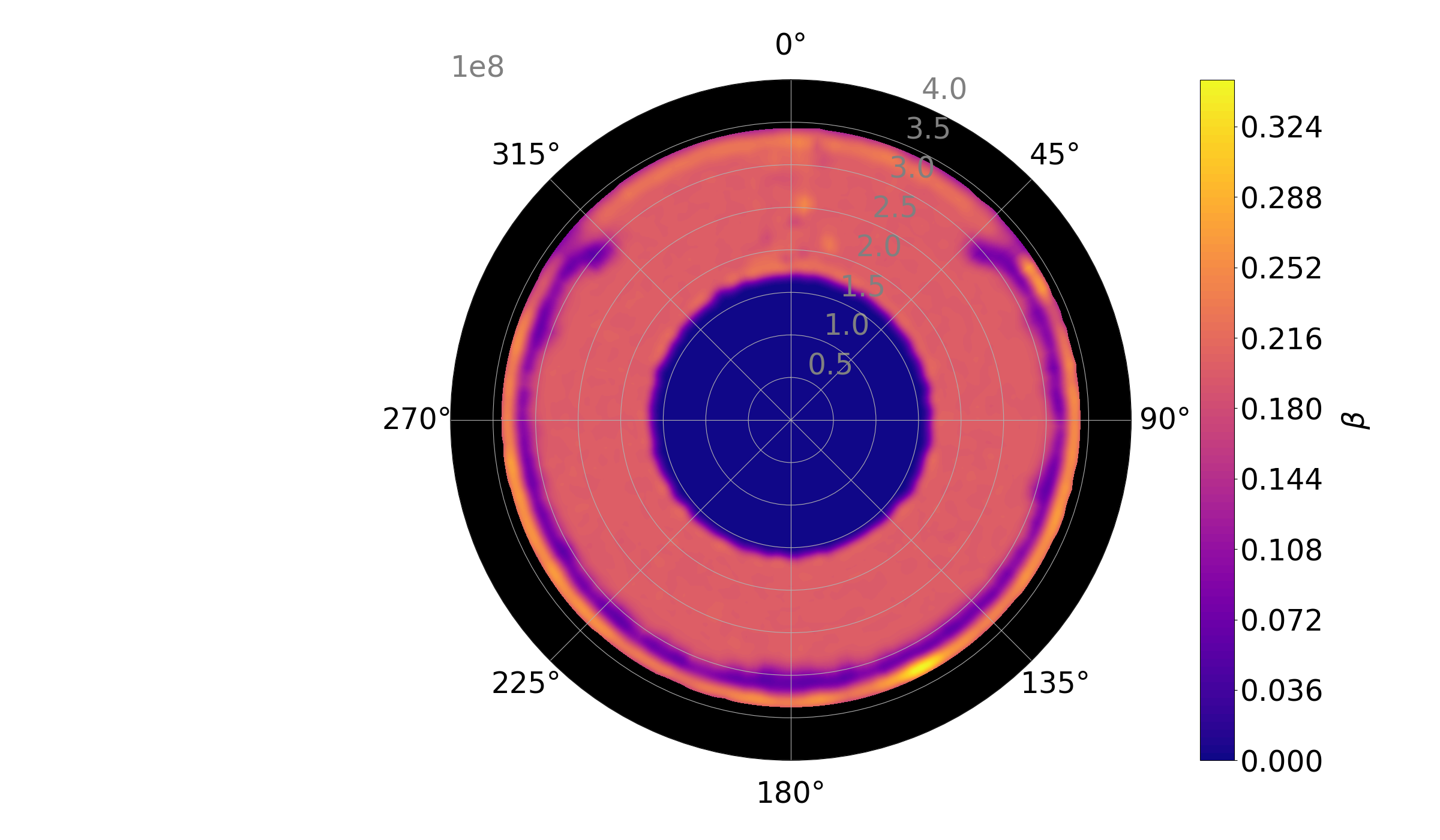}
    \end{subfigure}
    \begin{subfigure}{.49\textwidth}
        \adjincludegraphics[trim={{.15\width} 0 {.05\width} 0},clip,width=\linewidth]{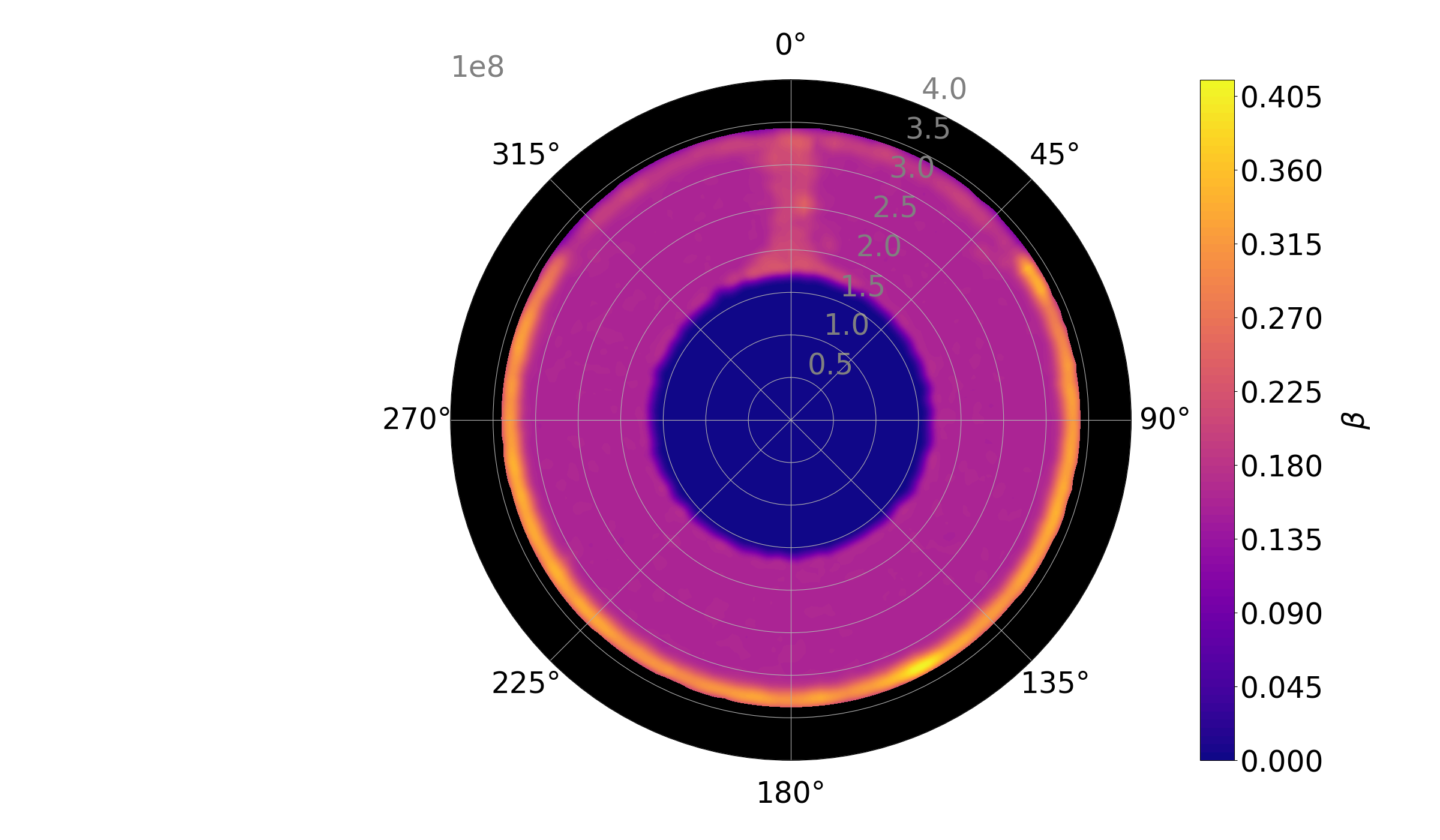}
    \end{subfigure}
    \caption{Various parameters related to our crust-production scheme displayed as a slice through the impact plane after the head-on classical impact with an initial fertility of $\beta_0 = 0.2$. The top row shows the crust contribution from primordial crust (left) and the melt fraction (right). The bottom row shows the fertility for the two mantle depletion schemes of fully-stratified (left) and fully-mixed (right).}
    \label{fig:mixedvsstrat_slice}
\end{figure}

In the fully mixed case, the crust distribution largely resembles those of Figure~\ref{fig:cumsum} (i.e. without any primordial crust calculations), but with a reduced contrast between the thickened impact site and the mostly untouched remainder of the planet. One distinct feature that emerges, however, is a ridge of thickened crust along the perimeter of the impact site (particularly visible in the $45\degree$ case). This corresponds to a crossover region where the impact shockwave is still energetic enough to melt a significant volume of mantle, but no longer capable of excavating the primordial crust at the surface. This primordial crust material therefore acts to increase the fraction of crust-bearing material in the melt, resulting in a thin region of particularly thick crust that abruptly drops off as the impact-induced melt fraction falls to zero. In the fully stratified case, the impact features are sharper and more concentrated; the centre of the impact-site has been amplified to even larger crustal thicknesses, but this region is smaller than that of the fully-mixed scheme, with the outer regions showing little to no crustal thickening, even displaying a very thin dip in crustal thickness at the immediate edge of the impact-thickened region. The reason for this dip is two-fold: the primordial crust has been stripped from this region, and a significant fraction of the impact-induced melt corresponds to the fully-depleted mantle material, thus contributing less crustal mass as it re-crystalises. The increase in crustal thickness at the impact site is caused by the depleted material initially present in the region being both displaced and replaced by the fertile mantle material of the impactor. The crust-producing melt therefore originates almost entirely from fully fertile mantle, with a negligible contribution from depleted material (see Figure~\ref{fig:mixedvsstrat_slice} for a visual representation of these effects in the head-on case).

\begin{figure}[htbp]
    \centering
    \begin{subfigure}{.49\textwidth}
        \adjincludegraphics[width=\linewidth]{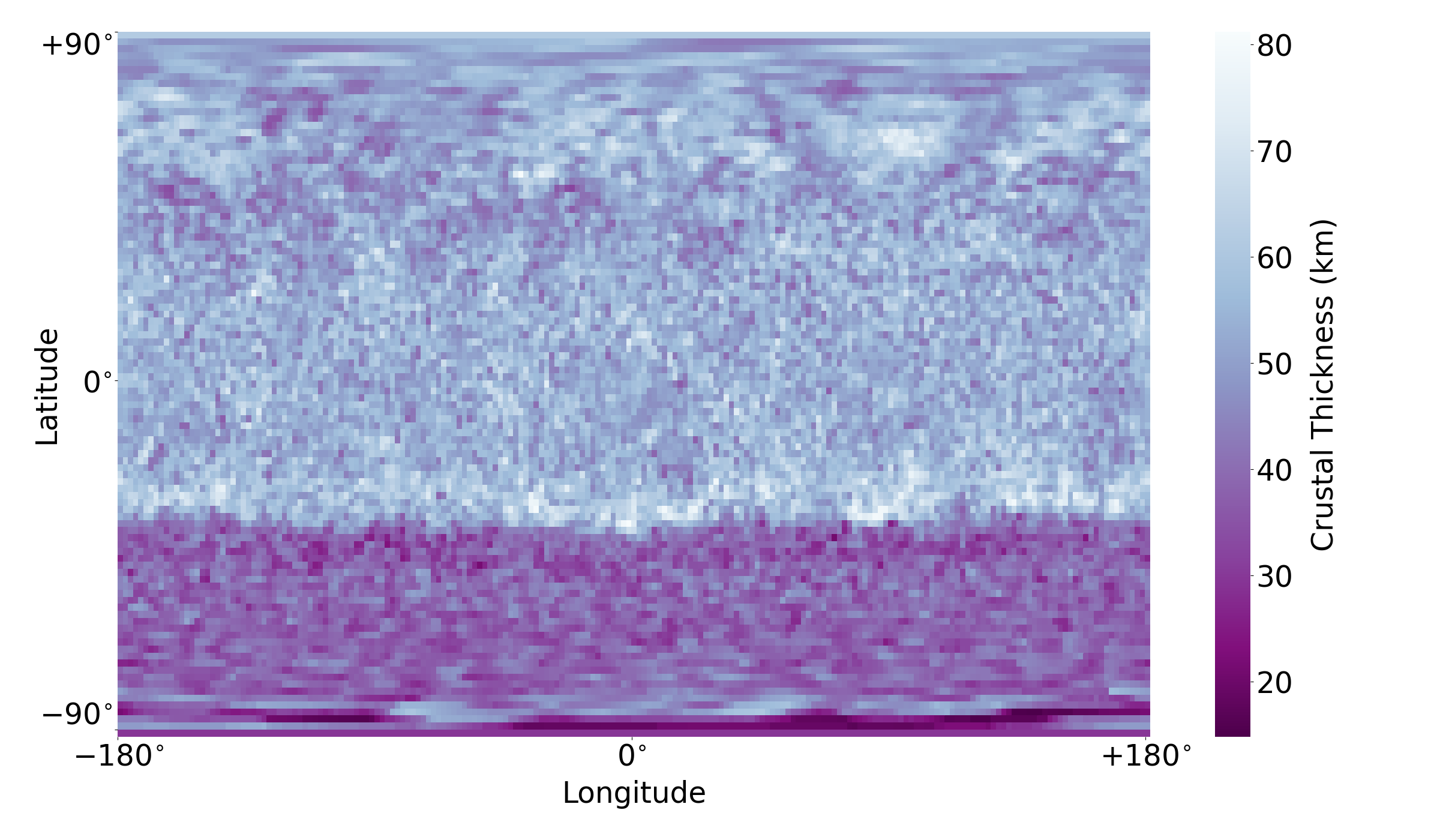}
    \end{subfigure}
    \begin{subfigure}{.49\textwidth}
        \adjincludegraphics[width=\linewidth]{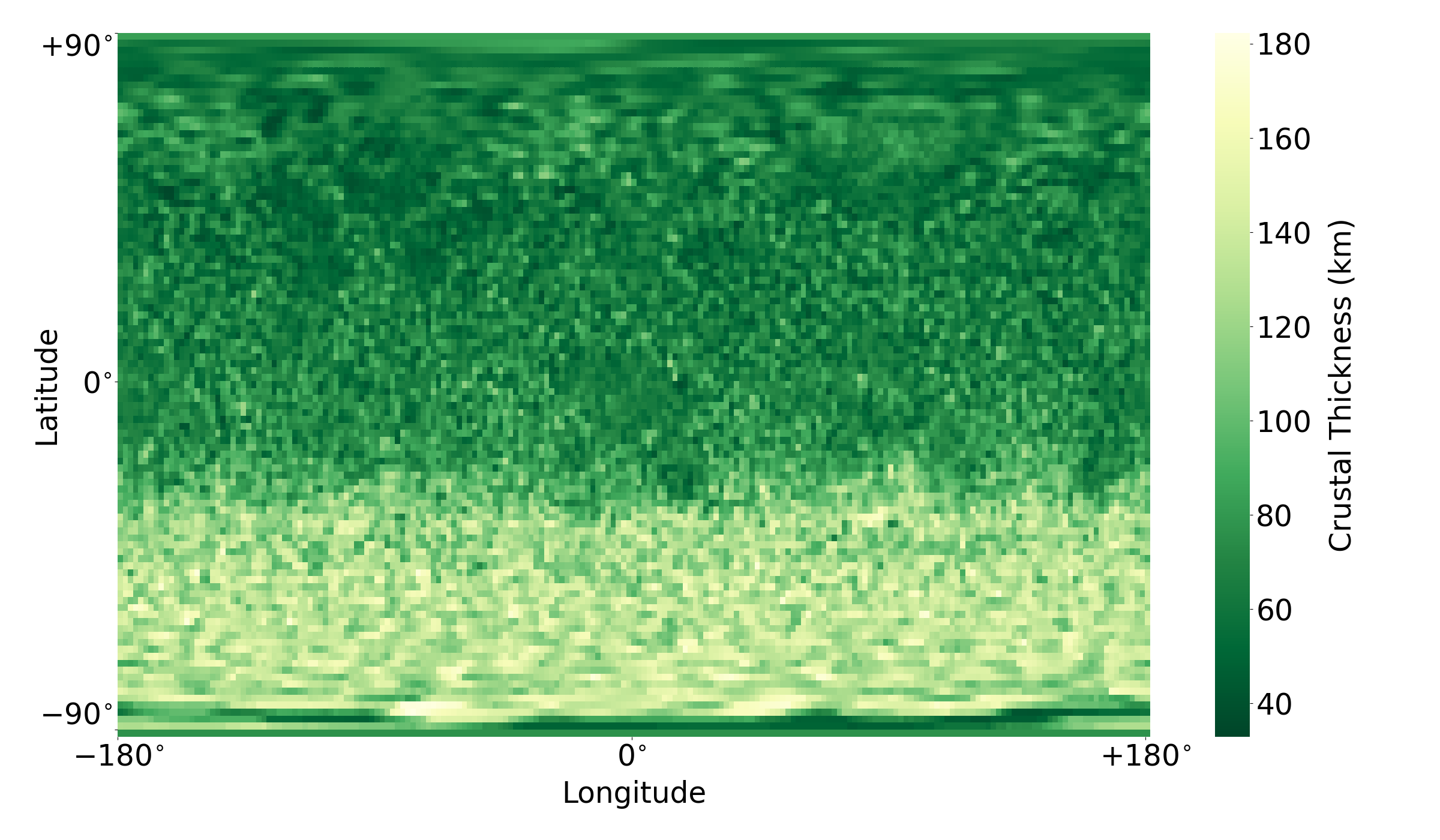}
    \end{subfigure}
    \caption{The crustal thickness distribution for the head-on classical case with a 40 km primordial crust and a initial mantle fertility of 0.10 (left) and 0.25 (right).}
    \label{fig:classicalmaps_ferts}
\end{figure}

The most significant variable in determining the precise impact-induced contrast in crustal thickness is the initial mantle fertility, $\beta_0$. This value relates to the silicate composition of Mars before crust-mantle differentiation (also known as the Bulk Silicate Mars or BSM). Higher values mean it is possible to produce more crust from a given mass of molten mantle. Lower values, on the other hand, lead to the contrary, with some parameters even producing an impact site with a lower crustal thickness than its surrounding area: an impact basin. For the same impact conditions, significantly different initial mantle fertilities can therefore imply very different crustal distributions, as illustrated in Figure~\ref{fig:classicalmaps_ferts}.

\subsection{Spherical Harmonic Analysis}
Clearly, with so many significant variables, studying each case individually is not feasible when investigating such a large dataset, and is not particularly effective when only comparing ``by eye'' rather than through a quantitative, scientific method. To resolve these issues, we utilise spherical harmonic analysis as a means to compare our computed crustal thickness maps to the observationally inferred crustal thickness map of Mars.

Spherical harmonics are the natural basis functions for the surface of a sphere, allowing any quantity that varies across this surface to be expressed as a series of such functions, i.e.
\begin{equation}
    f(\theta,\phi) = \sum^\infty_{l=0} \sum^l_{m=-1} f_{lm} Y_{lm}(\theta,\phi),
\end{equation}
where $\theta$ is latitude, $\phi$ is longitude, $l$ and $m$ are the spherical harmonic degree and order, respectively, $f(\theta,\phi)$ is the varying quantity, $f_{lm}$ is the spherical harmonic coefficient and $Y_{lm}$ is the corresponding spherical harmonic function. The total power of $f(\theta,\phi)$ is usually defined as the integral of the function squared, divided by the area, $\Omega$, of its domain\footnote{For any real, square-integrable function.}
\begin{equation}
    \frac{1}{4\pi}\int_\Omega f^2(\theta,\phi)~\mathrm{d}\Omega = \sum^\infty_{l=0} S(l)
    ,
\end{equation}
where Parseval’s theorem allows the total power to be expressed as a spectrum $S(l)$ related to the spherical harmonic coefficients by
\begin{equation}\label{powerspec}
    S(l) = \sum^l_{m=-l} f^2_{lm}
    .
\end{equation}

In this study, the varying quantity of interest is crustal thickness. Using the open-source SHTools Python library \citep{Wieczorek2018}, we calculate the spherical harmonic coefficients for real, 4$\pi$-normalised spherical harmonic functions that can accurately represent each crustal thickness longitude-latitude map up to degree $l_{max}=49$. Conveniently, the ``power spectrum'' defined in Equation~\ref{powerspec} is independent of coordinate-system orientation, thus providing us with an ideal tool to compare each case regardless of the pole-axis location relative to the impact site (which may differ between each case due to the impact-induced rotation of the target).

\begin{figure}[htbp]
    \centering
    \includegraphics[width=\textwidth]{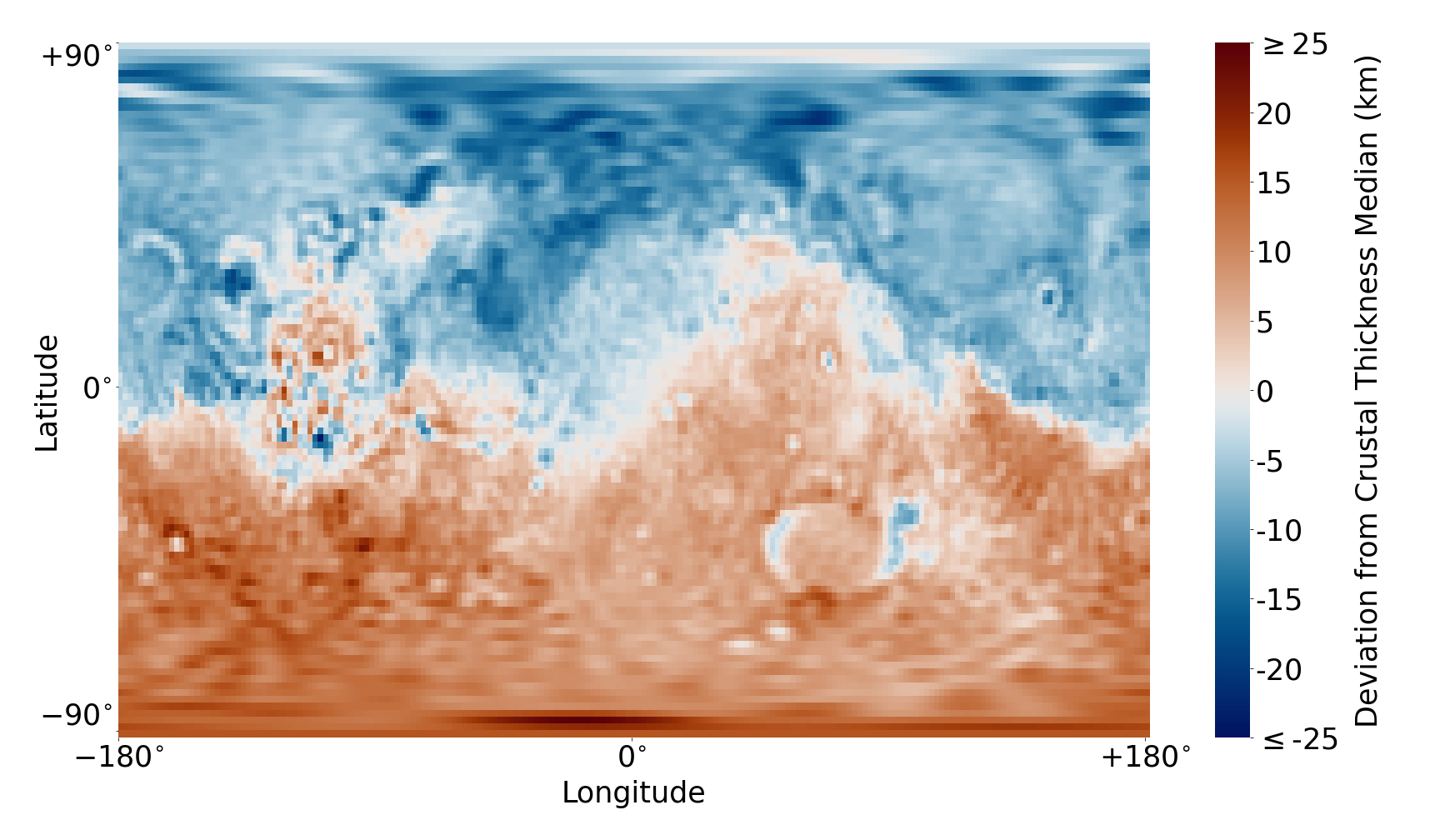}
    \caption{The crustal thickness distribution of Mars prior to any major geological events that occurred after the formation of the Martian Dichotomy, from \cite{Bouley2020}. This map shows the exact distribution used to compare to our simulation results, and as such it has been converted to the spherical grid resolution used in our simulation analysis via bivariate spline interpolation, and is displayed through the equirectangular projection.}
    \label{fig:mars_corrected}
\end{figure}

For comparison with the crustal thickness inferred by observations of Mars, one must be wary of the many aeons that have passed since the crustal dichotomy's emplacement. Although Mars is considered to have a stagnant lid and to be mostly geologically inactive (particularly when compared to Earth), there have been several significant, subsequent alterations to the distribution of crust across the planet's history in certain regions (e.g. the Tharsis region or the Hellas basin). To compare our results with the crustal thickness as it would have been prior to these major geologic events, we use the data provided in \cite{Bouley2020}, who developed a mass-conserving method to reconstruct the crustal thickness of Mars without the four largest impact basins (Hellas, Argyre, Isidis and Utopia) or the main volcanic provinces of Tharsis and Elysium, based on the crustal thickness model of \cite{Goossens2017}. This data is converted to the same resolution as our grids via bivariate spline interpolation\footnote{We make use of the interpolate.RectSphereBivariateSpline function of SciPy Python library for this purpose (documentation at \url{https://docs.scipy.org/doc/scipy/reference/generated/scipy.interpolate.RectSphereBivariateSpline.html})}, the result of which is shown in Figure~\ref{fig:mars_corrected}.

The power spectrum for this distribution is then calculated. Any differences between this spectrum and those of our impact simulations can be considered as deviations from the desired result. As such, we use the Euclidean norm of these differences as a measure of error
\begin{equation}
    \delta_{SH} = \sqrt{ \sum_{l=0}^{l_{max}} \left( S(l) - S_{\mars}(l) \right)^2 },
\end{equation}
where $S_{\mars}(l)$ is the power spectrum related to the observed crustal thickness of Mars. It should be noted that the median crustal thickness of the \cite{Bouley2020} data is subtracted from each crustal thickness distribution before performing the spherical harmonic analysis.

Using this method, we can identify the most promising areas of the parameter-space as those with the lowest $\delta_{SH}$. Conversely, a high $\delta_{SH}$ indicates a crust distribution that does not match well with observations, allowing us to rule out these cases as potential Dichotomy-forming events. A simple cutoff comes from the initial crust distributions on our pre-impact Mars. Using the same method as described, we can find a $\delta_{SH}$ corresponding to the distribution of crust on the pre-impact target for each primordial crust thickness investigated. The lowest of these values provides a maximum $\delta_{SH}$ cutoff for our final post-impact crust distributions, $\delta_{SH_{max}}$, as a greater value indicates that a uniform crust distribution better fits the observed data, thus implying no measurable gain from the impact.

\subsubsection{Impact Angle and Impactor Radius}\label{goodcases_angle_radius}
 
\begin{figure}[htbp]
    \centering
    \includegraphics[width=\textwidth]{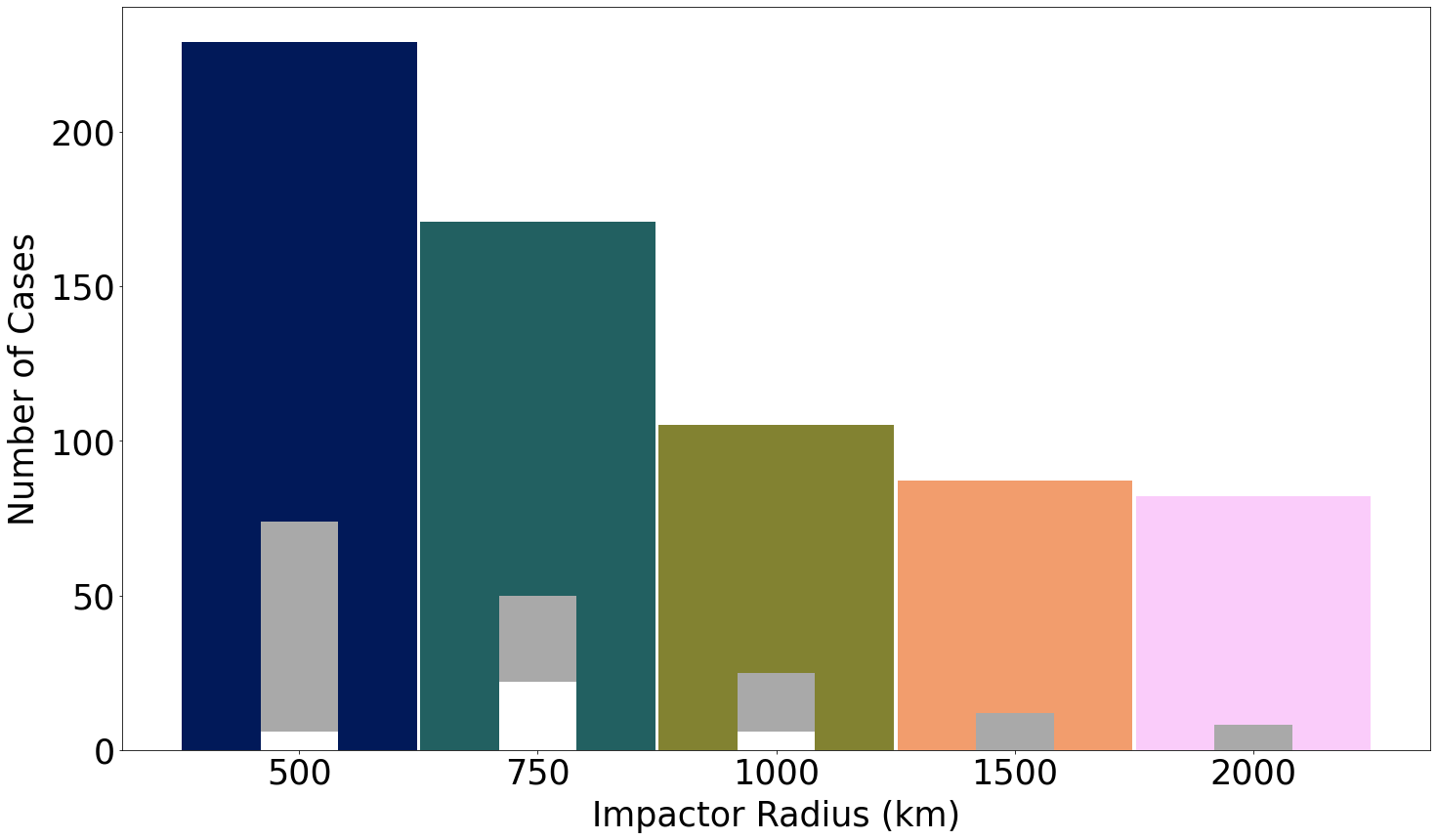}
    \caption{Histogram of cases within $\delta_{SH_{max}}$ for each impactor radius. The grey and white bars represent those within first quartile and 5th percentile of such cases, respectively.}
    \label{fig:goodcases_radius}
\end{figure}

Applying this logic to our entire parameter-space leads to the exclusion of many cases. In particular, a clear trend related to impactor radius is revealed, with a larger impactor indicating a lower number of feasible cases (Figure~\ref{fig:goodcases_radius}). This trend is still evident if we only consider those with a $\delta_{SH}$ below the first quartile of such cases, where the 1500 km and 2000 km radius impactors only represent 20 of the 169 remaining cases. If we focus on cases below the 5th percentile, however, the 750 km impactor proves to represent the most promising simulations, outnumbering the remaining 500 km and 1000 km impactor cases 22 to 12.

\begin{figure}[htbp]
    \centering
    \begin{subfigure}{.49\textwidth}
        \adjincludegraphics[width=\linewidth]{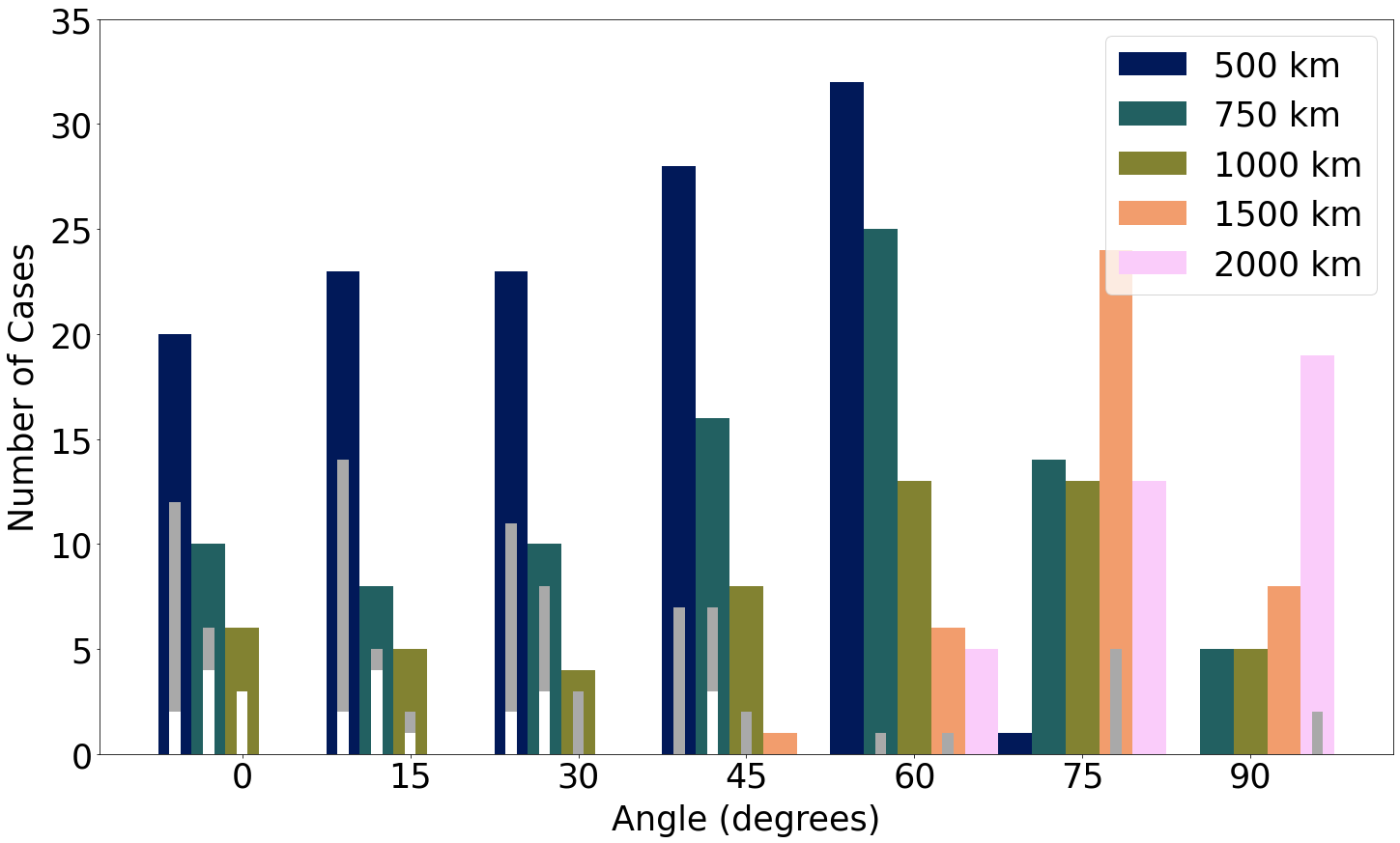}
    \end{subfigure}
    \begin{subfigure}{.49\textwidth}
        \adjincludegraphics[width=\linewidth]{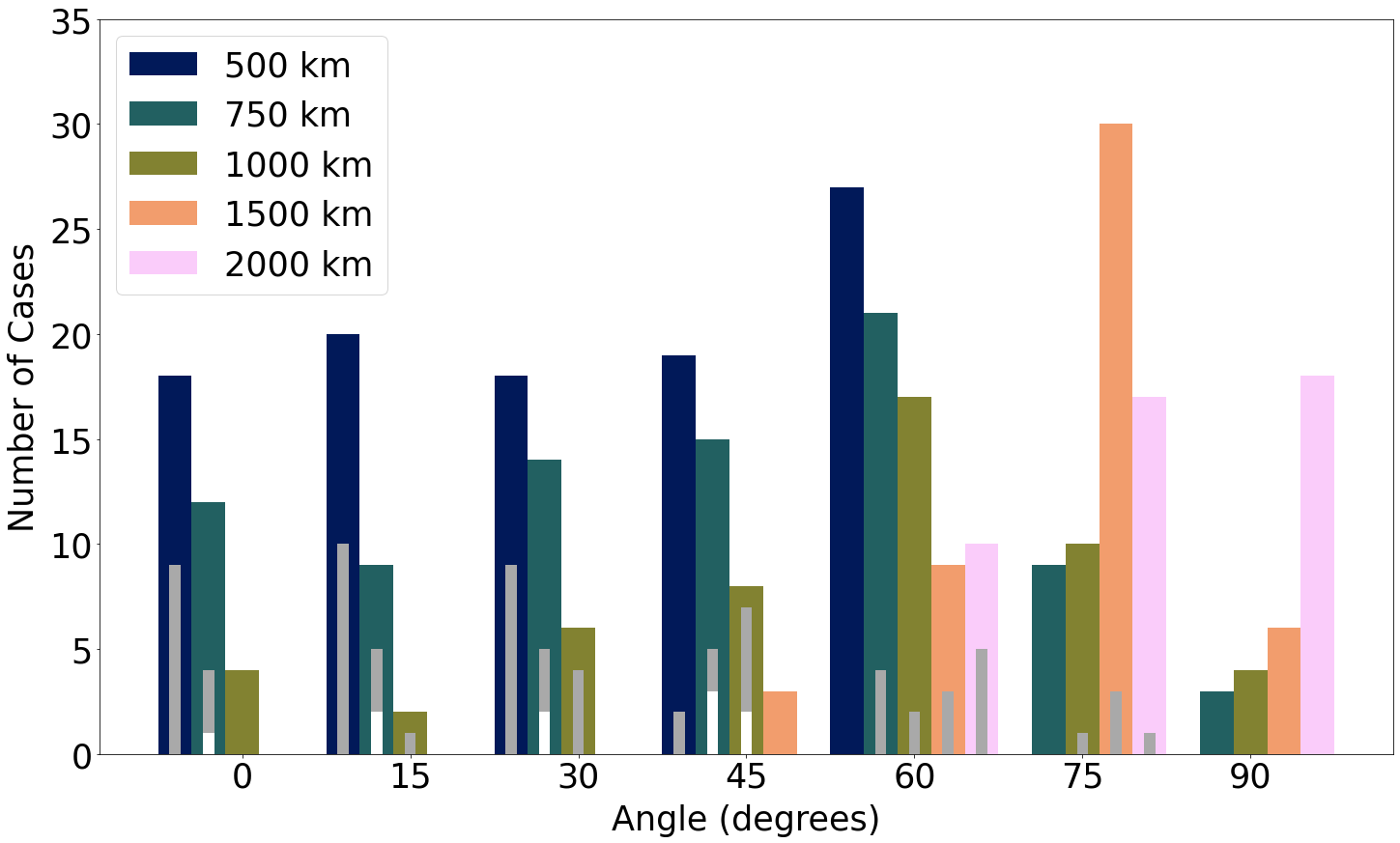}
    \end{subfigure}
    \caption{Histogram of cases within $\delta_{SH_{max}}$ for each impactor radius as a function of impact angle for the fully-mixed (left) and fully-stratified (right) mantle depletion schemes. The grey and white bars represent those within first quartile and 5th percentile of such cases, respectively.}
    \label{fig:goodcases_angle}
\end{figure}

Understanding this trend requires further investigation through the inclusion of other physical parameters. Figure~\ref{fig:goodcases_angle} again shows the distribution of cases with a $\delta_{SH}$ lower than the best pre-impact case, but now as a function of impact angle, divided into the two mantle depletion methods. Here we see that the large impactors only produced relevant cases with highly oblique impact angles, with $75\degree$ and $90\degree$ being the best impact angles for the 1500 km and 2000 km radius impactors respectively. This is to be expected if they are systematically producing too much crust (and thus melt), as higher impact angles mean more energy is carried away as ejecta or converted into target rotation rather than impact heating.

\begin{figure}[htbp]
    \centering
    \begin{subfigure}{.49\textwidth}
        \adjincludegraphics[width=\linewidth]{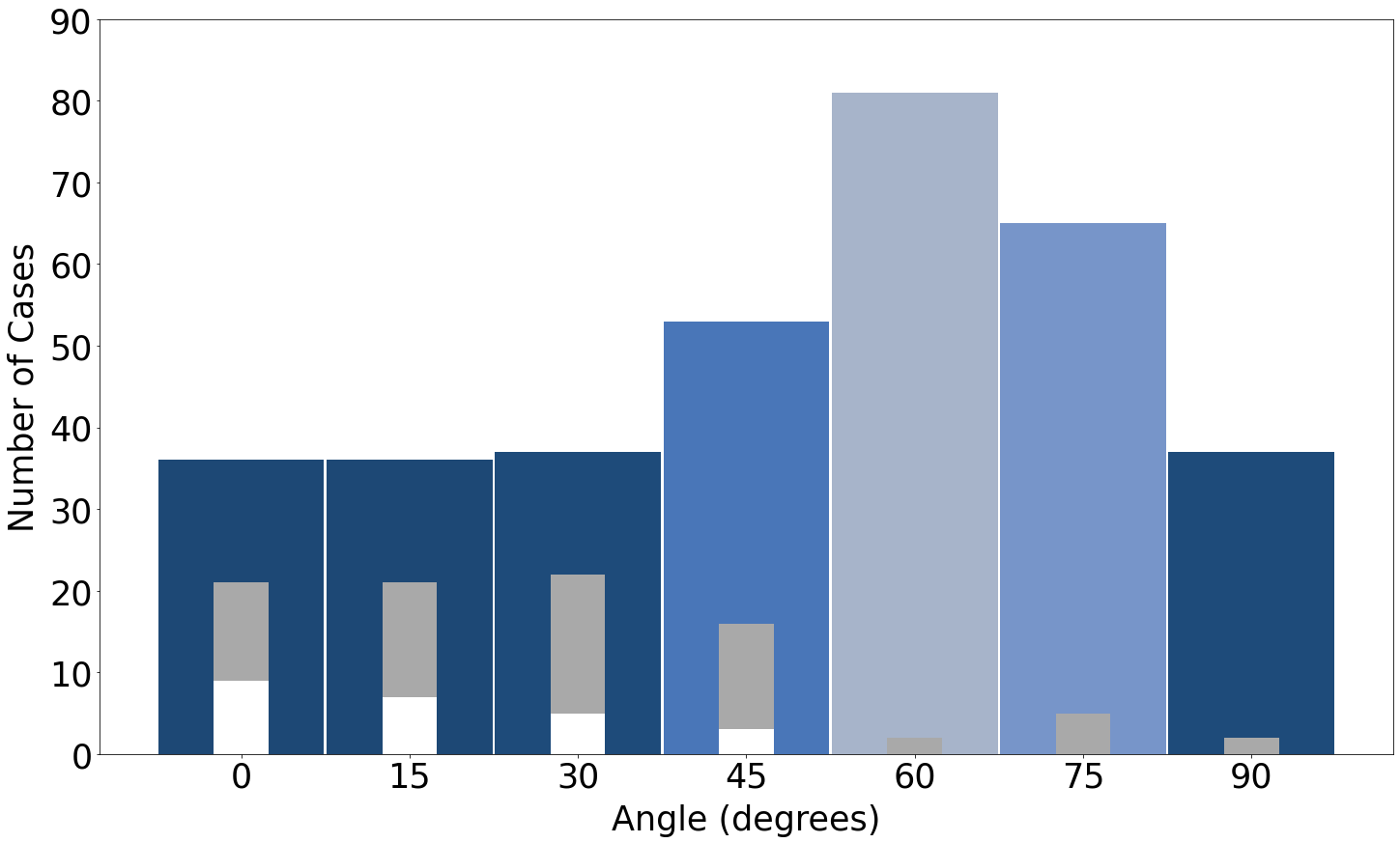}
    \end{subfigure}
    \begin{subfigure}{.49\textwidth}
        \adjincludegraphics[width=\linewidth]{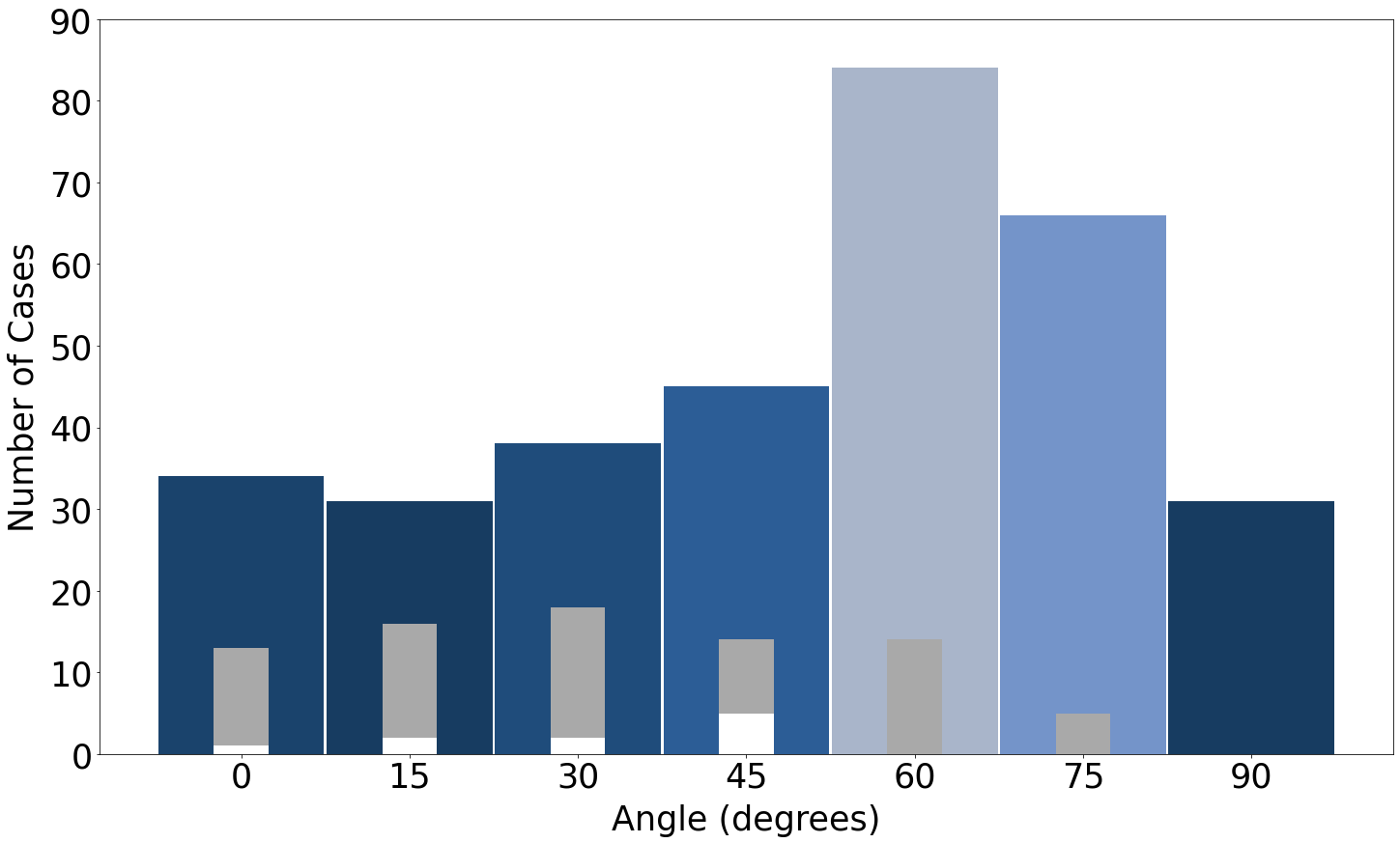}
    \end{subfigure}
    \caption{Histogram of cases within $\delta_{SH_{max}}$ summed across all impactor radius as a function of impact angle for the fully-mixed (left) and fully-stratified (right) mantle depletion schemes.}
    \label{fig:goodcases_angle_sum}
\end{figure}

For the smaller impactors (radii of 500-1000 km) the trend is less clear and, interestingly, significantly changes as we again restrict our cases to those below certain percentile thresholds, as well as when comparing the results of the two mantle depletion schemes. When including all cases with $\delta_{SH} < \delta_{SH_{max}}$, there is a positive trend towards larger impact angles for all radii, culminating in a peak at $60\degree$ impact angle that drops off for higher angles. For cases within the first quartile, however, this peak is no longer apparent, with almost all $60\degree$ cases being excluded for the fully-mixed mantle depletion model. As we restrict further to cases within the 5th percentile, a clear trend emerges for each depletion scheme, with the fully-mixed model favouring lower impact angles, and the fully-stratified model favouring higher impact angles up to a maximum of $45\degree$ (Figure~\ref{fig:goodcases_angle_sum}). The preference for near head-on impact angles in the fully-mixed model makes intuitive sense, as lower angles lead to a more distinct hemispherical contrast, reducing both the ellipticity of the impact site and the prominence iso-longitudinal ring of re-impacting ejecta. The positive trend in the fully-stratified model, on the other hand, is a consequence of the crustal thickness dip along the perimeter of the impact site (as described in Section~\ref{classical}), which causes a strong signal in the spherical harmonic analysis. At larger impact angles this feature becomes obscured, thus leading to a smaller $\delta_{SH}$ and giving rise to the trend observed.

\subsubsection{Mantle Fertility}
\begin{figure}[htbp]
    \centering
    \begin{subfigure}{.49\textwidth}
        \adjincludegraphics[width=\linewidth]{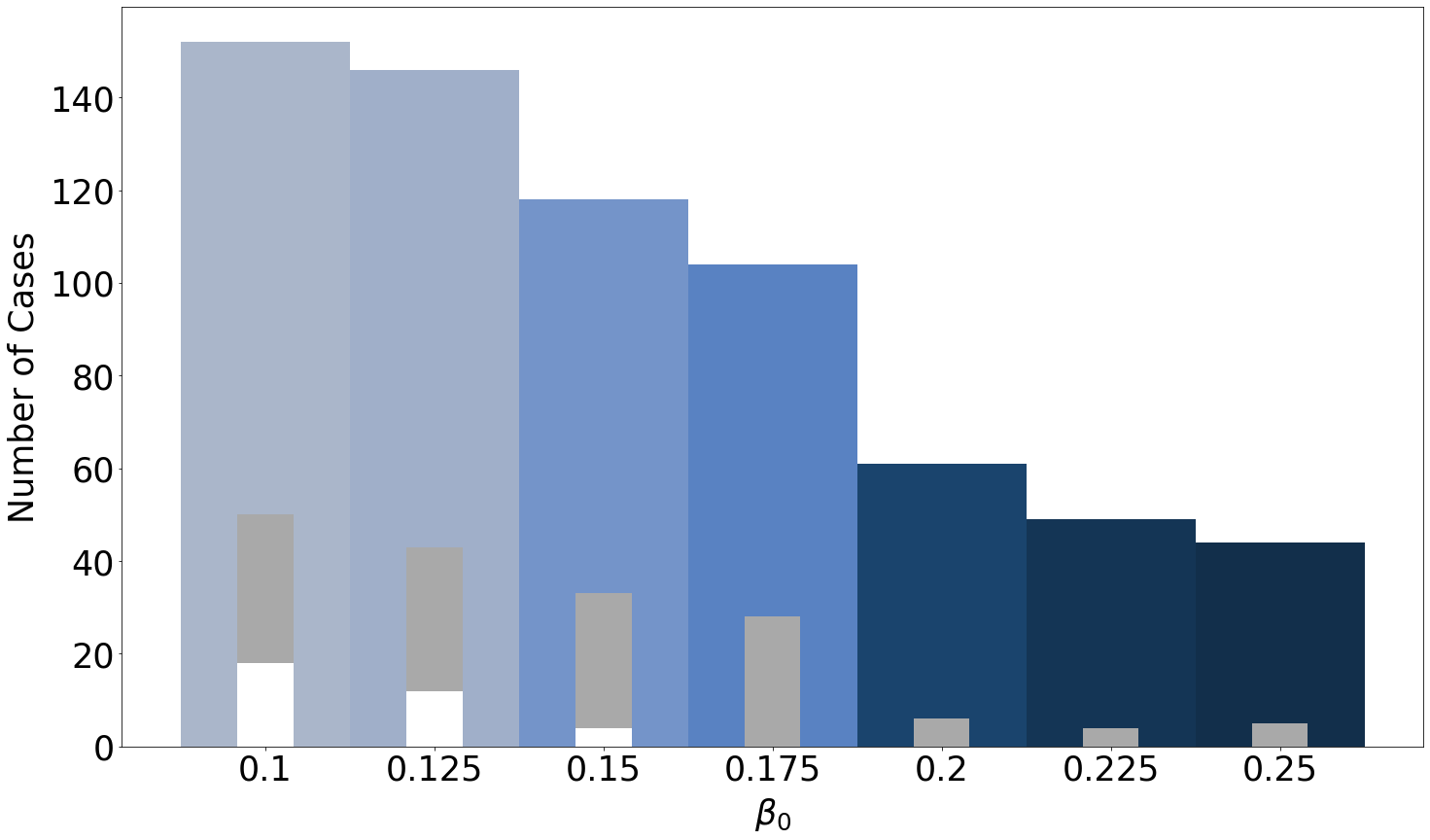}
    \end{subfigure}
    \begin{subfigure}{.49\textwidth}
        \adjincludegraphics[width=\linewidth]{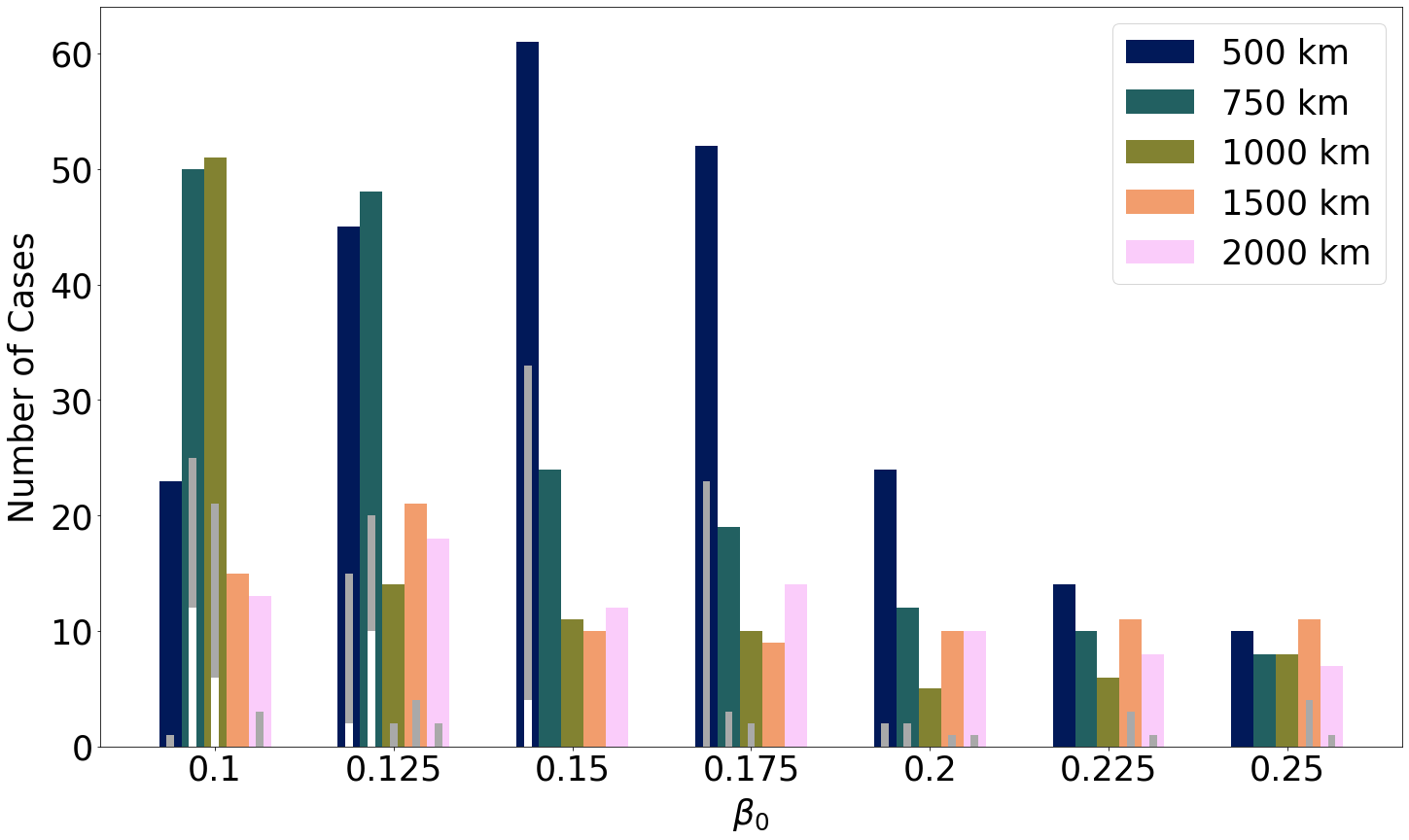}
    \end{subfigure}
    \caption{Histogram of cases within $\delta_{SH_{max}}$ for all impactor radius as a function of initial, pre-impact fertility $\beta_0$ either summed across all radii (left) or with each radius as a separate colour (right). The grey and white bars represent those within first quartile and 5th percentile of such cases, respectively.}
    \label{fig:goodcases_beta0}
\end{figure}

Figure~\ref{fig:goodcases_beta0} shows the distribution of our cases of interest ($\delta_{SH} < \delta_{SH_{max}}$) as a function of $\beta_0$. Here, we see a strong preference for a smaller $\beta_0$, with the majority of cases within the 5th percentile corresponding to the smallest value investigated (0.1). As we look to different impactor radii, however, the dominant $\beta_0$ changes. At 1000 km radius the general trend is reflected well, with 84\% (21/25) of such cases within 1st quartile corresponding to lowest fertility (0.1), and all within the 5th percentile. This value remains the most represented fertility in the 750 km case, but only by a small margin, with a fertility of 0.125 also producing a large number of promising cases. For a 500 km impactor, we find that a higher fertility of 0.15 produces the most cases of interest, particularly when we only consider cases within the 5th percentile. The largest impactors (1500 km and 2000 km) do not show any significant preference for a certain fertility.

\subsubsection{Impact Velocity}
\begin{figure}[htbp]
    \centering
    \includegraphics[width=\textwidth]{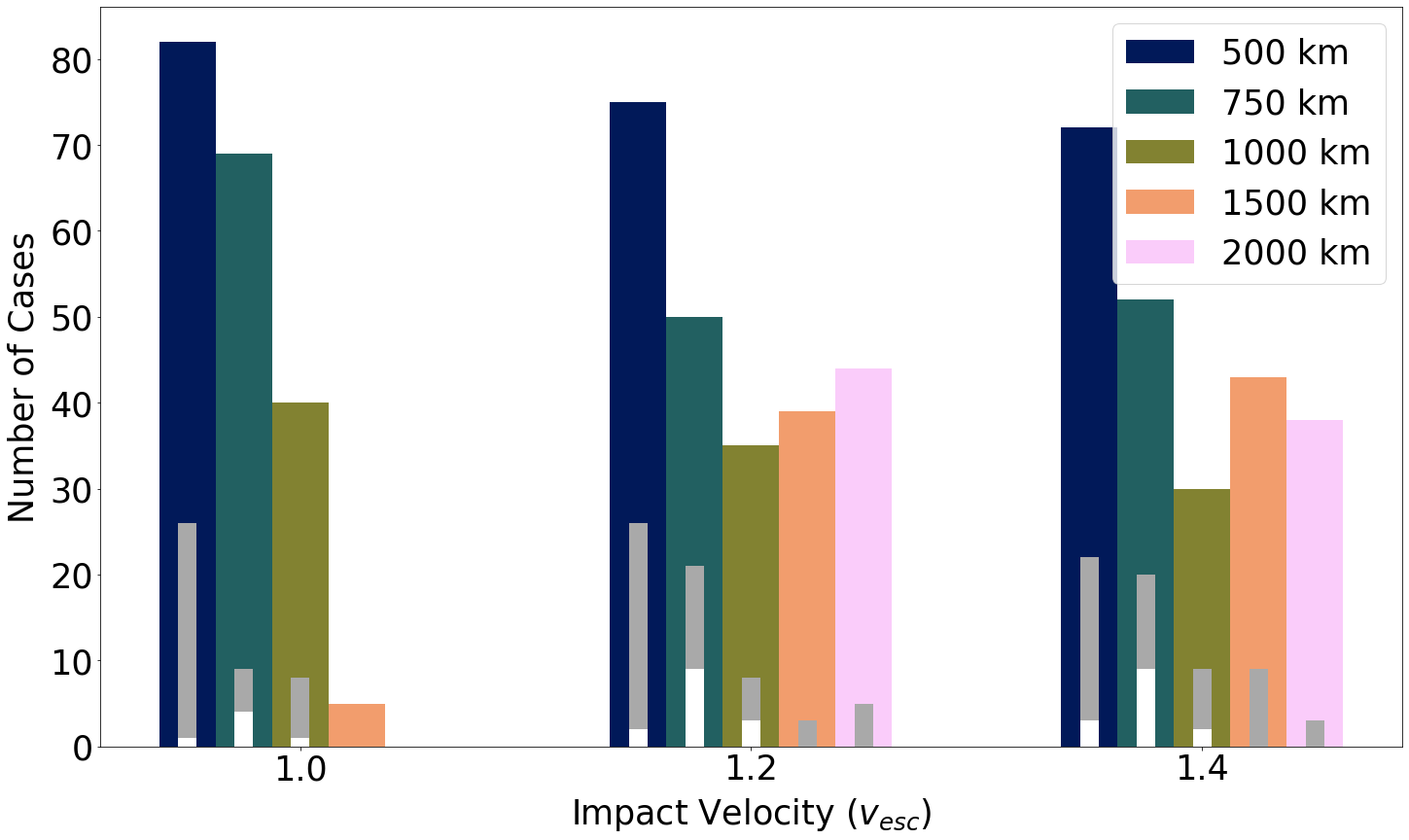}
    \caption{Histogram of cases within $\delta_{SH_{max}}$ for each impactor radius as a function of impact velocity. The grey and white bars represent those within first quartile and 5th percentile of such cases, respectively.}
    \label{fig:goodcases_speeds}
\end{figure}

As visible in Figure~\ref{fig:goodcases_speeds}, the smaller impactors (radii $\leq$ 1000 km) that represent the best cases do not show any strong preference in impact velocity, other than a slight trend towards higher values for cases within the 5th percentile, which disfavour the mutual escape speed cases in particular. For the larger impactors, this preference is far more significant, with almost no mutual escape speed cases surviving our $\delta_{SH_{max}}$ threshold. Without further information from other parameters, this trend seemingly contradicts the intuition that larger impactors that systematically induce too much crust production (due to excessive impact-induced melt) should favour the less energetic, lower impact velocities in producing the best matches to the Dichotomy. As discussed in Section~\ref{goodcases_angle_radius}, however, the large impactor cases strongly favour high impact angles. The combination of high velocities at very oblique angles lead to grazing collisions in which the impactor retains enough kinetic energy to remain gravitationally unbound to the target-impactor system and continue on its hyperbolic path relatively unimpeded (a scenario known as ``hit-and-run''). It is this regime that produces the best cases for our largest impactors.

\subsubsection{Primordial Crust Thickness}
\begin{figure}[htbp]
    \centering
    \includegraphics[width=\textwidth]{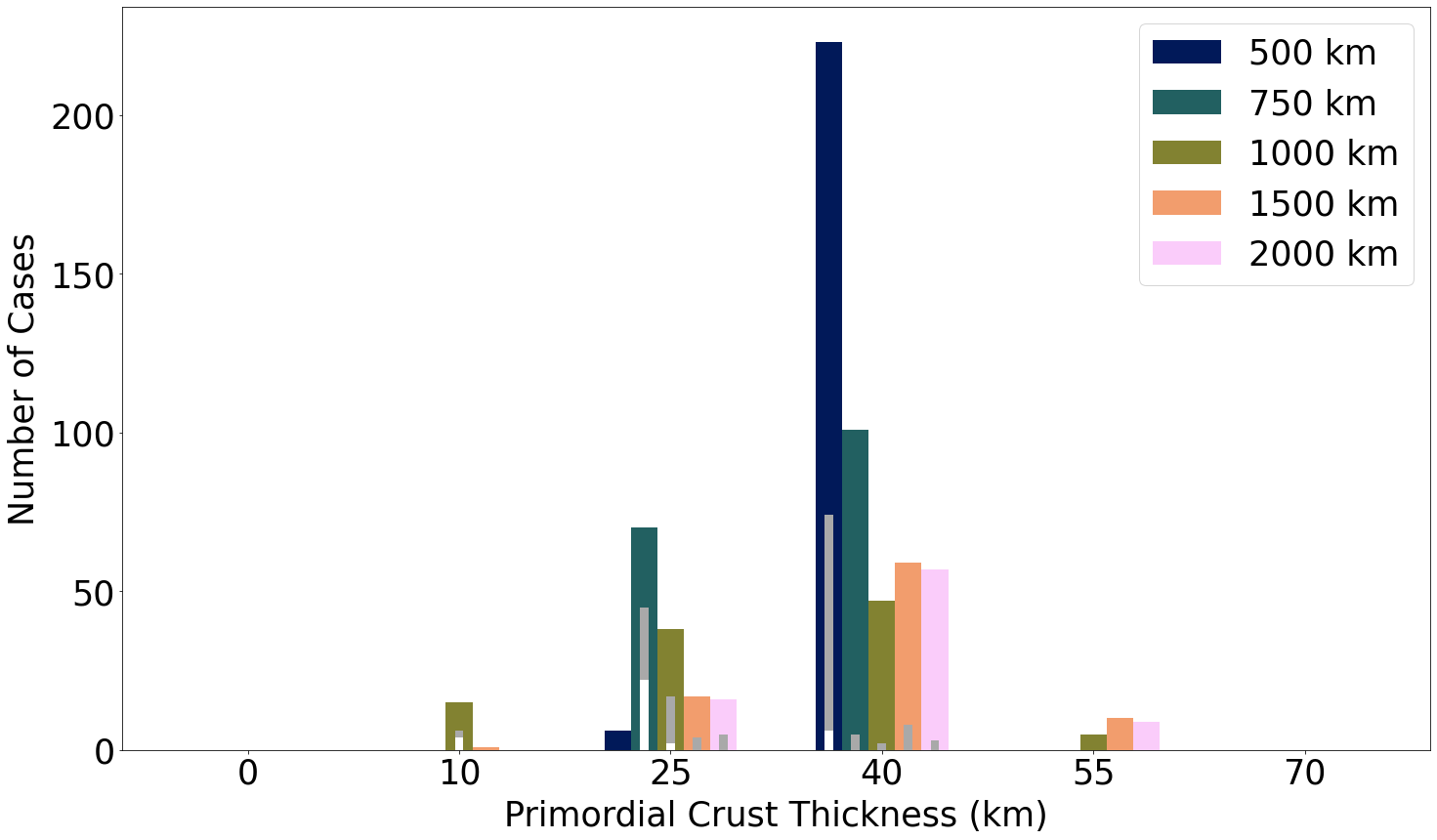}
    \caption{Histogram of cases within $\delta_{SH_{max}}$ for each impactor radius as a function of primordial crust thickness. The grey and white bars represent those within first quartile and 5th percentile of such cases, respectively.}
    \label{fig:goodcases_primcrust}
\end{figure}

Figure~\ref{fig:goodcases_primcrust} illustrates the influence of primordial thickness in producing a Dichotomy-like crust distribution. If we include all cases below our minimum threshold of $\delta_{SH_{max}}$, a clear peak emerges in all impactor radii for a 40 km crust prior to impact. For cases below the first quartile and/or the 5th percentile thresholds, however, distinct preferences appear. The 1000 km impactor produced most of its sub-quartile cases with a 25 km primordial crust, although the majority of its cases within the 5th percentile came instead from a 10 km crust, being the sole impactor radius with any significant cases associated with such a thin pre-impact crust. The 750 km impactor shows a strong preference for a 25 km crust in its best cases, with almost all of its sub-quartile cases corresponding to this thickness. The 500 km impactor has a very strong preference for the 40 km primordial crust thickness within all $\delta_{SH}$ thresholds. The largest impactors do not show a strong preference for a single thickness within the first quartile, with a near-equal number of cases being represented by the 25 km and 40 km crustal thicknesses, and even some 55 km cases with $\delta_{SH} < \delta_{SH_{max}}$.

\subsubsection{Mantle Depletion Scheme and Core Fraction}
\begin{figure}[htbp]
    \centering
    \begin{subfigure}{.49\textwidth}
        \adjincludegraphics[width=\linewidth]{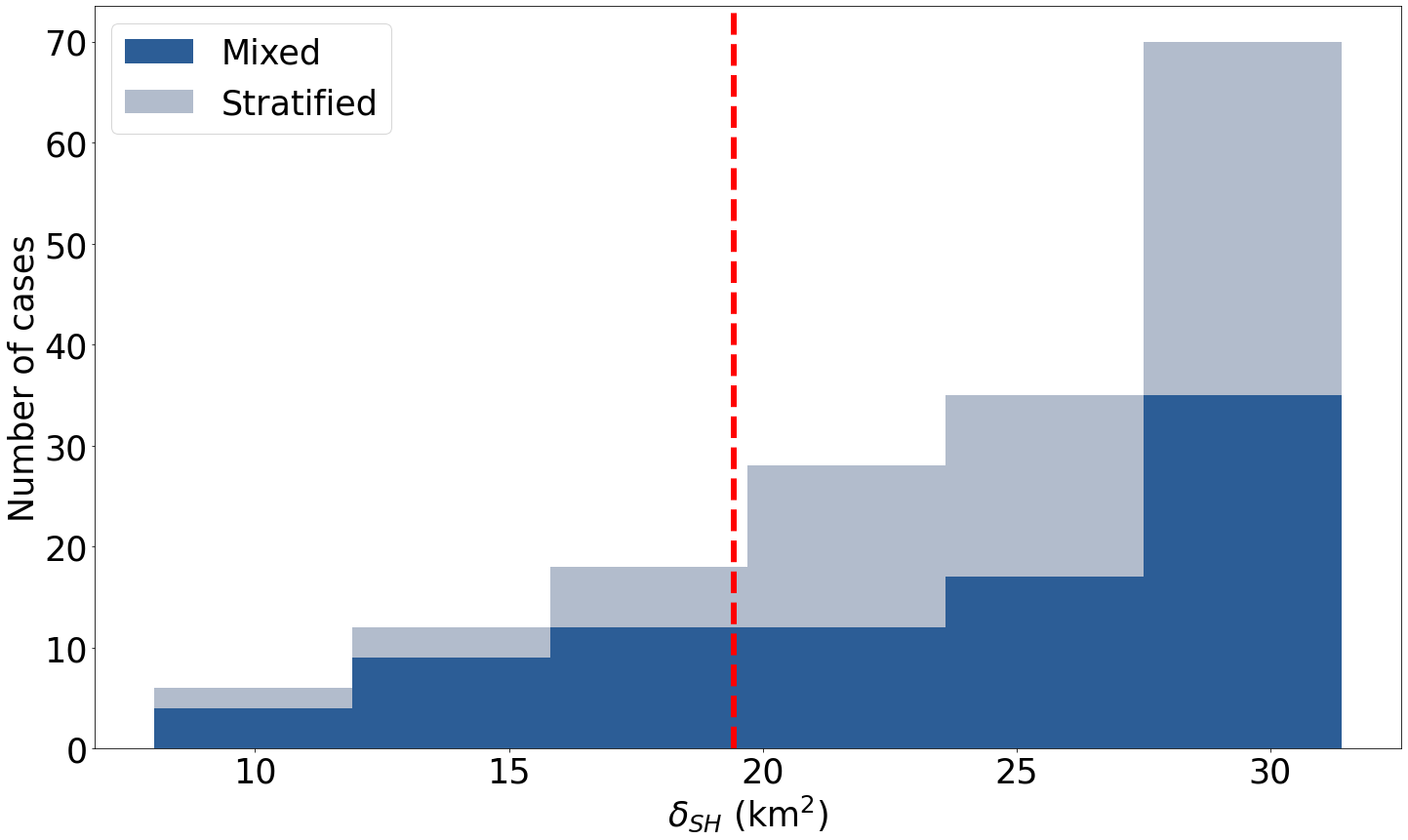}
    \end{subfigure}
    \begin{subfigure}{.49\textwidth}
        \adjincludegraphics[width=\linewidth]{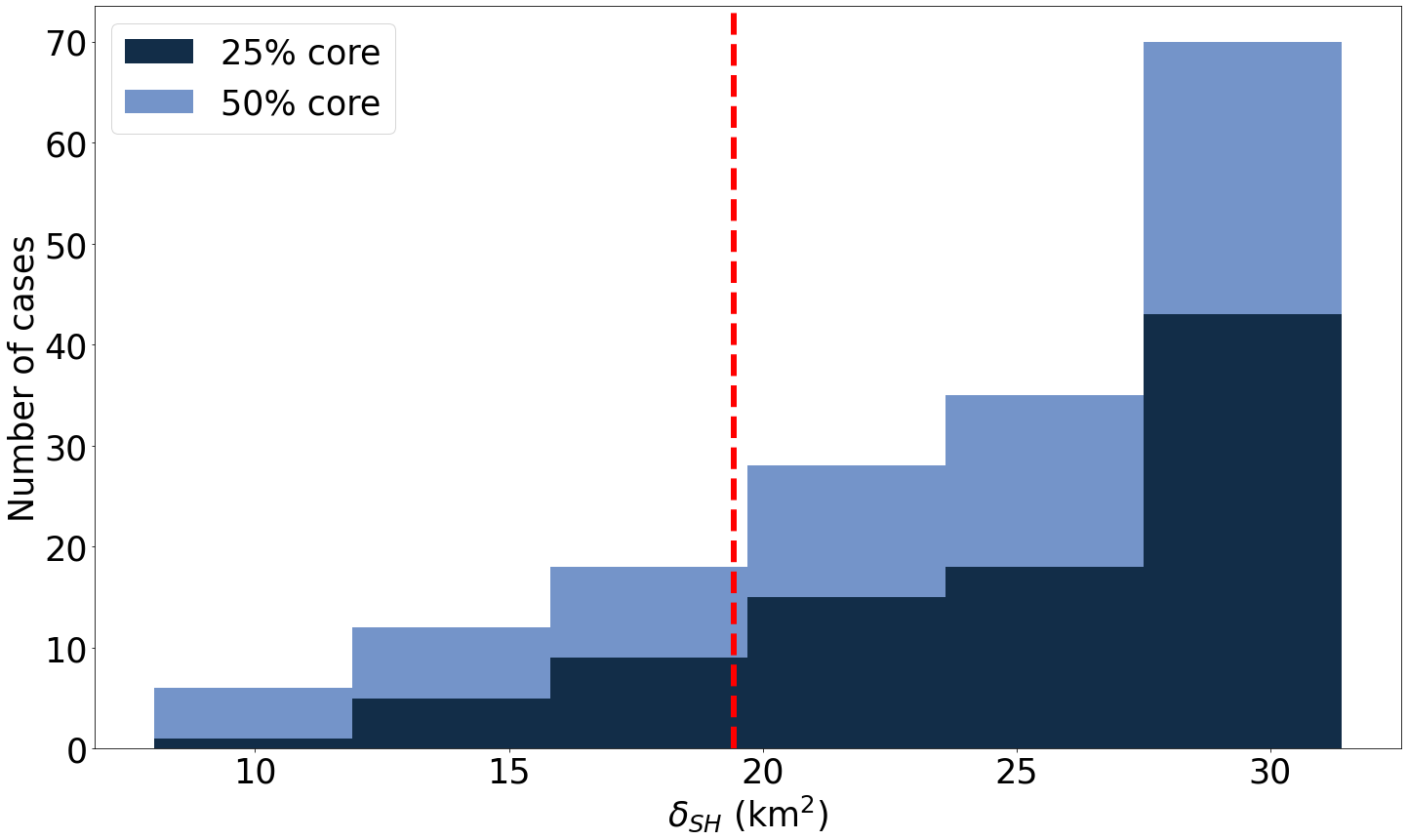}
    \end{subfigure}
    \caption{Cases within the 1st quartile of cases with $\delta_{SH} < \delta_{SH_{max}}$ for the two mantle depletion schemes (left) and the two impactor core fractions (right). The red line depicts the 5th percentile of cases with $\delta_{SH} < \delta_{SH_{max}}$.}
    \label{fig:goodcases_deltaSH_mantle_core}
\end{figure}

When comparing the two mantle depletion schemes for cases with $\delta_{SH}$ within $\delta_{SH_{max}}$, we see a very small preference for the fully-mixed approach over the stratified (345 compared to 329 cases, respectively). If we look at the distribution of these cases within the first quartile, however, we see that the majority of the best cases correspond to the fully-mixed cases, with this imbalance between the two schemes evening out at higher $\delta_{SH}$ (Figure~\ref{fig:goodcases_deltaSH_mantle_core}). Although there appears to be a preference for the 50\% core fraction for the very best cases (8 out the 10 best-fitting cases have this core fraction), this trend quickly disappears, even switching to a minor excess in 35\% cases (91 against 78 50\% core fraction cases within the first quartile).

\subsection{Best Cases}
\begin{figure}[htbp]
    \centering
    \includegraphics[width=\textwidth]{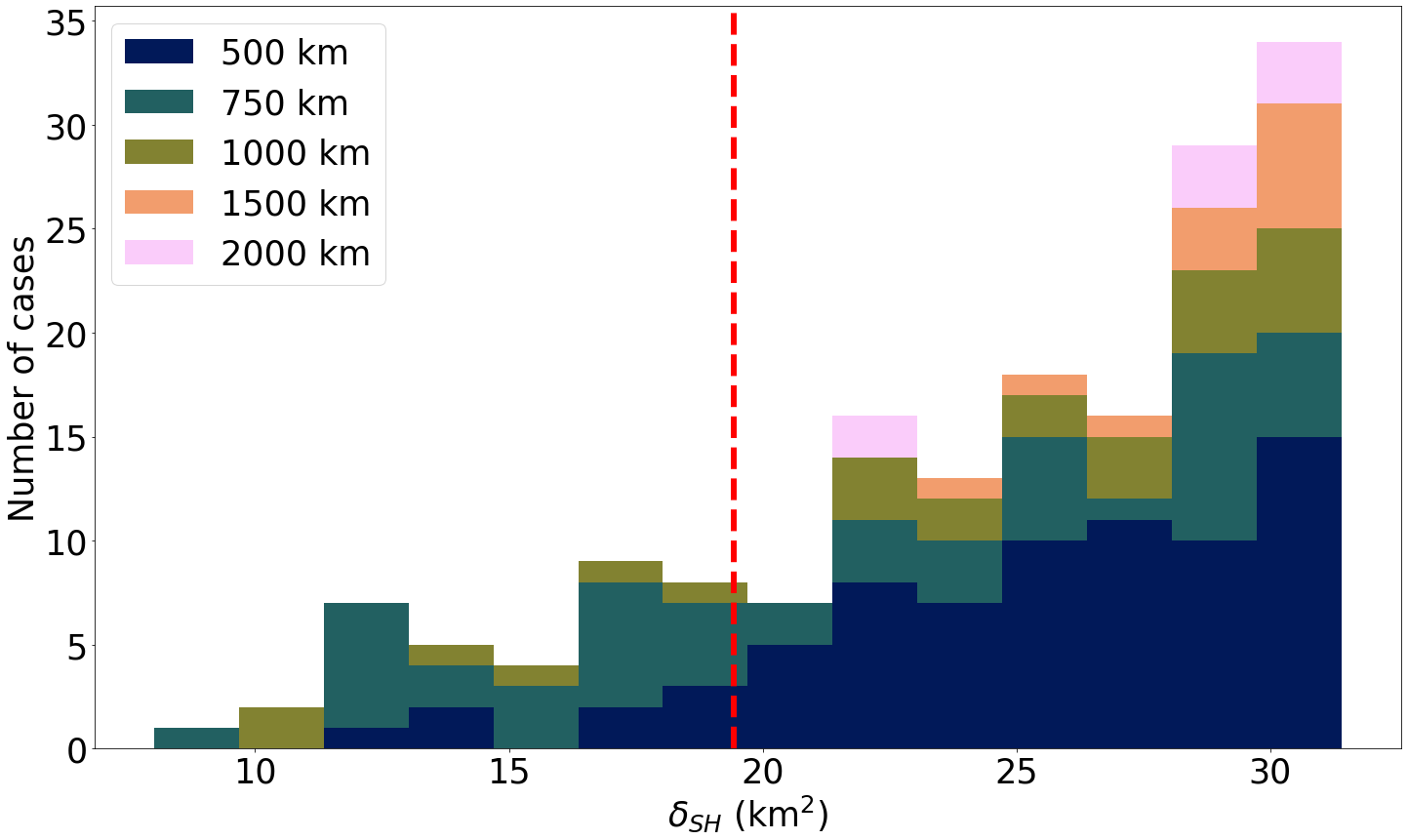}
    \caption{Cases within the 1st quartile of cases with $\delta_{SH} < \delta_{SH_{max}}$ where colour depicts impactor radius. The red line depicts the 5th percentile of cases with $\delta_{SH} < \delta_{SH_{max}}$.}
    \label{fig:goodcases_deltaSH_radius}
\end{figure}

As illustrated in Figure~\ref{fig:goodcases_deltaSH_radius}, the majority of the best cases (i.e. those within the 5th percentile) are associated with a 750 km radius impactor, including our best fit to the Martian Dichotomy data. All impactor radii are represented at least by some cases within the first quartile, however. We will therefore use this section to clarify the conditions necessary for the best fitting cases in the context of each impactor radius.

\subsubsection{The largest impactors (1500 km and 2000 km)}
\begin{figure}[htbp]
    \centering
    \begin{subfigure}{.49\textwidth}
        \adjincludegraphics[width=\linewidth]{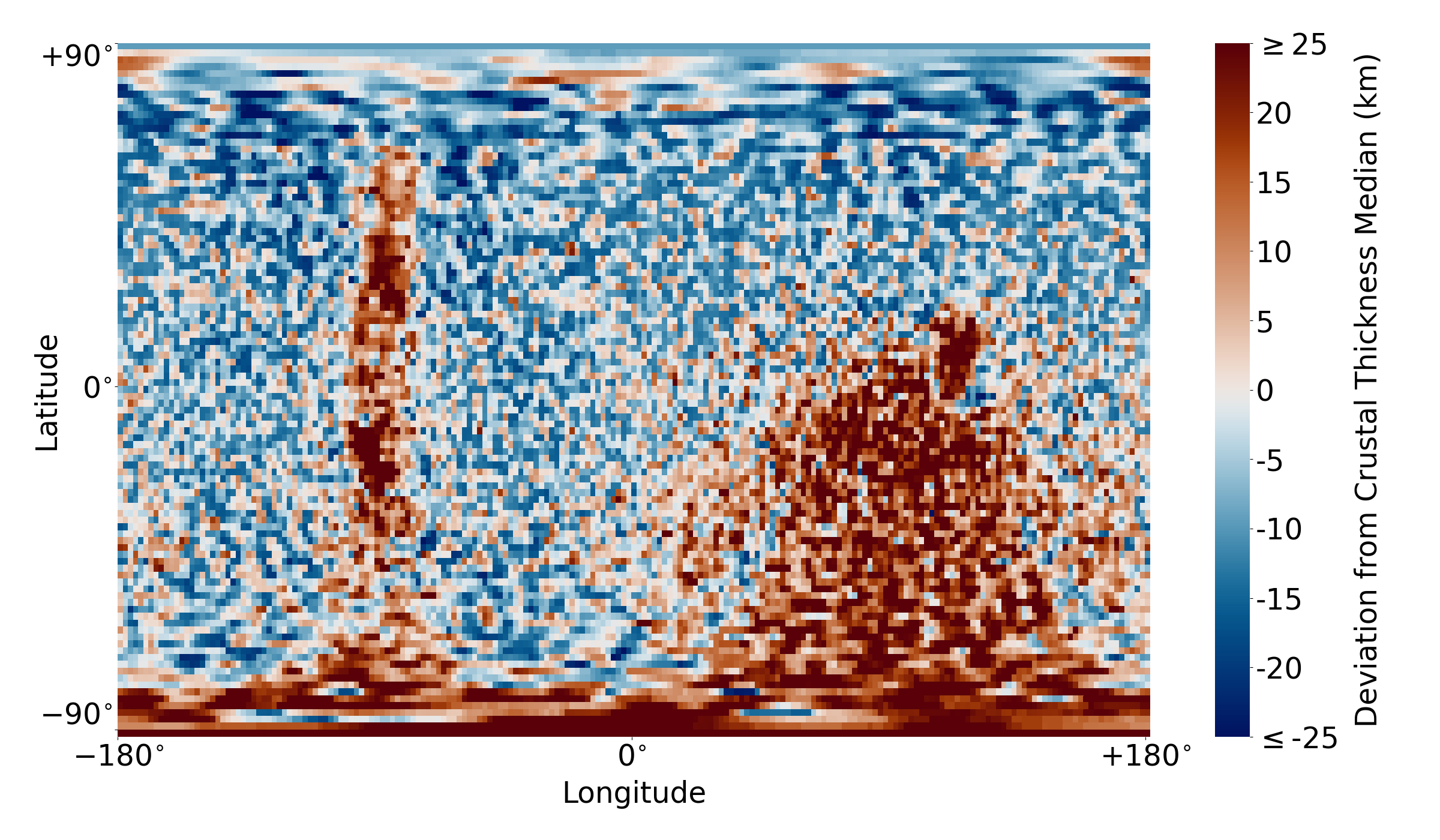}
    \end{subfigure}
    \begin{subfigure}{.49\textwidth}
        \adjincludegraphics[width=\linewidth]{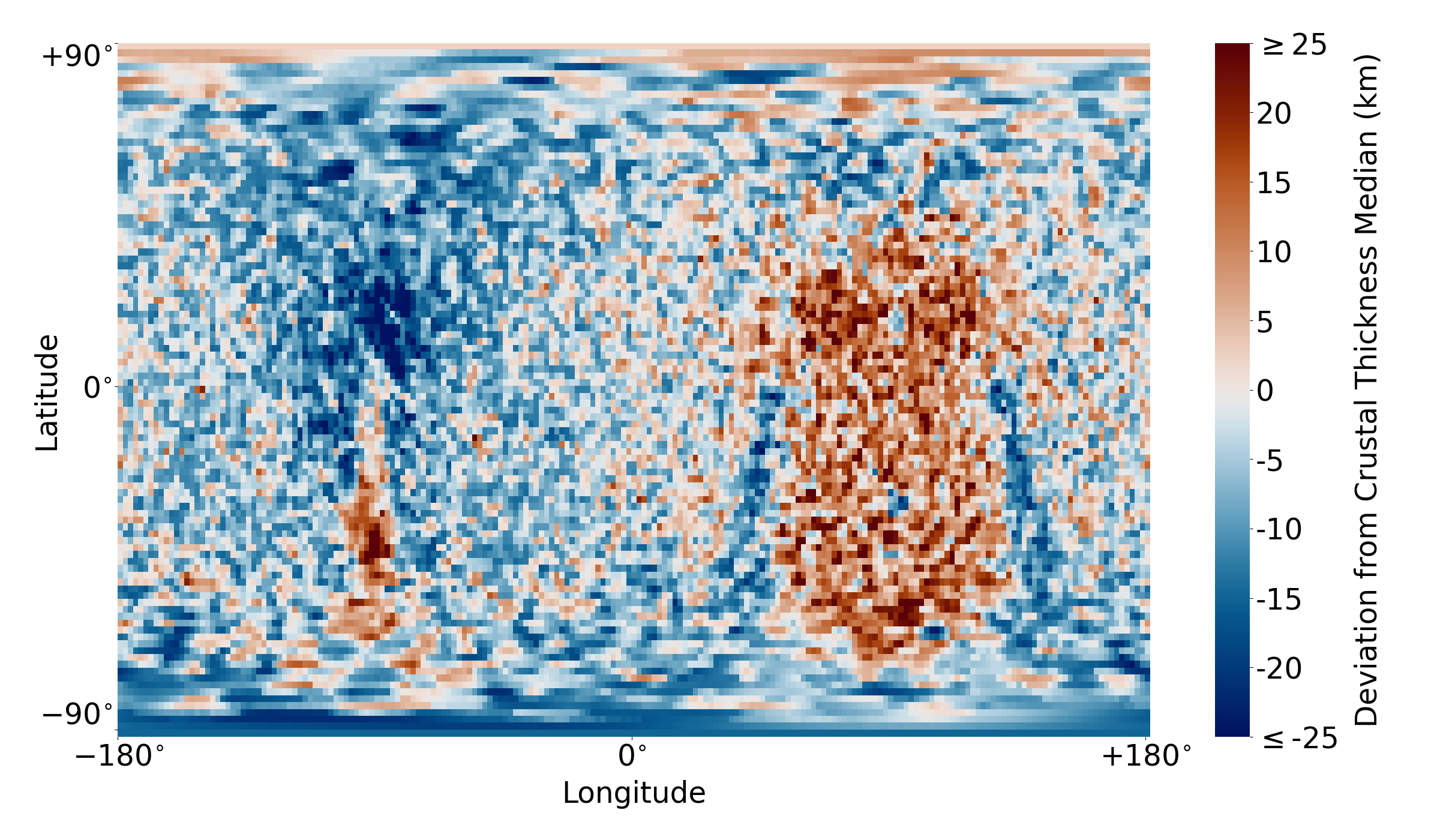}
    \end{subfigure}
    \caption{Equirectangular projection of post-impact crustal thickness for the best-fitting cases of the 1500 km-radius (left) and 2000 km-radius (right) impactors. The median crustal thickness of the \cite{Bouley2020} distribution has been subtracted for easy comparison to Figure~\ref{fig:mars_corrected}. Both of these cases correspond to a 50\% impactor core fraction, 1.2 $v_{\textrm{esc}}$ impact velocity, 60$\degree$ impact angle, the fully-stratified depletion scheme and a 25 km primordial crust, but differ in their initial mantle fertilities of 0.125 (left) and 0.1 (right).}
    \label{fig:best_largest}
\end{figure}

Although there are no cases within the 5th percentile for the largest impactor radii of 1500 km and 2000 km, there are 20 cases within the first quartile, the top 5 of which stand out as the best cases with significantly lower $\delta_{SH}$, particularly for those of the 2000 km radius impactor (Figure~\ref{fig:goodcases_deltaSH_radius}). All of these cases have an impact angle of $60\degree$ and an impact velocity that is greater than mutual escape, placing them in the ``hit-and-run'' regime, and giving them a very similar crust distribution. This consists of an elongated ellipse at the immediate grazing impact site along with a ring of re-impacting material in the impact plane that results in an iso-longitudinal band of thickened crust antipodal to the main site (Figure~\ref{fig:best_largest}).

All of these cases also correspond to a 25 km primordial crust, and interestingly, the stratified mantle depletion scheme. This is due to a large portion of the crust production from the re-impacting material corresponding to melt within the depleted layer (as the melt is only present at relatively shallow depths compared to more head-on, merging collisions). The unwanted iso-longitudinal band of crust is therefore greatly reduced in the stratified scheme when compared to the mixed scheme (\ref{fig:best_largest_fert}). Additionally, the lack of deep melt also reduces the effect of the crustal thickness dip associated with the stratified scheme (introduced in Section~\ref{classical}), as this corresponds to concentrated areas of depleted mantle at the perimeter of the impact site that have been forced to lower depths (see Figure~\ref{fig:mixedvsstrat_slice}). These cases also correspond to a low fertility of either $\beta_0 = 0.125$ for the 1500 km impactor or $\beta_0 = 0.1$ for the 2000 km impactor.

\subsubsection{1000 km Impactor}\label{1000km_results}
\begin{figure}[htbp]
    \centering
    \begin{subfigure}{.49\textwidth}
        \adjincludegraphics[width=\linewidth]{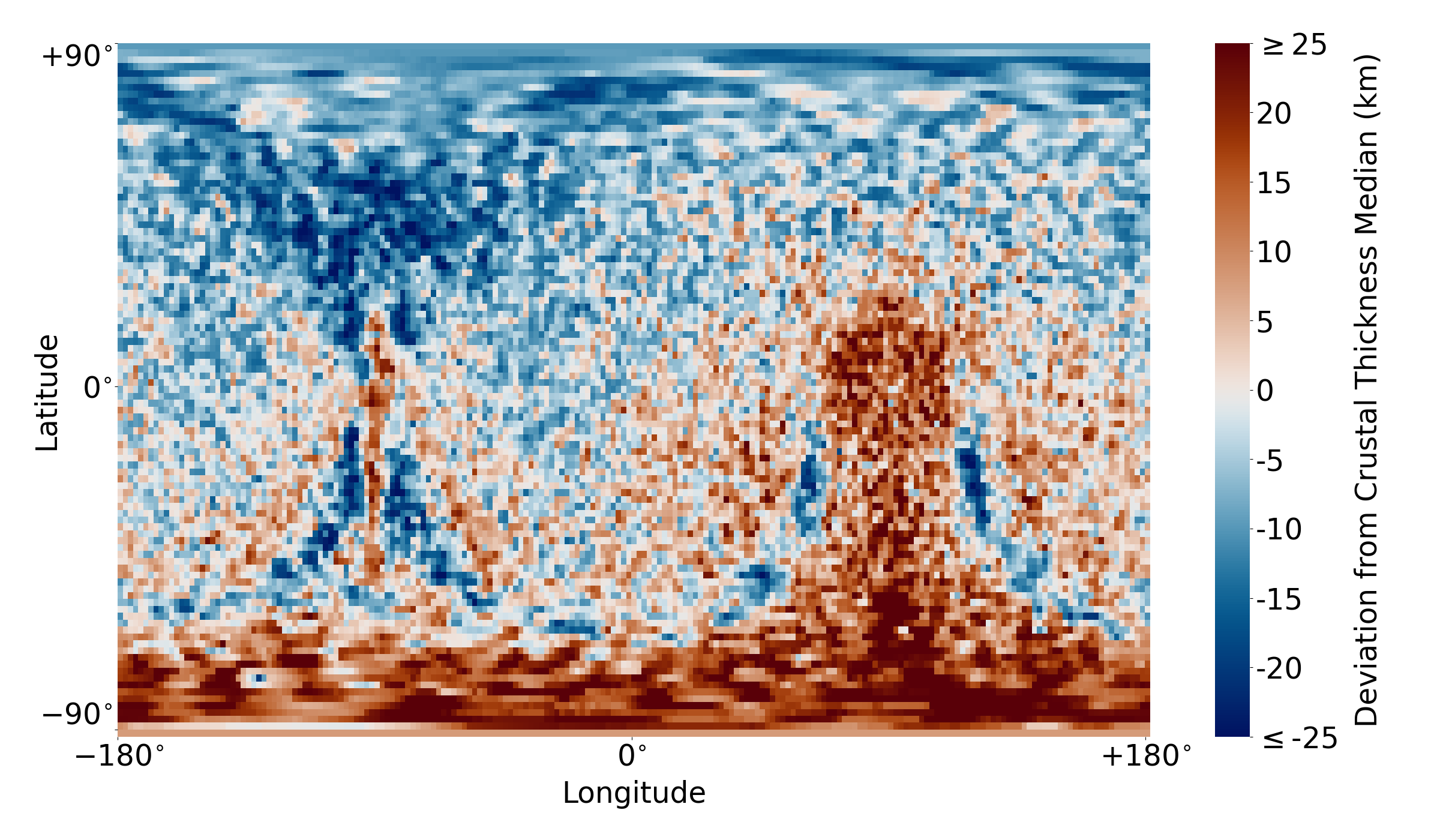}
    \end{subfigure}
    \begin{subfigure}{.49\textwidth}
        \adjincludegraphics[width=\linewidth]{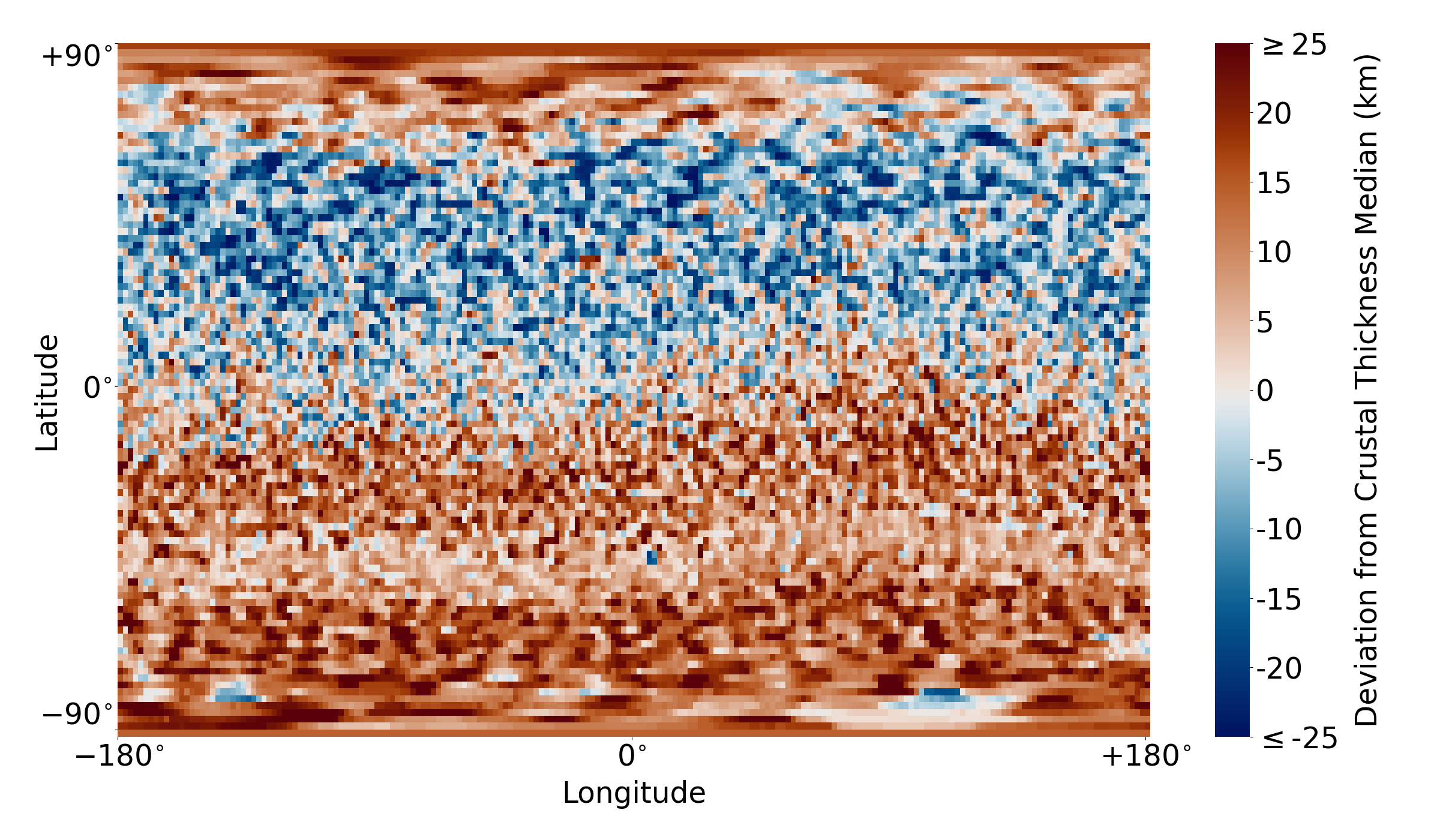}
    \end{subfigure}
    \caption{Equirectangular projection of post-impact crustal thickness for the first (left) and second (right) best-fitting cases of the 1000 km radius impactor. The median crustal thickness of the \cite{Bouley2020} distribution has been subtracted for easy comparison to Figure~\ref{fig:mars_corrected}. The corresponding parameters for these cases are: (left) 1000 km radius, 50\% impactor core fraction, 1.0 $v_{\textrm{esc}}$ impact velocity, 45$\degree$ impact angle, fully-stratified depletion scheme, 25 km primordial crust and 0.1 initial mantle fertility; and (right) 1000 km radius, 50\% impactor core fraction, 1.4 $v_{\textrm{esc}}$ impact velocity, 0$\degree$ impact angle, fully-mixed depletion scheme, 10 km primordial crust and 0.1 initial mantle fertility.}
    \label{fig:best_1000 km}
\end{figure}

While there are fewer 1000 km impactor cases within the 5th percentile than that of the 500 km or 750 km impactors, they represent some of the best-fitting cases of our entire parameter space. Two cases stand out in particular, with our spherical harmonic analysis placing them as the second and third best fits overall (Figure~\ref{fig:best_1000 km}). Both of these cases require the lowest fertility of $\beta_0 = 0.1$; however, their other parameters differ greatly. The most important difference is the impact angle, with the lowest $\delta_{SH}$ case having a $45\degree$ angle, while the other is head-on ($0\degree$). Although the two cases differ in their deviation from the Martian Dichotomy (the $45\degree$ case is too non-uniform while the head-on case produces too much antipodal crust), a clear issue in both cases is a crustal thickness contrast between the two hemispheres that is too large. This is a trend across all 1000 km cases, and explains their low representation within the 5th percentile as they only give a reasonable match to the Dichotomy with highly specific parameters.

\subsubsection{500 km Impactor}
\begin{figure}[htbp]
    \centering
    \begin{subfigure}{.49\textwidth}
        \adjincludegraphics[width=\linewidth]{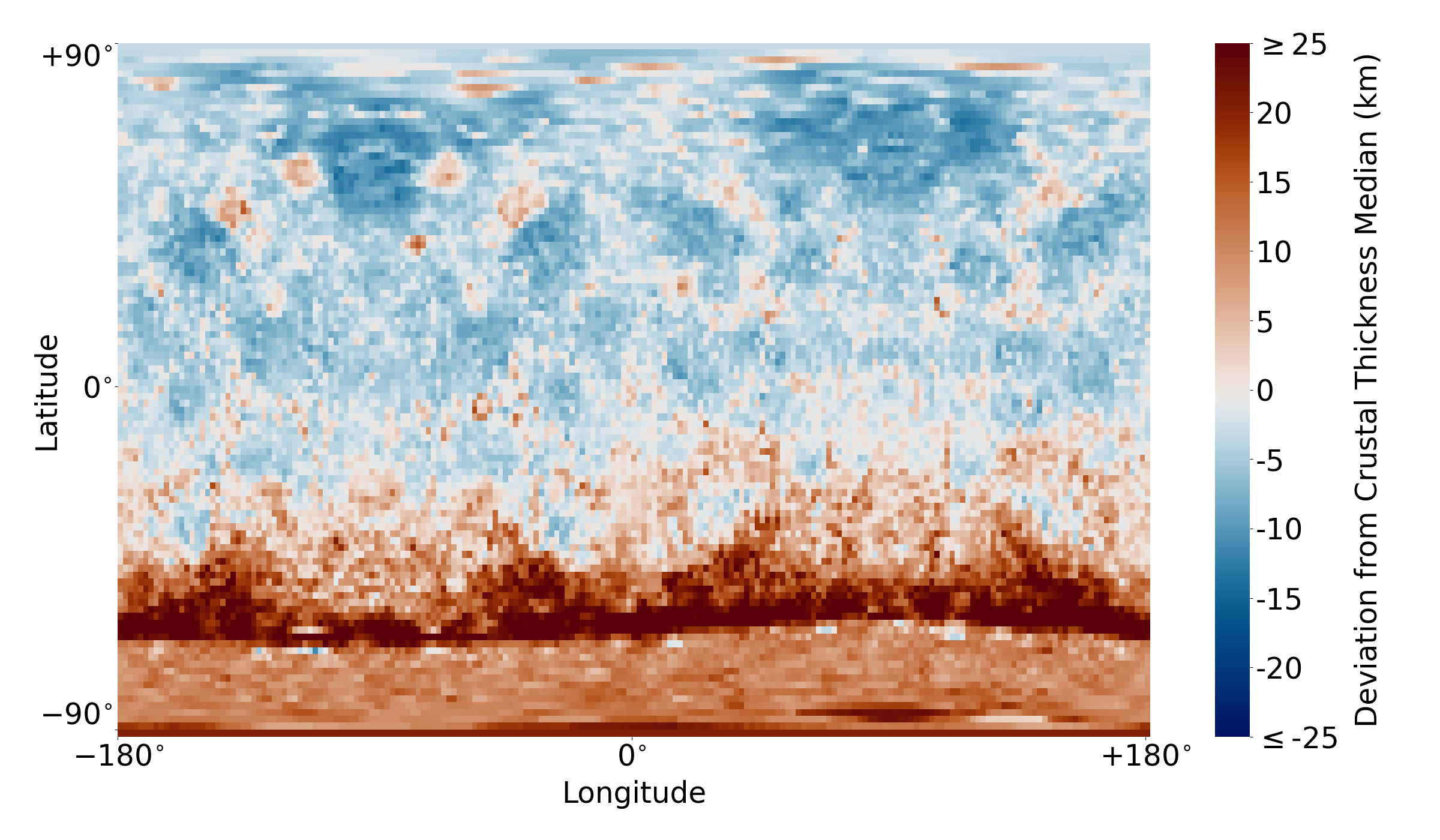}
    \end{subfigure}
    \begin{subfigure}{.49\textwidth}
        \adjincludegraphics[width=\linewidth]{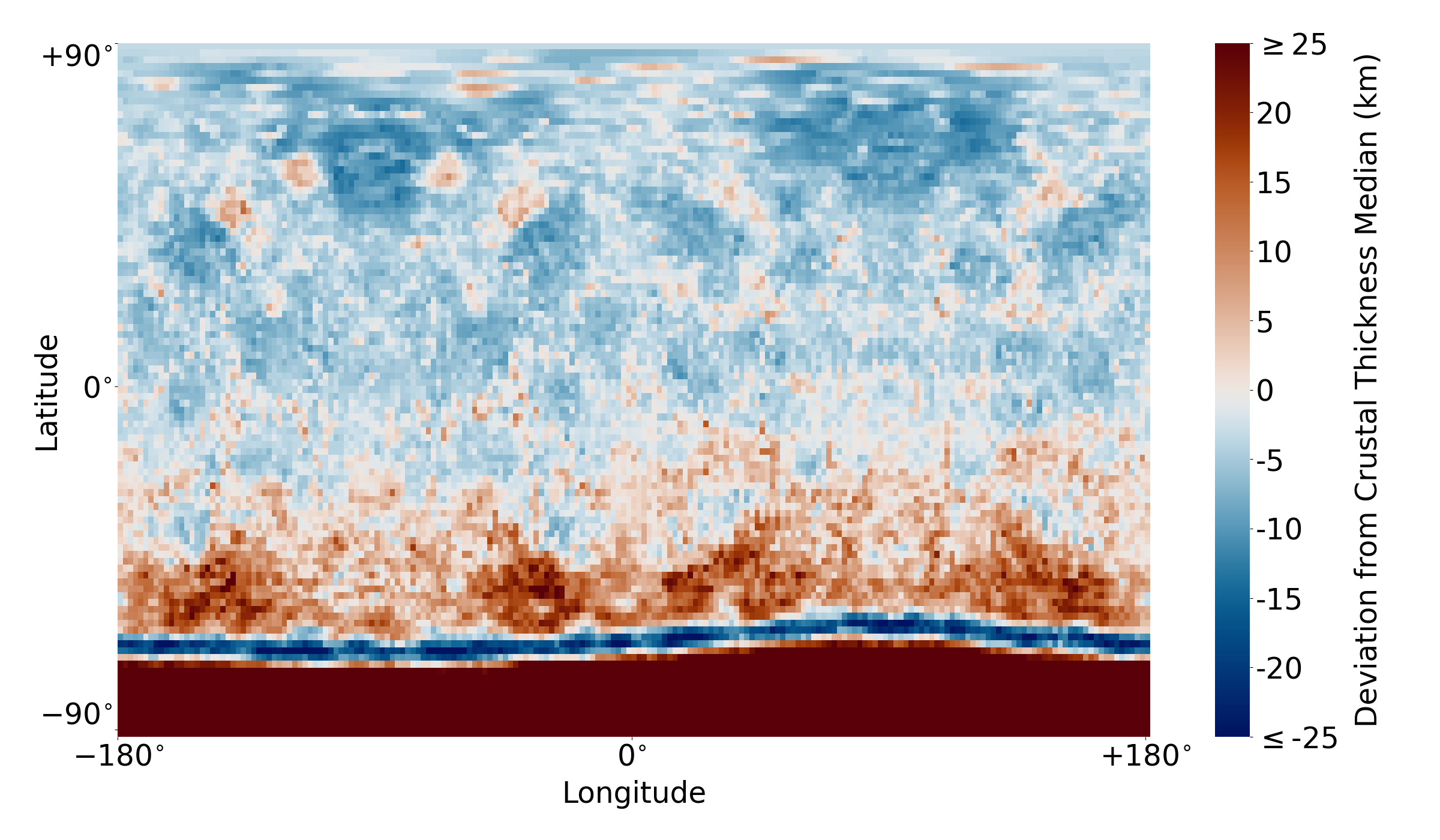}
    \end{subfigure}
    \caption{Equirectangular projection of post-impact crustal thickness for the best-fitting case of the 500 km radius impactor (left) and the equivalent case but with the fully-stratified scheme instead (right). The median crustal thickness of the \cite{Bouley2020} distribution has been subtracted for easy comparison to Figure~\ref{fig:mars_corrected}. The corresponding parameters for these cases are: 500 km radius, 25\% impactor core fraction, 1.4 $v_{\textrm{esc}}$ impact velocity, 0$\degree$ impact angle, 40 km primordial crust and 0.125 initial mantle fertility. The case using the stratified mantle depletion scheme is the 35th best-fitting case overall (the 5th percentile corresponds to 34th best case).}
    \label{fig:best_500 km}
\end{figure}

As noted in Section~\ref{goodcases_angle_radius}, the 500 km-radius impactor represents the most cases within the first quartile (44\%), yet only 18\% of those within the 5th percentile. If we look at the best case for this radius (and 7th best-fitting across all radii), we see that the average crustal thickness in each hemisphere is a much closer fit than the cases previously discussed (Figure~\ref{fig:best_500 km}, left). In comparison to the lower $\delta_{SH}$ 1000 km impactor cases of Figure~\ref{fig:best_1000 km}, however, the region of impact-induced crust falls significantly short of the equator. Additionally, the effects associated with crustal stripping at the impact site are greatly amplified, with the fully-mixed scheme leading to a sharp band of thickened crust where significant melt is present beneath un-excavated crust, while the stratified scheme leads to a band of crust with a thickness below pre-impact values where crust has been stripped and the impact-induced melt lies predominantly in the depleted mantle layer. The underlying cause of these effects being more significant in the 500 km cases compared to the large impactors also accounts for the smaller contrast between the two hemispheres: the impact-induced melt only reaches to relatively low depths, meaning crust-production is reduced overall but any remaining primordial crust represents a larger fraction of the crust-producing melt.

\subsubsection{750 km Impactor}
\begin{figure}[htbp]
    \centering
    \includegraphics[width=\textwidth]{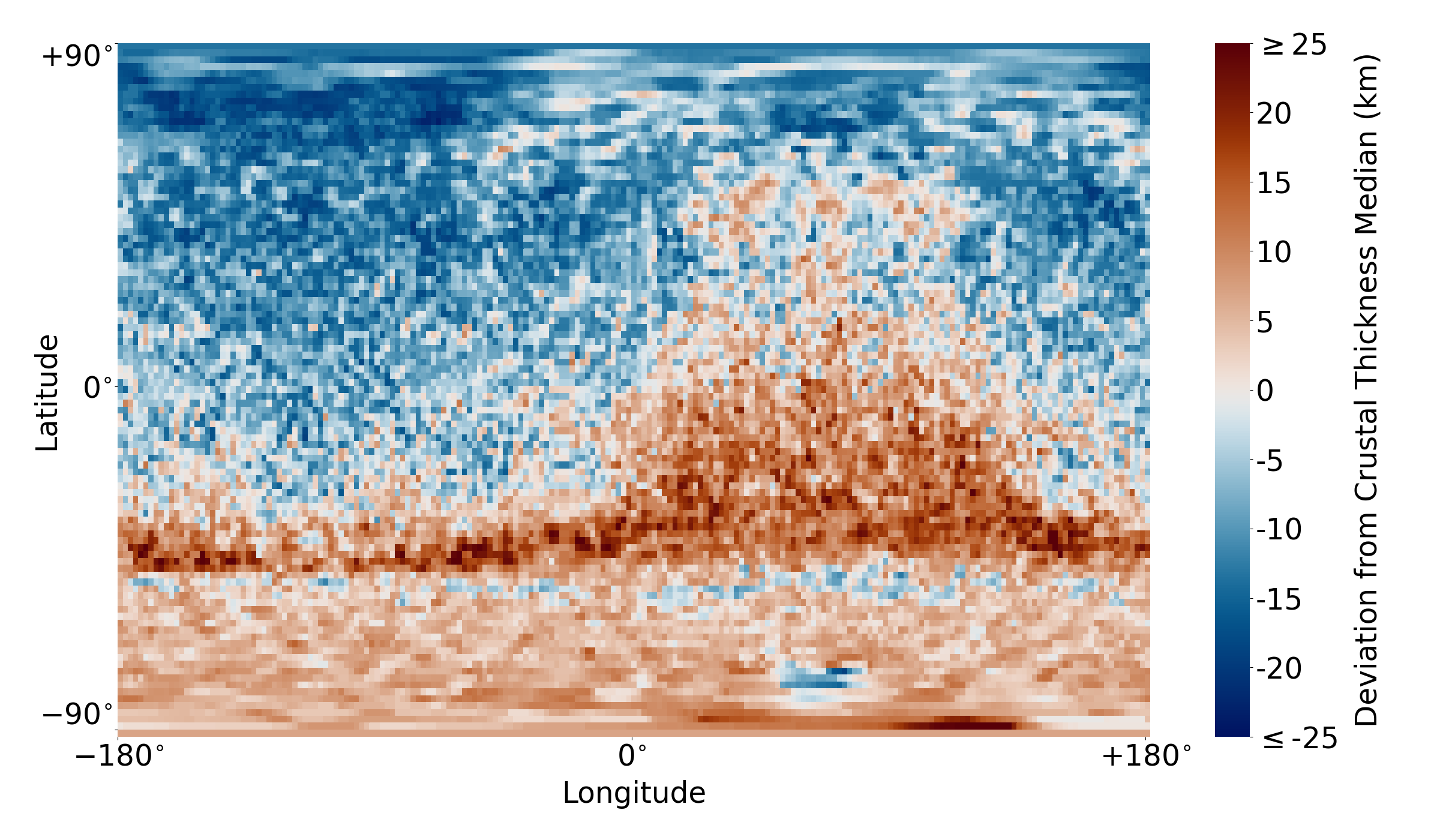}
    \caption{Equirectangular projection of post-impact crustal thickness for the best-fitting case of the 750 km radius impactor and the best-fitting case overall. The median crustal thickness of the \cite{Bouley2020} distribution has been subtracted for easy comparison to Figure~\ref{fig:mars_corrected}. The specific parameters of this case are: 750 km radius, 25\% impactor core fraction, 1.4 $v_{\textrm{esc}}$ impact velocity, 15$\degree$ impact angle, 25 km primordial crust and 0.1 initial mantle fertility.}
    \label{fig:best_750 km}
\end{figure}

The overwhelming majority of the best-fitting cases come from the 750 km impactor (65\% of those within the 5th percentile, see Figure~\ref{fig:goodcases_deltaSH_radius}), including the best case overall (Figure~\ref{fig:best_750 km}). After reviewing the drawbacks of both the 500 km and 1000 km impactors this comes as no surprise; the impactor is large enough to produce crust close to (or sometimes beyond) the equator at low impact angles, while being small enough to avoid excessive crust production and/or strong antipodal effects. Impact melting is still increased relative to 500 km cases, however, pushing the best cases to the lowest fertilities and a primordial crust thickness of 25 km.

Although there are cases with impact angles ranging from $0\degree$ to $45\degree$ within the 5th percentile, the head-on impacts fall short of the very best-fitting cases (the best head-on fit is only the 17th best-fitting case overall). Similar to the 500 km cases, the head-on cases suffer from an inability to produce melt all the way to the Martian equator. In addition, the oblique cases produce a closer match to the equator-crossing regions of the southern highlands north of the Hellas basin such as Terra Sabaea. The signal of this feature in the spherical harmonic analysis is strong, causing some cases to yield low $\delta_{SH}$ values with a visibly incorrect fit for large fractions of the Martian surface. The $45\degree$ cases suffer from this effect in particular, displaying significant iso-longitudinal bands of thickened crust from re-impacting ejecta yet corresponding with some of the lowest $\delta_{SH}$ values of the entire parameter space due to their reasonable match with the region described. Such a distinct feature would probably be visible today, making the $45\degree$ cases difficult to justify as potential Dichotomy-forming impacts.

The best case has an impact velocity of 1.4 times the mutual escape speed, and the top cases are shared between between cases with impact velocities of 1.2 $v_{\textrm{esc}}$ and 1.4 $v_{\textrm{esc}}$. 1.2 $v_{\textrm{esc}}$ cases are still represented within the 5th percentile; however, it is clear that they do not have enough kinetic energy to produce the required area of melt.

\subsection{Northern Excavation}
\begin{figure}[htbp]
    \centering
    \begin{subfigure}{.49\textwidth}
        \adjincludegraphics[width=\linewidth]{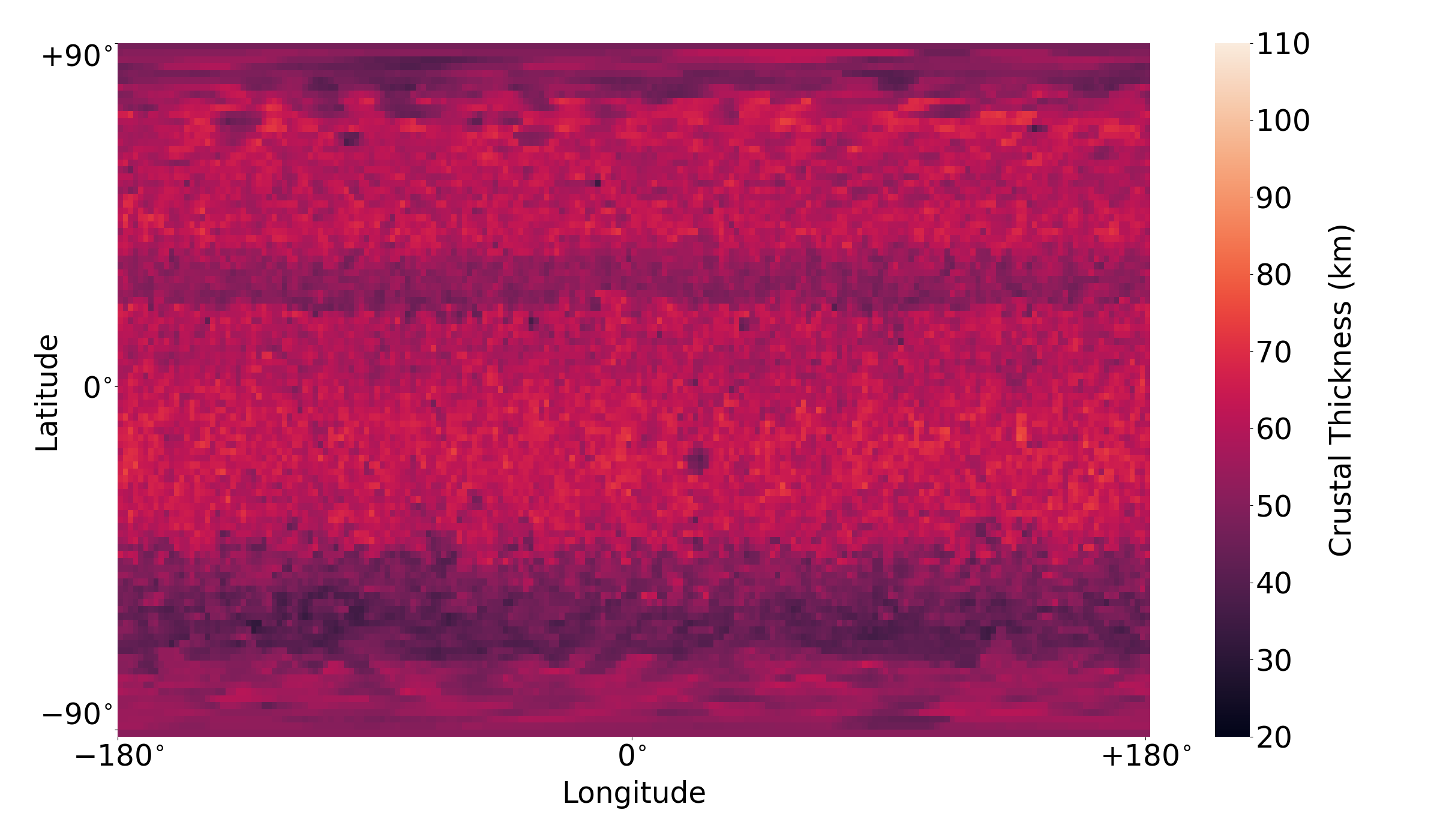}
    \end{subfigure}
    \begin{subfigure}{.49\textwidth}
        \adjincludegraphics[width=\linewidth]{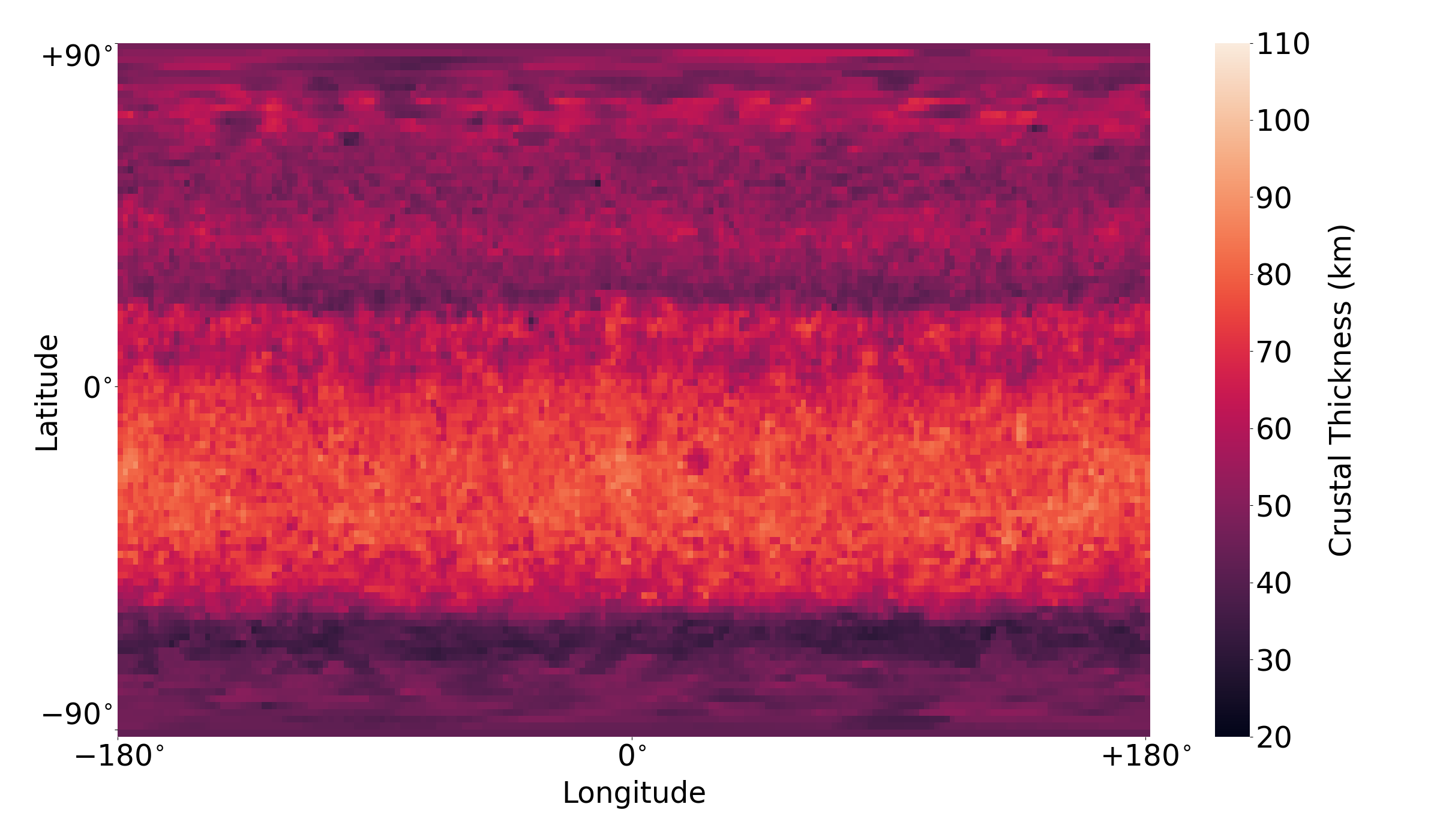}
    \end{subfigure}
    \begin{subfigure}{.49\textwidth}
        \adjincludegraphics[width=\linewidth]{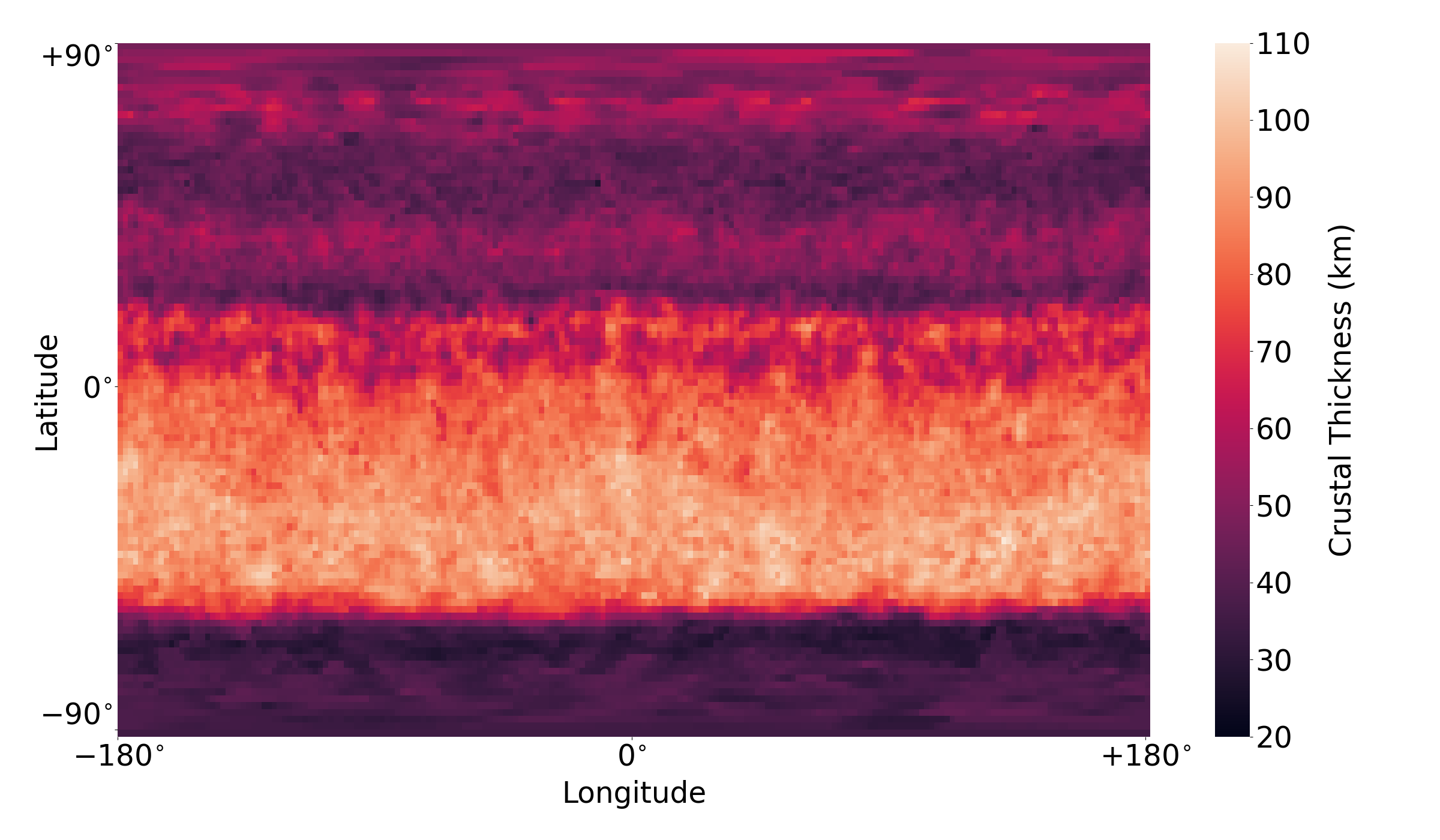}
    \end{subfigure}
    \begin{subfigure}{.49\textwidth}
        \adjincludegraphics[width=\linewidth]{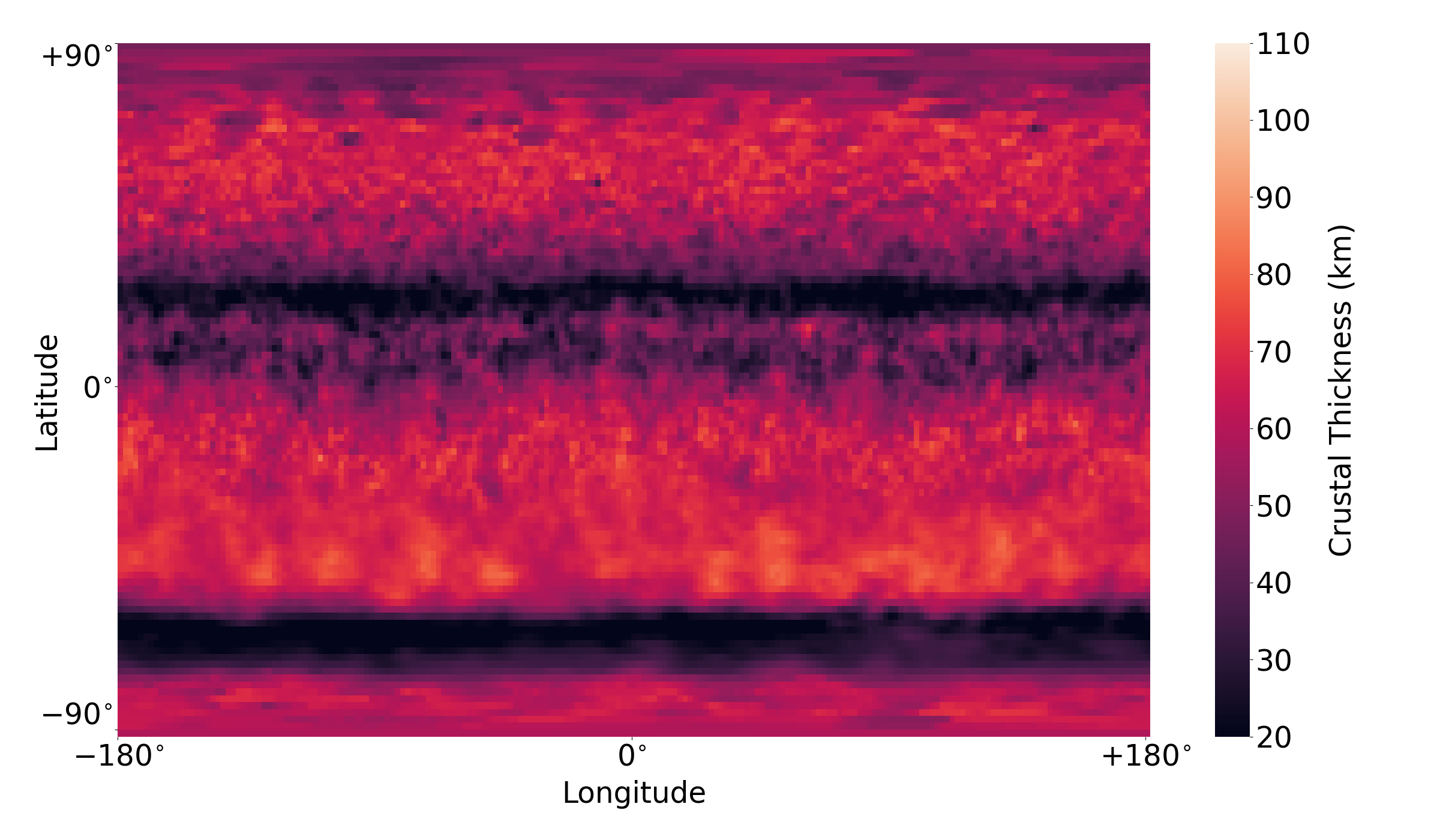}
    \end{subfigure}
    \caption{Equirectangular projection of post-impact crustal thickness for an example northern excavation case. The top row shows this case with a 10 km (left) and a 25 km (right) primordial crust thickness using the fully-mixed mantle depletion scheme. The bottom row shows this case with a 40 km primordial crust thickness for the fully-mixed (left) and fully-stratified (right) mantle depletion schemes. The other parameters are: 1500 km radius, 25\% impactor core fraction, 1.0 $v_{\textrm{esc}}$ impact velocity, 0$\degree$ impact angle and 0.1 initial mantle fertility. The orientation of the projection has been inverted compared to the other crustal thickness maps of this study such that the impact occurred in the northern hemsiphere, as to represent the ``Borealis Basin'' hypothesis for the Martian Dichotomy.}
    \label{fig:excavation}
\end{figure}

The spherical harmonic analysis did not identify any Dichotomy-like cases that resulted in a thinner crust within the impact site than outside of it. The only cases with $\delta_{SH} < \delta_{SH_{max}}$ corresponding to such a distribution are a handful of highly oblique, high velocity hit-and-run impacts with the largest impactors; however, these produce basins that are far too small in comparison to the northern depression of Mars. This issue of an undersized impact basin is widespread across our parameter space. Even at our largest impactor radii, the region of excavated crust falls short of the Dichotomy boundary. In addition, such energetic collisions cause strong antipodal effects not seen in Mars' southern hemisphere today (Figure~\ref{fig:excavation}). These effects result from the impact-induced shockwave travelling through the planet until converging at the antipodal free-surface boundary, leading to significant disruption of this antipodal material. If the antipodal stresses are greater than the tensile strength of the material then fracturing will occur \citep{Schultz1975}, and for planetary-scale impacts the shock is energetic enough to excavate the antipodal material. In our simulations, these effects result in a crustal thickness of this antipodal region that is lower than that of the impact site, as the crust is entirely excavated but with far less melt. As expected, these cases require the lowest fertility ($\beta_0 = 0.1$) to avoid too much crust production at the impact site; however, further depletion through primordial crust formation is necessary to reduce impact-induced thickness to the desired range. This additional mantle depletion causes a trend in crustal thickness distribution, whereby larger primordial thicknesses lead to deeper impact basins, due to both reduced crust-production at the impact site and a thicker pre-impact crust. This therefore pushes the northern excavation scenario to larger primordial crust thicknesses which, with the addition of any global interior melting caused by the impact shockwave, result in a crust outside the impact site that is significantly too thick.

\section{Discussion}\label{discussion}
\subsection{Material Strength in the Giant Impact Regime}
{It has already been established that material strength plays a significant role in planetary-scale impacts on a Mars-like planet, through independent studies utilising both SPH simulations \mbox{\citep{Emsenhuber2018}} and the Simplified Arbitrary Lagrangian Eulerian hydrocode, iSALE-2D \mbox{\citep{Monteux2019}}. More recently, strength has also been shown to be important at Pluto-scales \mbox{\citep{Denton2021,Ballantyne2022a}}. Prior to this work, however, there has been no thorough investigation into the effect material strength has on the antipodal features that occur in such large-scale impacts.

\mbox{\citep{Marinova2008,Marinova2011}} found that antipodal heating and crust excavation can occur for a near head-on giant impact on Mars. These results were determined using an SPH model that did not include any strength, treating materials as inviscid fluids (with the exception of the artificial viscosity used to handle shocks in SPH). In such simulations, the shock dissipates only through the irreversible compression intrinsic to shock events. When strength is included, additional energy is partitioned into plastic deformation, causing the shock wave to dissipate faster for a given distance travelled. The larger the shear strength, the more energy contributes to this effect and thus the faster the shock decays. This has already been demonstrated in previous impact simulations at smaller scales \mbox{\citep{Leinhardt2009}}. For planetary-scale impacts capable of inducing significant antipodal disruption, such as those investigated in this study, material strength therefore acts to reduce the energy carried to the impact antipode. This explains the lack of significant antipodal melting seen in the ``classical'' head-on impact case of Figure~\mbox{\ref{fig:fluidcases}} that included material strength in the simulation. This work therefore provides clear evidence that material strength is important for collisions between bodies (at least) up to the mass-scales considered here, further supporting the aforementioned studies.}

\subsection{Borealis Basin or Southern Magma Ocean?}\label{Borealis}
A clear result of this work is a lack of support for the Martian Dichotomy to have formed through the canonical northern impact basin hypothesis (the so-called ``Borealis Basin''), contrary to previous works \citep{Marinova2008, Andrews-Hanna2008, Nimmo2008, Marinova2011}. We attribute our alternative findings to our method of crust production calculation, which uses separate solidus and liquidus models relevant for the Martian mantle to quantify melt along with a modern estimate of crustal density (see Section~\ref{crustproduction}), rather than a simplified approach based on internal energy and/or pressure limits. We also use the more sophisticated ANEOS equation of state than the \cite{Tillotson1962} semi-empirical formula used in previous studies, and include material strength in our models which we show to be important on these scales. Furthermore, we employed a more sophisticated model of primordial crust than that of previous work. These differences lead to our results yielding much more significant crustal thickening from the impact, largely due to more widespread melt. 

One possibility that could reduce the prominence of this effect is that Mars was far colder than considered in this study prior to the impact, with interior temperatures significantly below the solidus temperature. Additional cooling time may be permitted if the Martian magma ocean crystallised extremely rapidly within 20 Myr after Solar System formation, as suggested by the radiometric dating of Martian meteorite zircons \citep{Bouvier2018}. For efficient cooling to occur, Mars would need an alternative tectonic regime to its current insulating stagnant lid, such as plate tectonics \citep{Breuer2003}. However, there is no strong evidence for a plate tectonics episode on early Mars \citep{Pruis1995,Zuber2001,Halliday2001}. Moreover, the precise time of Martian magma ocean crystallisation remains a topic of debate, with several studies proposing significantly longer timescales \citep[e.g.][]{Borg2016,Debaille2007,Kruijer2017,Kruijer2020a}. Alternatively, Mars may have started cool from birth if it formed predominantly via ``pebble accretion'' rather than more traditional planetesimal accretion-based methods \citep{Johansen2021}; however, this may prohibit any silicate melting, making planetary differentiation and primordial crust formation difficult. 

The excavation of crust antipodal to the impact site associated with our basin-forming impacts is an aspect that is particularly difficult to reconcile with present-day observations. If the northern hemisphere truly is the result of such an impact, an extremely deep canyon should be present near the Martian south pole, which is clearly not the case. Although the presence of solid strength in our model does act to reduce antipodal effects, the crustal stripping remains prominent. With all these considerations in mind, we consider the Borealis Basin case very unlikely.

\subsection{An HPE-induced Crustal Dichotomy}
\cite{Citron2018} suggested that the Martian Dichotomy may have been caused by a giant impact in the north that led to degree-1 mantle convection with an upwelling beneath the southern hemisphere, antipodal to the impact site. This results from heat-producing elements (HPE) expected to be concentrated in a primordial crust \citep{Elkins-Tanton2003,Elkins-Tanton2005} being stripped from the impact site, leading to a southern hemisphere crust that is both enriched in HPE and thicker than its northern counterpart. The insulating effect of these properties would cause a temperature increase in the mantle below the southern highlands, causing the aforementioned upwelling and subsequent crust-production in the southern hemisphere. Such a scenario relies upon the same assumptions as a Borealis-forming impact: crustal excavation of the entire northern lowlands region with very little melt and thus crust-production at the impact site. For the same reasons detailed in Section~\ref{Borealis}, therefore, our results do not support this hypothesis.

\subsection{Explaining low fertility}\label{discussion:lowfert}
At all impactor radii in our model, the best cases have Martian bulk silicate fertilities of 0.1-0.125. These values are significantly lower than that of the bulk silicate Earth which, if we use the pyroxene fraction as a proxy for potential crust fraction due to its prominence as the main component of a basaltic to andesitic crust, is predicted to be $\sim0.2$ \citep{Mckenzie1988}. Predictions for the bulk silicate composition of Mars are consistently different to those of Earth \citep[][and references therein]{Taylor2013,Yoshizaki2020}, but whether these differences could account for such a lower crust-production budget is unclear.

{Throughout this study, the neutral buoyancy pressure and melt extraction threshold were held constant at 7.4 GPa and 4\%, respectively. While these values do not affect the distribution of melt indicated by the SPH impact simulations, changing them would naturally alter the amount of crust produced from a given region of melt, thus potentially leading to a different favoured range of bulk silicate fertilities. A larger melt extraction threshold, for instance, would lead to less crust being produced by the impact, but only in regions dominated by low melt fractions. For the majority of the impacts investigated in this study, the main impact site displays very high melt fractions (near-100\%) down to depths well below the neutral buoyancy depth, and so a change in the extraction threshold should not change the volume of crust expected to form in the region. Instead, the regions beyond the main impact site would be most affected, although the changes should not be large for melt thresholds within a reasonable range as most regions of melt still have relatively large melt fractions ( $>$10\%). This was confirmed by re-analysing the most promising case using melt extraction thresholds of 2--8\%, the results of which can be seen in Figure~\mbox{\ref{fig:melt_thresh}}. A different neutral buoyancy pressure, on the other hand, most significantly affects the main impact site, although any regions with melt at depths below the nominal neutral buoyancy depth (and thus higher pressures) are affected, which includes the majority of the impact-induced magma ponds encountered in this work. Figure~\mbox{\ref{fig:buoyP}} shows the crustal thickness distribution for the most promising case when assuming a different neutral buoyancy pressure (7 or 8 GPa). Here, we see that a relatively small change in neutral buoyancy depth has a large consequence on crustal thickness. This is because a small change in depth (or equivalently pressure) leads to a large change in mantle volume and thus mass of applicable melt for crust production. Many other modelling approaches involving melt extraction from the Martian mantle also use a constant neutral buoyancy pressure of 7.4 GPa (or equivalently a depth of $\sim$600 km) \mbox{\citep[e.g.][]{Breuer2006,Keller2009,Turbet2020,Drilleau2022}}, indicating that further work in accurately constraining this value is necessary in the future.}

Some effects during magma ocean crystallisation not included in our simplified model may reduce crust production efficiency, such as permeability barriers \citep{Schools2018}, but these effects are generally only significant at lower melt fractions than those associated with this study. One effect that may be of significance is the tendency for iron-enriched cumulates to solidify last during magma ocean solidification \citep{Elkins-Tanton2003,Elkins-Tanton2005}. As the mantle liquidus is steeper than its adiabat, the magma ocean should freeze ``bottom-up'' with lower depths solidifying first \citep[e.g.][]{Abe1997}. This therefore leads to a compositional gradient of increasing iron-enrichment and hence density towards the surface; a gravitationally unstable configuration susceptible to Rayleigh-Taylor overturns that may drive much of this material down into the deep mantle \citep{Samuel2021,Boukare2018,Ballmer2017}, thus removing a significant portion of crustal material. Whether or not a magma ocean would truly solidify in this way is a matter of debate, however, with various studies suggesting that magma oceans may in fact freeze ``middle-out'' whereby intermediate depths solidify first, creating an expanding solid region between two simultaneously-crystallising magma oceans \citep{Caracas2019,Labrosse2007,Nomura2011}, although such studies generally consider depths (and thus pressures) much greater than those of Mars. Moreover, the hot material that would rise from depth during such an overturn may experience melting due to adiabatic decompression \citep{Elkins-Tanton2003}, making it difficult to estimate the final influence of this phenomenon on our results.

\subsection{Primordial crust thickness}
The most promising primordial crust thickness in this study was found to be 25 km; however, reasonable fits were still found with 10 km and 40 km thicknesses, depending on the impactor radius. Thicknesses of 55 km and above, however, did not yield any cases of interest. These findings agree with geochemical analyses of Martian meteorites, which place an upper limit of $\sim45$ km \citep{Norman1999}, and suggest a most probable value of 20-30 km which matches with geophysical models \citep{Morschhauser2011}. A primary crust as thin as 10 km may be difficult to justify; however, all cases of interest with such a crust correspond to the 1000 km-radius impactor, which still result in far too much crust at the impact site (and often its antipode) and can likely be ruled out (as discussed in Section~\ref{1000km_results}).

\subsection{Additional features of the simulations}
Strong antipodal effects are not the only features present in some of our best-fitting cases that may be difficult to reconcile with observational knowledge of Mars. In all of our 45$\degree$ cases, for instance, we find a strong band of thickened crust that forms an iso-longitudinal ring of thickened crust in the impact plane around the target body; something not seen on Mars today. However, as our crust-production model is based on the immediate position of post-impact melt and is unable to simulate smaller-scale spreading at the surface that would occur over longer timescales, such a sharp feature may relax into a much shallower region that covers a large area in reality. The extent of this spreading is strongly dependent on post-impact topography, however, and our simulations display significant excavation in these regions due to the re-impacting material (\ref{excavation}). Thus, the majority of this melt should be contained within this excavated valley until crystallisation ultimately freezes the feature's morphology to the present day. Considering Mars' low level of resurfacing due to its lack of tectonic activity, we therefore consider such cases to be highly unlikely candidates for the Dichotomy-forming impact.

Another feature that is present in almost all of our cases, and particularly those using the fully-mixed mantle depletion scheme, is a thickened band of crust at the perimeter of the impact site. As described in Section~\ref{classical}, this is caused by a crossover region whereby the impact shock wave is no longer energetic enough to strip the region of crust, but is still capable of non-negligible melting, thus resulting in a region that receives crust contributions from both the undisturbed primordial crust and any melting of the mantle below. Unlike the iso-longitudinal rings, the spreading of this melt as it cools would likely be very important in determining the final distribution of crust. With little excavation, there is no reason to think that this melt would remain contained to its immediate longitude-latitude coordinates and would likely spread out considerably across the surface, leading to a much weaker feature in the final crust distribution. This impact-perimeter thickening is most prevalent in the 500 km-radius impactor cases, and such melt-spreading may in fact improve their likelihood in explaining the Dichotomy, as they all suffer from an impact site that is too small, but may increase in size considerably if such spreading was accounted for. On the other hand, cases that under-produced melt, such as many of the 500 km cases with the lowest fertility, would experience even thinner crusts than predicted here, making them unlikely candidates.

\subsection{Depletion schemes}
The nature of Mars' mantle differentiation is still under scientific debate. Unlike Earth or the Moon, ancient chemical heterogeneities are preserved within Mars' mantle which provide evidence for both the presence of an early magma ocean phase and a lack of vigorous whole mantle convection and/or active plate tectonics, although some degree of mixing through convection immediately after the magma ocean stage is likely \citep[][and references therein]{Mezger2013}. The depth of the magma ocean during this phase is unclear, with estimates ranging from 700 km to the entire mantle down to the core-mantle boundary \citep[e.g.][]{Richter1998, Righter2011,Debaille2008,Elkins-Tanton2003,Elkins-Tanton2005}; however, a fully molten mantle is not consistent with with observed depletions of moderately siderophile elements \citep{Mezger2013,Righter2011}. A relatively shallow magma ocean and lack of whole mantle convection implies that stratification within the Martian mantle is possible. Indeed, even the interior of Earth has been proposed to contain an undepleted layer of primordial mantle below $\sim$1000 km depth \citep{Xiang2021}. A primordial-depleted boundary is not the only form of compositional stratification that can occur, however: the mantle overturn described in Section~\ref{discussion:lowfert} may cause a build-up of heat-producing elements at Mars' core-mantle boundary that can have a dramatic influence on the planet's thermo-chemical evolution \citep{Samuel2021}. On the other hand, if the magma ocean phase lasts sufficiently long ($\gtrsim 1$ Myr, which may be the case due to the build-up of a thick, insulating atmosphere generated by volatile degassing \citep{Lebrun2013,Zahnle2007,Nikolaou2019}), then solid-state convection may occur in the mantle during magma ocean crystallisation, which would avoid any global overturn and lead to a much more homogeneous, well-mixed mantle \citep{Maurice2017}. Clearly, the Martian interior is very unlikely to be described by a ``one-size fits all'' style of approach, and the reality of its compositional layering prior to impact was most likely at some intermediate between our two end-member cases.

While our results favour the fully-mixed mantle depletion scheme over the fully-stratified model, this is predominantly due to the latter model producing a crustal pattern at the immediate impact site that generally consists of too much crust at the centre encircled by a region of very thin crust. All of these features are contained within the excavated region of the impact site, meaning that such a sharp contrast would likely disappear if we were to take turbulent mixing of the entirely molten magma ocean into account. The overall excess in post-impact crustal thickness relative to the fully-mixed model is a robust result, however, as we see that the impact causes large-scale displacement of the depleted material away from the impact site (\ref{fig:mixedvsstrat_slice}). Our analysis should therefore still identify the most likely parameters for a Dichotomy-forming impact, regardless of mantle depletion scheme; however, any potential mantle stratification that involves less depleted material at depth will act to increase crust production at the impact site, thus suggesting that our pre-impact fertility values may in fact need to be lower than those suggested in our fully-mixed cases.

\subsection{Impact Statistics}
The probability distribution of impacts on Mars in the early Solar System is not well understood. Various N-body models have been used to investigate the formation and accretionary evolution of the Solar System, but these generally cannot resolve individual impactors smaller than $\sim$1000 km radius, instead using tracers representative of swarms of smaller bodies \citep[e.g.][]{Jacobson2014,Rubie2015,OBrien2006}. In addition, classical models systematically produce planets at Mars' semi-major axis that are 5-10 times too massive \citep{Wetherill1991,Chambers2001,Raymond2009}, leading to various proposed scenarios such as particularly eccentric \citep{Raymond2009,Morishima2010} or migrating \citep{Walsh2011,Raymond2014} gas giants, making concrete predictions even more difficult. Nonetheless, it is certain that smaller bodies are more common, as this is both predicted by models \citep[][and references therein]{Raymond2014a} and verified by observations of near-Earth objects \citep{Michel2007}, the main asteroid belt \citep{Bottke2005,Ryan2015} and the Kuiper belt \citep{Bernstein2004,Fraser2014,Adams2014,Morbidelli2021}. For impact velocities, values of $\sim$1.4 $v_{\textrm{esc}}$ are expected according to classical arguments considering self-stirring of similar-sized bodies \citep{Safronov1969}, and N-body simulations confirm this, with \cite{Emsenhuber2020,Emsenhuber2021} finding a median impact velocity of 1.6 $v_{\textrm{esc}}$ but with an increased probability for lower values (the first and third quartiles are at 1.22 $v_{\textrm{esc}}$ and 2.12 $v_{\textrm{esc}}$, respectively). Impact angles are also stochastic in nature, and are usually considered to follow the \cite{Shoemaker1962} probability distribution $dp(\theta_{coll}) = \sin{2\theta_{coll}}~d\theta_{coll}$ which predicts 50\% of impacts to occur within angles of 30-60$\degree$, centred on the most probable angle of 45$\degree$.

Using this statistical knowledge, we can further speculate on which of our cases represent the most-likely impact scenario to form the Dichotomy. The largest impactors (radii $\geq$ 1500 km) can almost certainly be ruled out, as their sizes make them highly unlikely impactors and they do not represent any significant cases. Their ability to produce large, shallow magma ponds that leave very little compositional traces from the target are certainly an interesting regime that should be further examined in future studies, however. As already stated, the 1000 km impactor is difficult to justify, and its size only weakens its case further. Apart from their differing radii, the remaining impactor sizes (radii of 500 km and 750 km) are remarkably similar in their impact angle and velocity preferences. Most notably they both favour velocities of 1.2-1.4 $v_{\textrm{esc}}$ and angles of 15-30$\degree$, in contrast to previous studies \citep{Golabek2018,Marinova2011,Marinova2008}. As the 750 km impactor proved to be the best-fitting radius in our analysis, we still consider it to be the most likely size to form the Dichotomy in spite of the probability trend towards smaller impactors; however, we predict that many cases would be favourable at radii in the intermediate of our two smallest impactors.

\subsection{Pre-Impact Rotation}
{All simulations were set up with bodies that did not have any spin prior to impact. At present, Mars rotates with a period of 24.6 hours. It has been shown that a 1000 km radius body impacting Mars at the mutual escape speed and with a 45$\degree$ impact angle (i.e. the ``classical'' oblique case) can reproduce this rotation rate when no pre-impact spin was present \mbox{\citep{Marinova2011,Emsenhuber2018}}. The most promising cases identified in this study (impactor radii of 500 km and 750 km), on the other hand, do not carry enough angular momentum to introduce such a spin rate from a single impact. However, the spin of Mars prior to the impact is not well understood, meaning that the angular momentum introduced by the Dichotomy-forming impact may only be the final contribution to an already spinning Mars, leading to the Martian rotation rate observed today.

Planet formation via planetesimal accretion alone suggests rotational periods much longer than those of the terrestrial planets \mbox{\citep{Lissauer1991,Dones1993,Miguel2010}}. Pebble accretion may produce shorter periods in some regimes \mbox{\citep{Johansen2010,Visser2020}}, although protoplanetary envelopes could weaken this effect if taken into account. Giant impacts are capable of providing the required angular momentum \mbox{\citep[e.g.][]{Wetherill1985}}, but they should produce an isotropic distribution of obliquities rather than the preferentially prograde spins seen in the Solar System today \mbox{\citep{Safronov1966,Miguel2010,Visser2022}}. In fact, the high likelihood for the classical oblique case to introduce a retrograde spin could be reason to favour a smaller, less oblique impact within the most-promising parameter-space identified in this study; in this case, if there is some hidden mechanism that led to the fast, prograde spins of the terrestrial planets, the stochastic Dichotomy-forming impact would not be capable of causing a significant deviation from the trend.

The shock wave caused by the impact passes through the target body much faster than the rotational period, and thus introducing a pre-impact spin would not significantly affect the results of this study. This can be demonstrated through a simple, back-of-the-envelope calculation: the sound speed of forsterite is $\sim$5.6 km/s \mbox{\citep{Suzuki1983}}, meaning that the shock must pass through the $\sim$6779 km diameter of Mars in a minimum of 6779/5.6 = 1210.5 s, which is only $\sim$1\% of Mars' current rotational period. An additional effect of rotation is the deviation of a planets geometry from a perfect sphere to an ellipsoid. For smaller scale collisions such as the Chixculub impact, this effect has been shown to weaken the antipodal effects of an impact \mbox{\citep{Meschede2011}}. However, for the planetary-scale impacts considered in this study, such effects are not significant. The observed contrast in the equatorial radius and average polar radius of Mars is 20 km \mbox{\citep{Archinal2018}}, whereas the impacts that display significant antipodal effects experience surface deformations of 100s--1000s of kilometres.}

\section{Conclusions and future outlook}\label{conclusions}
\subsection{Conclusions}
Through a large suite of SPH impact simulations coupled to a simplified crust-production model, we have identified the most favourable parameters for an impact-induced Martian Dichotomy. The most important conclusions can be summarised as follows:
\begin{enumerate}
    \item The northern lowlands of Mars cannot have formed through a classical impact basin forming impact event, as it is not possible to produce such a large crater without excessive crust production due to a deep magma ocean at the impact site and/or strong antipodal effects. If the Martian Dichotomy formed through an impact it must therefore have occurred in the southern hemisphere of Mars, where the impact-induced melt crystallised to form a thicker crust at the impact site than its relatively undisturbed antipode.
    \item We find the most likely candidate for the projectile in such an impact to be a 500-750 km radius body impacting at an impact angle of 15-30$\degree$ and impact velocity of 1.2-1.4 times mutual escape speed ($\sim$6-7 km/s). 
    \item To avoid an over-thickened post-impact southern highlands, the bulk silicate fertility of Mars (i.e. maximum mass fraction of crust that can be extracted from a given mass of primordial mantle before silicate differentiation) must be less than or equal to 12.5\%. The reason for such a low crust-production efficiency is unclear.
    \item Solid strength plays a significant role in the distribution of melt after a giant impact on a body as large as Mars, and must therefore be included in any future impact studies on these scales.
\end{enumerate}

\subsection{Outlook}\label{outlook}
This work serves as an in-depth insight into the immediate consequences of a giant impact on Mars and its feasibility in forming the Martian Dichotomy; however, it is not without limitations and thus motivates future studies on the topic. One clear improvement that could be made is an increase in resolution (i.e. more particles if using the SPH method). This would lead to more accurate results overall, with the calculations of primordial crust being improved in particular due to their under-resolved nature. All of these results represent Mars directly after the crystallisation of the impact-induced magma ocean and neglect any long-term interior evolution. A natural follow-up to this work would therefore involve a sophisticated geodynamic model to investigate any further crust production due to decompression melting associated with convection processes in the mantle, and would potentially allow for direct comparison to present-day observations of Mars (such a study utilising the StagYY mantle convection code \citep{Tackley2008} is currently underway (Cheng et al. in prep.)). A more sophisticated geophysical model could also include the effects of heat-producing elements that are expected to be concentrated in the primordial crust. Finally, the parameter space explored could be further refined with a particular focus on impactors with radii within our preferred range of 500-750 km impacting at intermediate 15-30$\degree$ angles, and with a larger range of impact velocities.

\subsection*{Acknowledgements}
We thank David Baratoux for providing us with the \cite{Bouley2020} crustal thickness data. Supported by SNSF grant 200021\_175630.

\appendix
\section{Cubic Spline Function}
\subsection{Standard Form}\label{bspline}
\begin{equation}
    W(r,h) = 
    \alpha
    \begin{cases} 
          1 - \frac{3}{2} q^2 + \frac{3}{4} q^3 & 0 \leq q \leq 1 \\
          \frac{1}{4}(2 - q)^3 & 1 < q < 2 \\
          0 & q \geq 2
    \end{cases}
    ,
\end{equation}
where $q = r/h$ and $\alpha$ is $\frac{2}{3h}$, $\frac{10}{7 \pi h^2}$ or $\frac{1}{\pi h^3}$ in the 1, 2 and 3 dimensional cases, respectively.
\subsection{Integrated 1D Form}\label{bsplineprime}
\begin{equation}
    \int_{0}^{r} W_{\textrm{1D}}(r,h)~\mathrm{d}r = W'_{\textrm{1D}}(r,h) = 
    \begin{cases} 
          \frac{2}{3}q - \frac{1}{3}q^3 + \frac{1}{8}q^4 & 0 \leq q \leq 1 \\
          \frac{1}{2} - \frac{1}{24}(2-q)^4 & 1 < q < 2 \\
          0.5 & q \geq 2
    \end{cases}
    ,
\end{equation}
where $W_{\textrm{1D}}(r,h)$ is the 1-dimensional form of $W(r,h)$.

\section{Improved SPH Neighbour Symmetry}\label{neighlist}
\begin{figure}[htbp]
\centering
\includegraphics{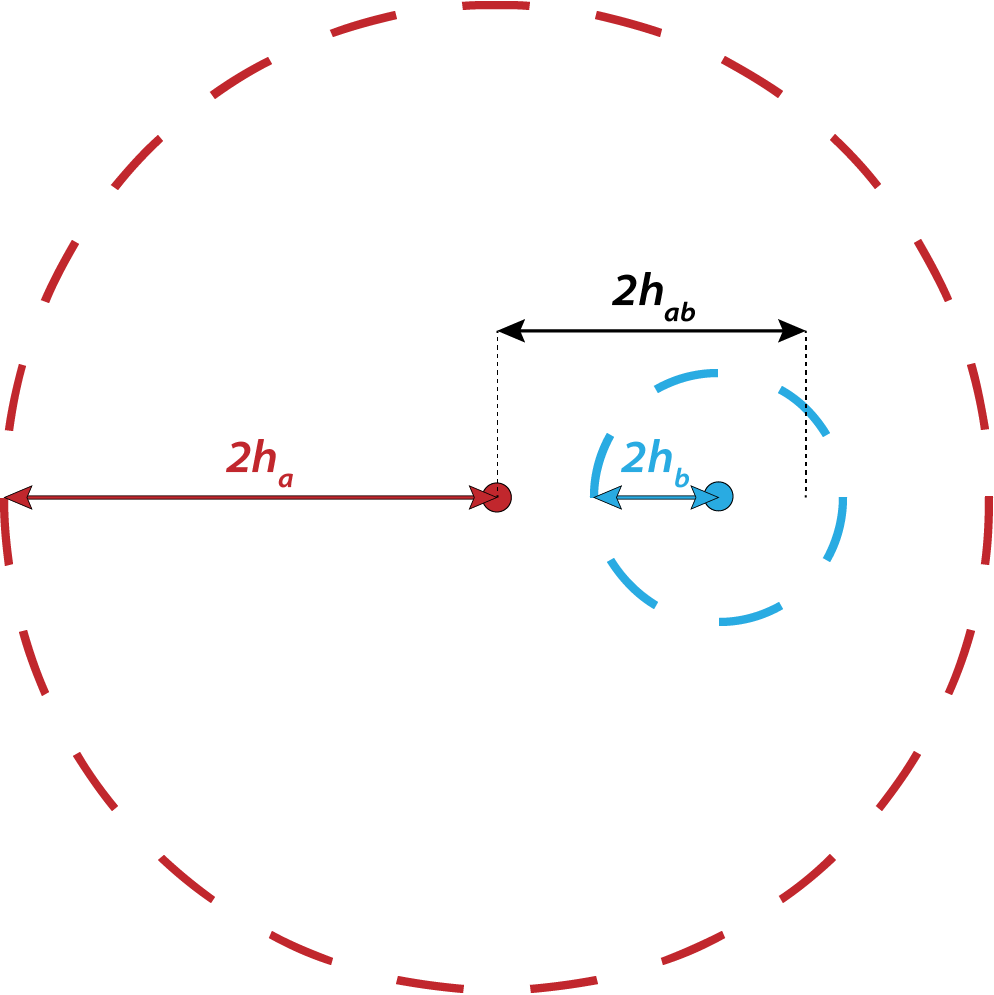}
\caption{A schematic diagram illustrating the asymmetry in the SPH sum calculations. Here, the distance between the two particles is less than {$2h_{ab}$}, meaning they will contribute to each other's SPH sums; however, particle $a$ will not be found in particle $b$'s neighbour search within {$2h_b$}. }
\label{fig:neighlist}
\end{figure}

The smoothing kernel is almost always chosen to have compact support, with the function (usually) decaying to zero at distances greater than $2h$, as is the case for our function in \ref{bspline}. Because of this convenient property, when calculating the SPH sums for a given particle, $a$, the previous versions of SPHLATCH neglected any neighbours at distances greater than twice the smoothing length, $h_a$, of said particle. This allowed for very efficient neighbour searching via the Barnes-Hut octree method. However, in order to maintain symmetric forces between particles with different smoothing lengths, the ``true'' smoothing length, $h_{ab}$, used for the force calculations is in fact the arithmetic mean of the central particle smoothing length and the contributing neighbour smoothing length, $h_b$, i.e. $h_{ab} = \frac{h_a + h_b}{2}$. Of course, for this symmetry to hold, particle $b$ must be included in particle $a$'s SPH sum \textit{and} vice versa. Unfortunately, this is not always the case when our search radius is only {$2h_a$ rather than $2h_{ab}$}. If the distance between the two particles is $x_{ab}$, the asymmetry occurs when {$x_{ab}>2h_b$ but $x_{ab}<2h_a$ and $x_{ab}<2h_{ab}$}. Here, particle $b$ will contribute to particle {$a$'s} SPH sum, but not the other way round, as is illustrated in Figure~\ref{fig:neighlist}.

To resolve this in the updated SPHLATCH code, we create a list of neighbours for each particle when performing the tree walk, and add the central particle to the neighbour's neighbour-list if the above conditions for asymmetry are met. These lists are then used for any future SPH sums, which has the additional benefit of a reduced number of tree-walks and thus a reduced computational cost.

\section{Primordial Crust Excavation at the Impact Site}\label{excavation}
\begin{figure}[H]
    \centering
    \begin{subfigure}{.49\textwidth}
        \adjincludegraphics[width=\linewidth]{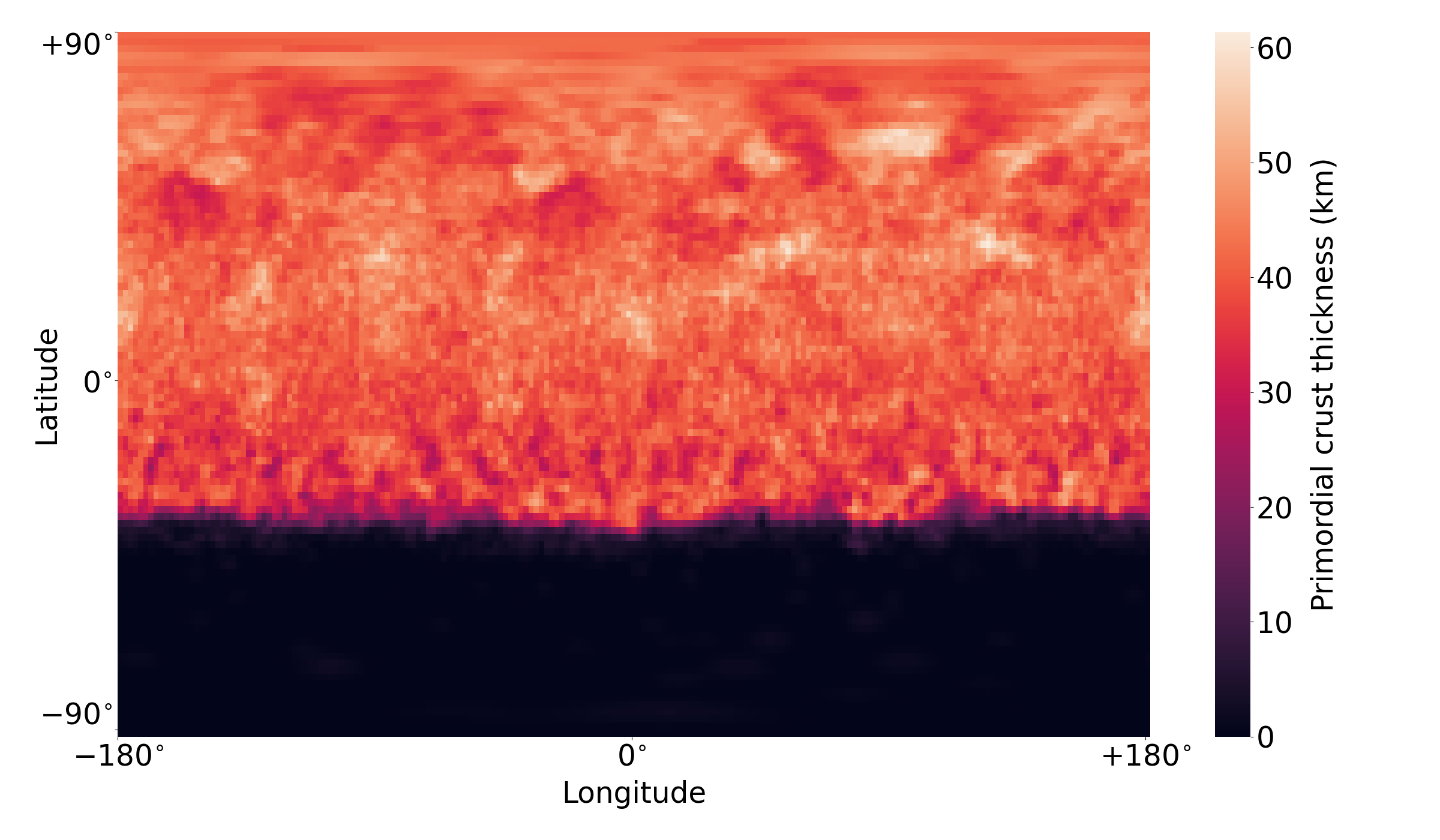}
    \end{subfigure}
    \begin{subfigure}{.49\textwidth}
        \adjincludegraphics[width=\linewidth]{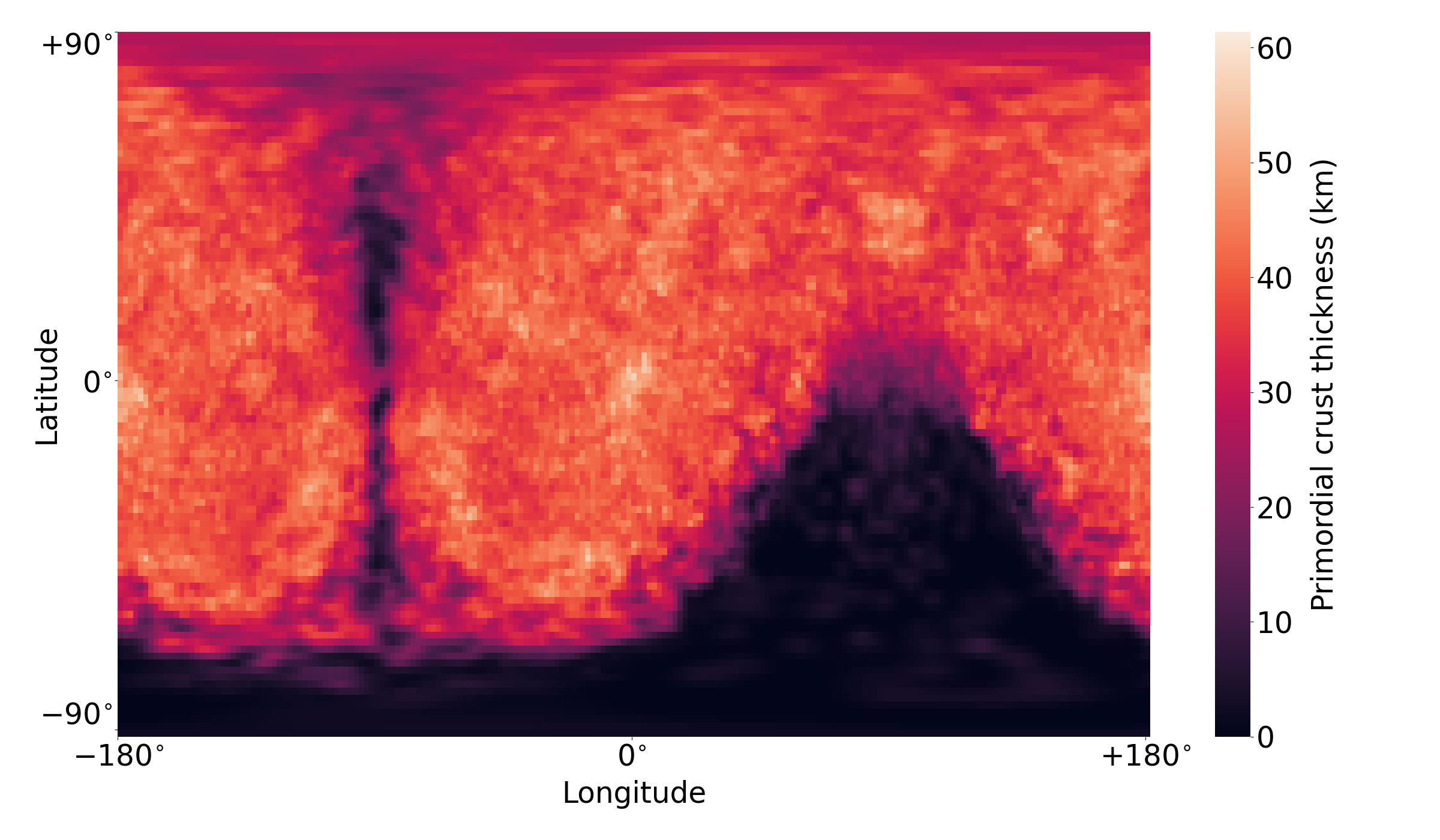}
    \end{subfigure}
    \caption{Crustal thickness contribution from primordial crust alone for the classical cases of a 1000 km radius body impacting at the mutual escape speed with an impact angle of 0$\degree$ (left) and 45$\degree$ (right).}
    \label{fig:classicalmaps_onlycrust}
\end{figure} 

\begin{figure}[H]
    \centering
    \begin{subfigure}{.49\textwidth}
        \adjincludegraphics[width=\linewidth]{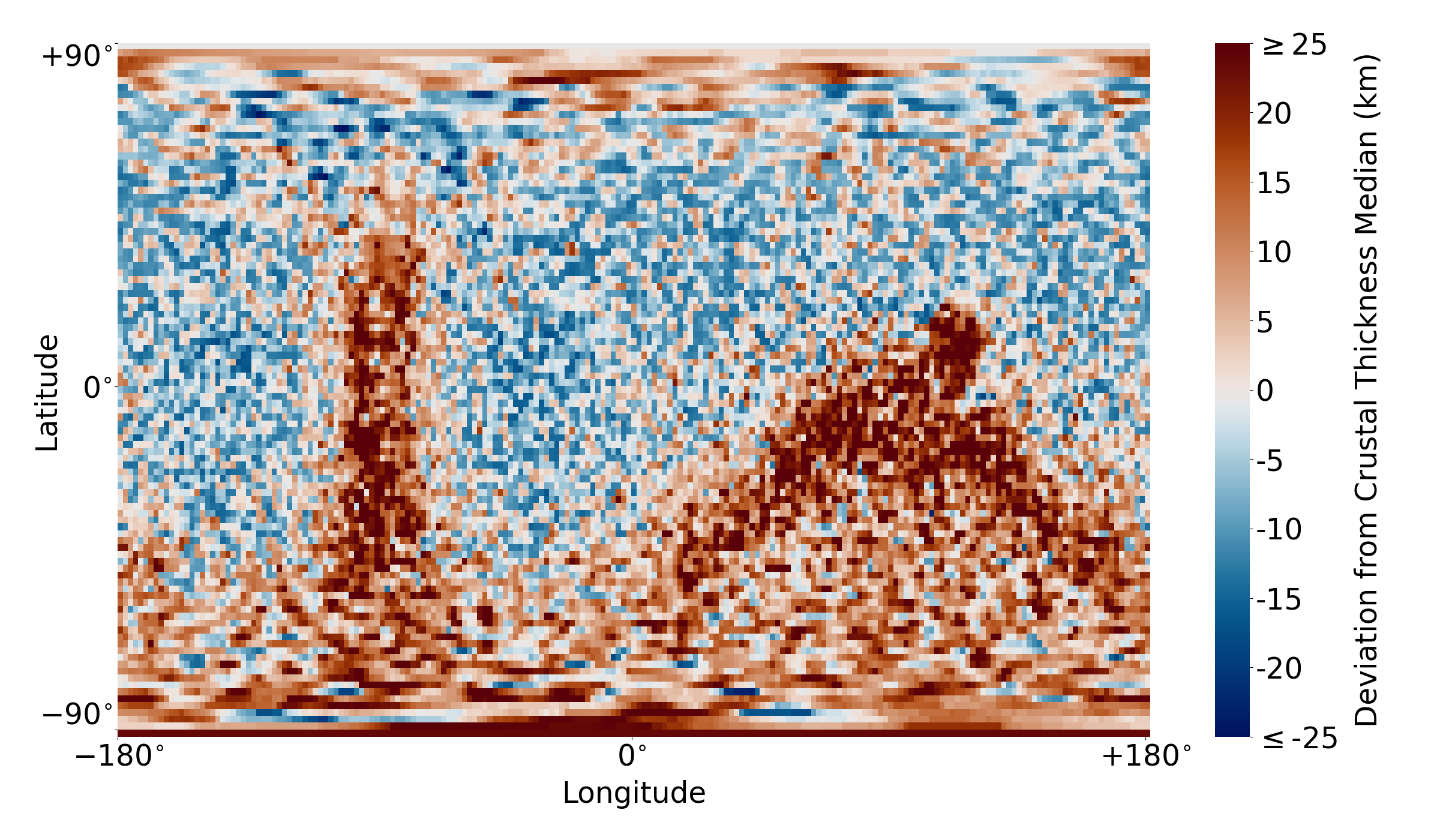}
    \end{subfigure}
    \begin{subfigure}{.49\textwidth}
        \adjincludegraphics[width=\linewidth]{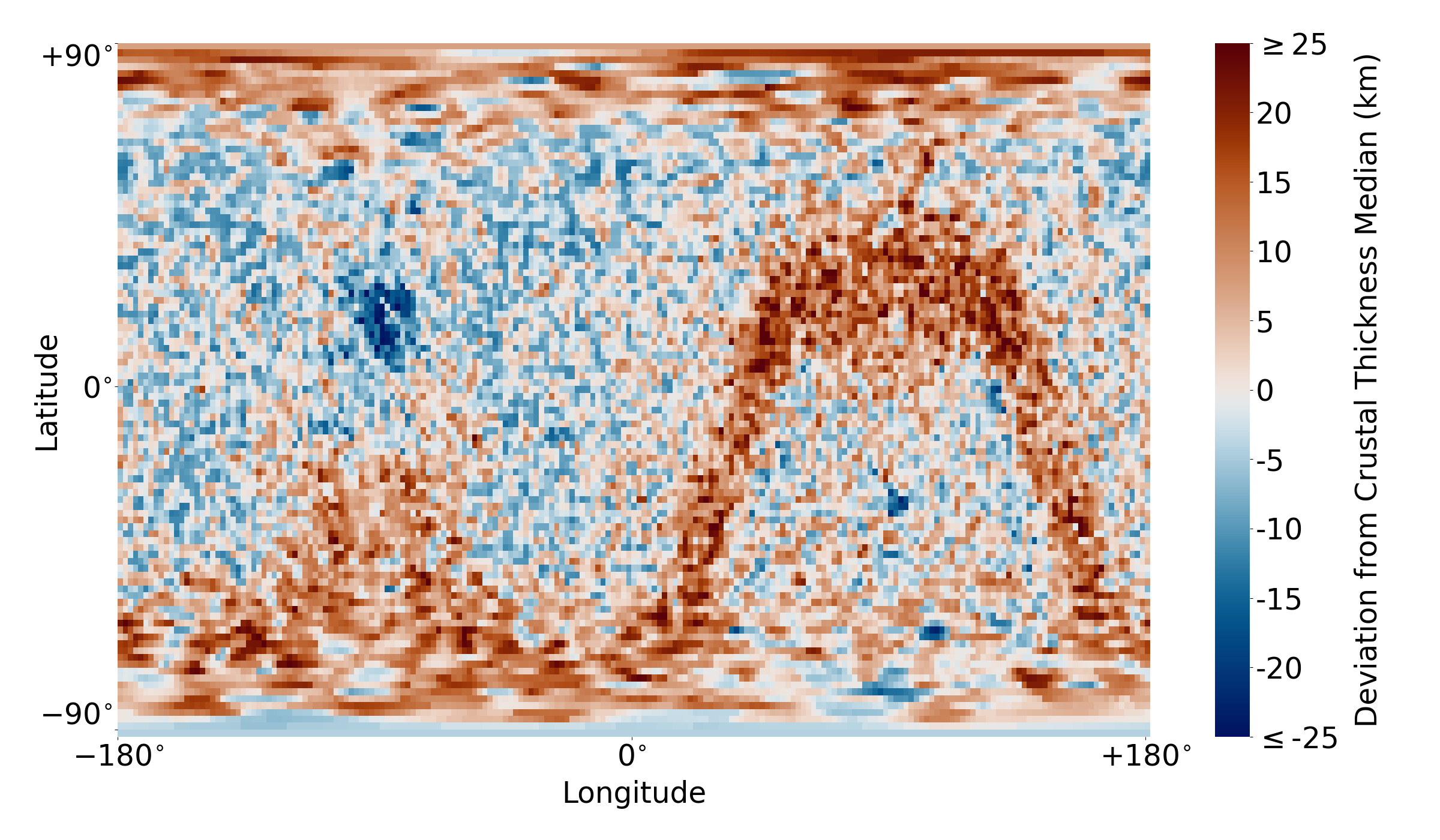}
    \end{subfigure}
    \caption{The best-fitting cases for impactor radii of 1500 km (left) and 2000 km (right) but with the fully-mixed mantle depletion scheme. The same cases with the fully-stratified scheme can be seen in Figure~\ref{fig:best_largest}.}
    \label{fig:best_largest_fert}
\end{figure}

\section{Core Fraction}
\begin{figure}[H]
    \centering
    \includegraphics[width=\textwidth]{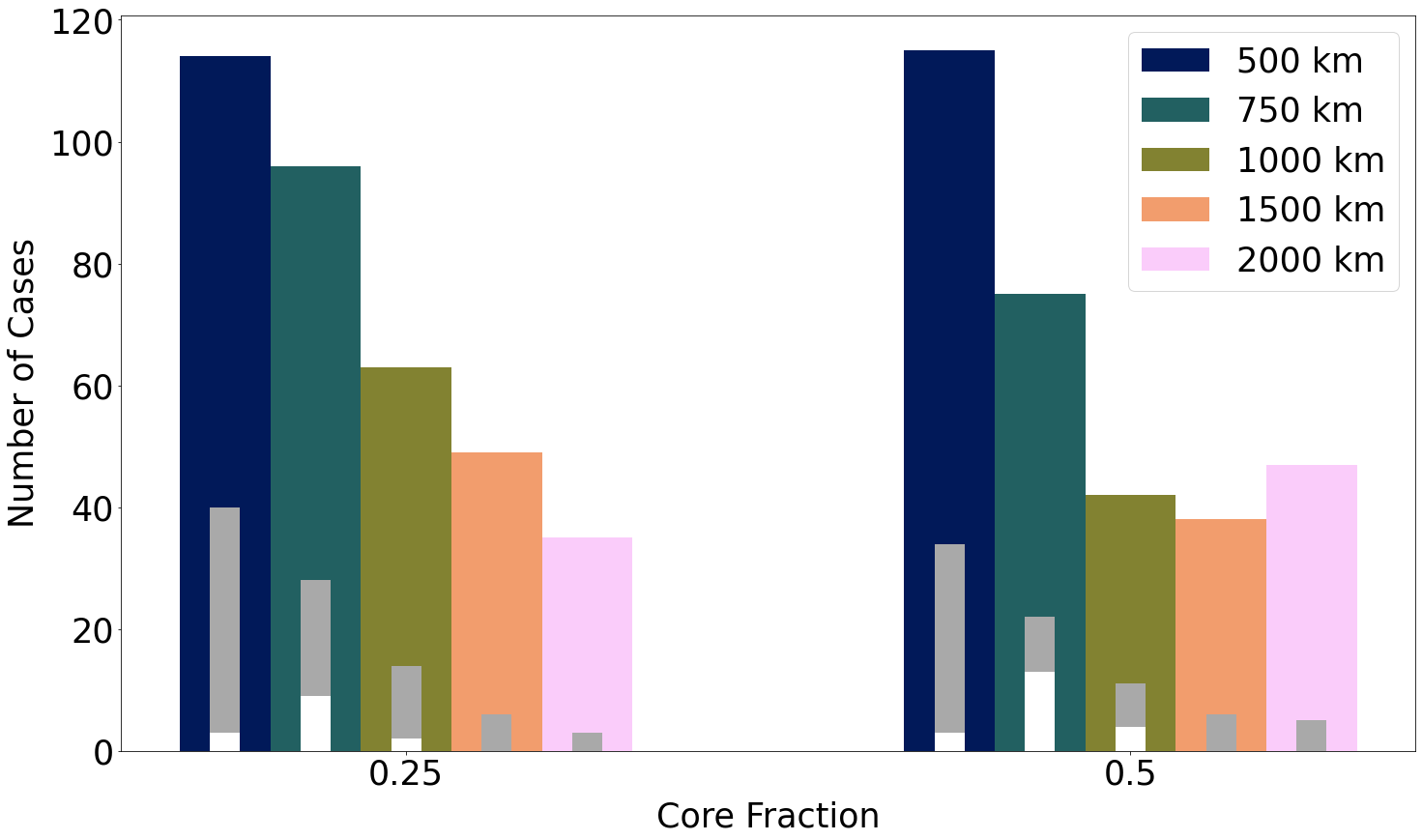}
    \caption{Histogram of cases within $\delta_{SH_{max}}$ for each impactor radius as a function of impactor core fraction.}
    \label{fig:goodcases_core}
\end{figure}

\section{Angular Momentum Conservation}
\begin{figure}[H]
    \centering
    \includegraphics[width=\textwidth]{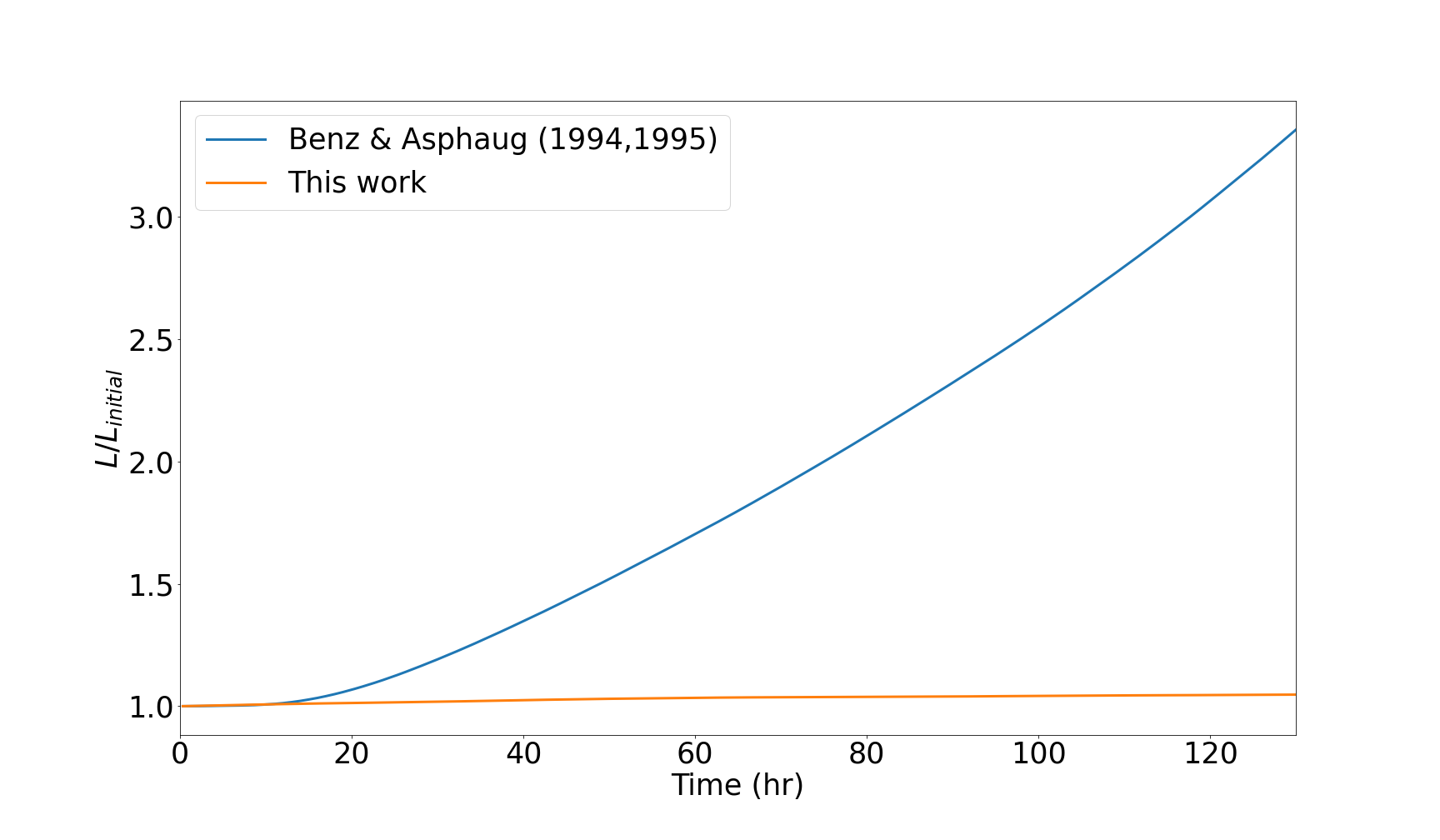}
    \caption{Angular momentum as a function of time with and without the rotation rate tensor corrections described in Section~\ref{strength}. These values correspond to a Mars-like body rotating with a 10 hr period. Angular momentum is given relative to its value at the beginning of the simulation.}
    \label{fig:conservation}
\end{figure}

\section{Pre-Impact Crust Distributions}
\begin{figure}[H]
    \centering
    \begin{subfigure}{.49\textwidth}
        \adjincludegraphics[width=\linewidth]{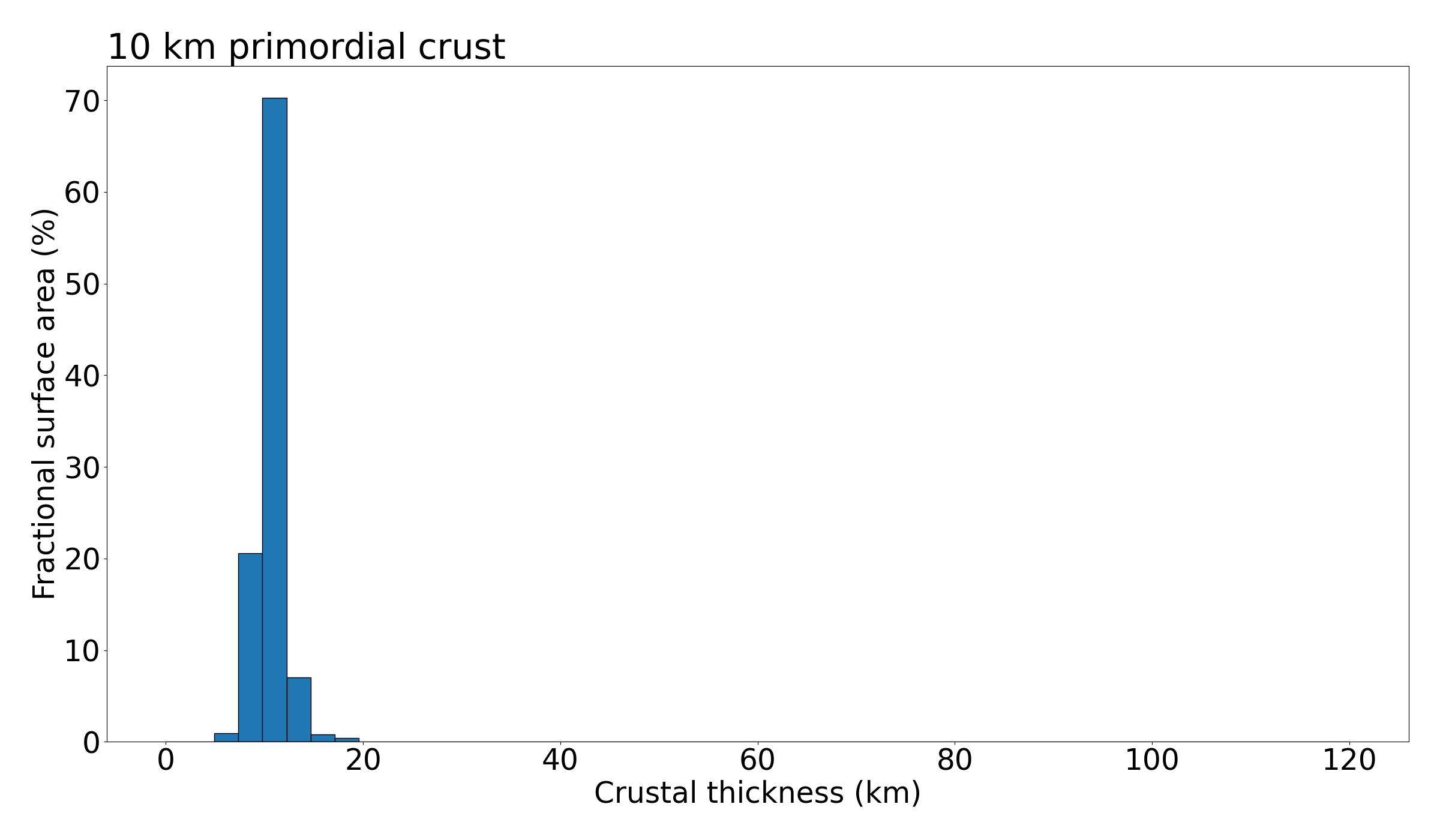}
    \end{subfigure}
    \begin{subfigure}{.49\textwidth}
        \adjincludegraphics[width=\linewidth]{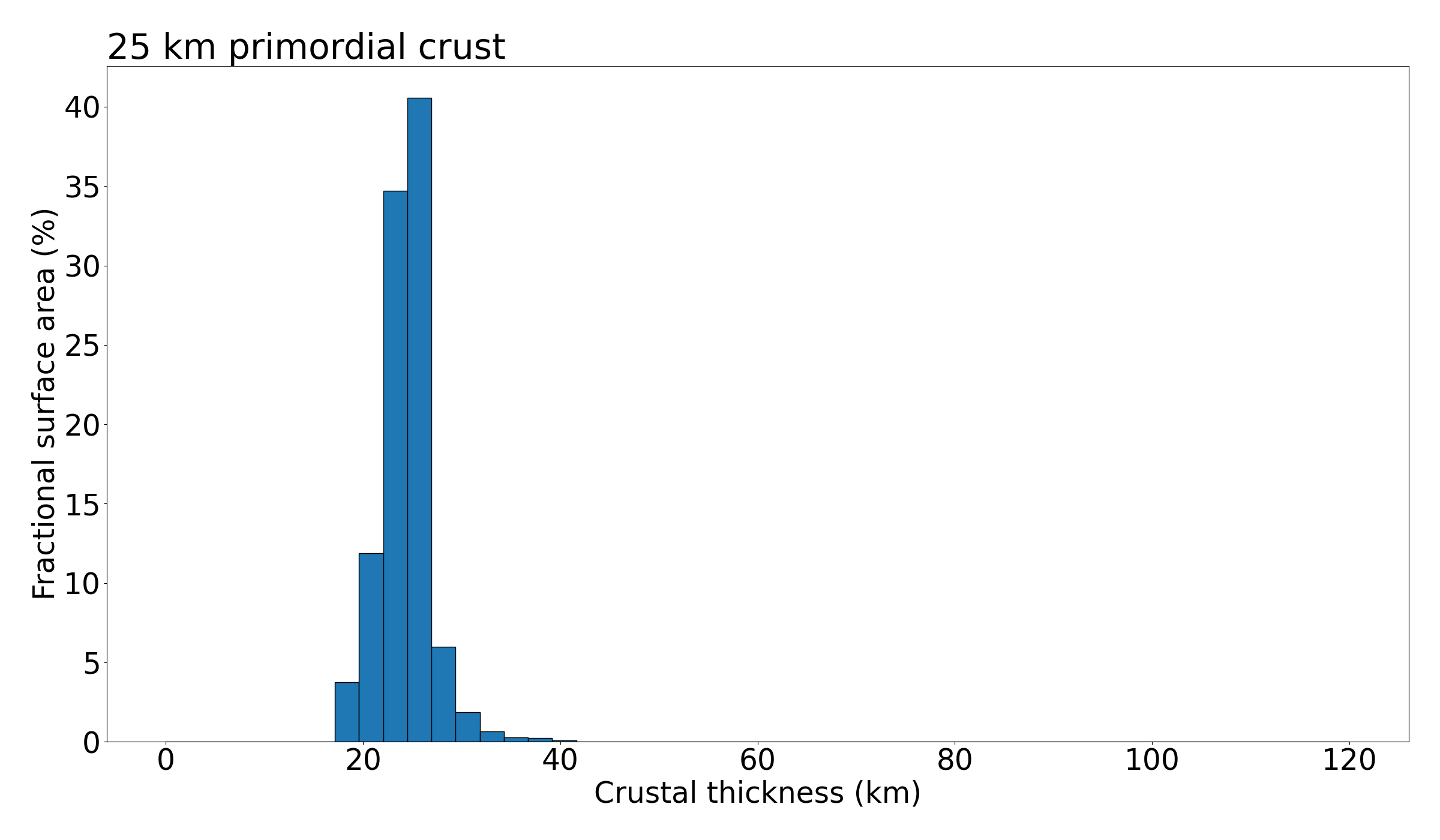}
    \end{subfigure}
    \begin{subfigure}{.49\textwidth}
        \adjincludegraphics[width=\linewidth]{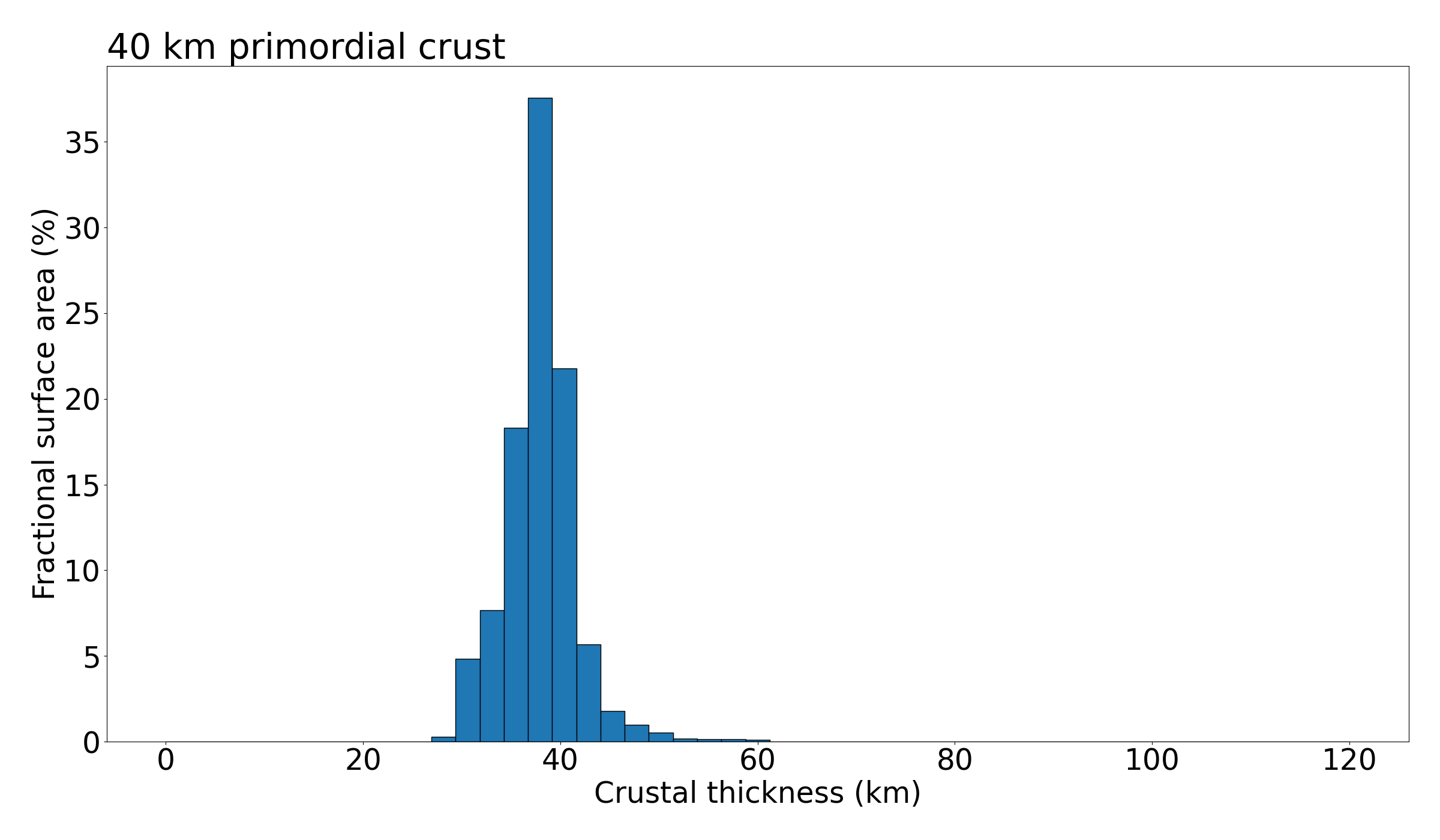}
    \end{subfigure}
    \begin{subfigure}{.49\textwidth}
        \adjincludegraphics[width=\linewidth]{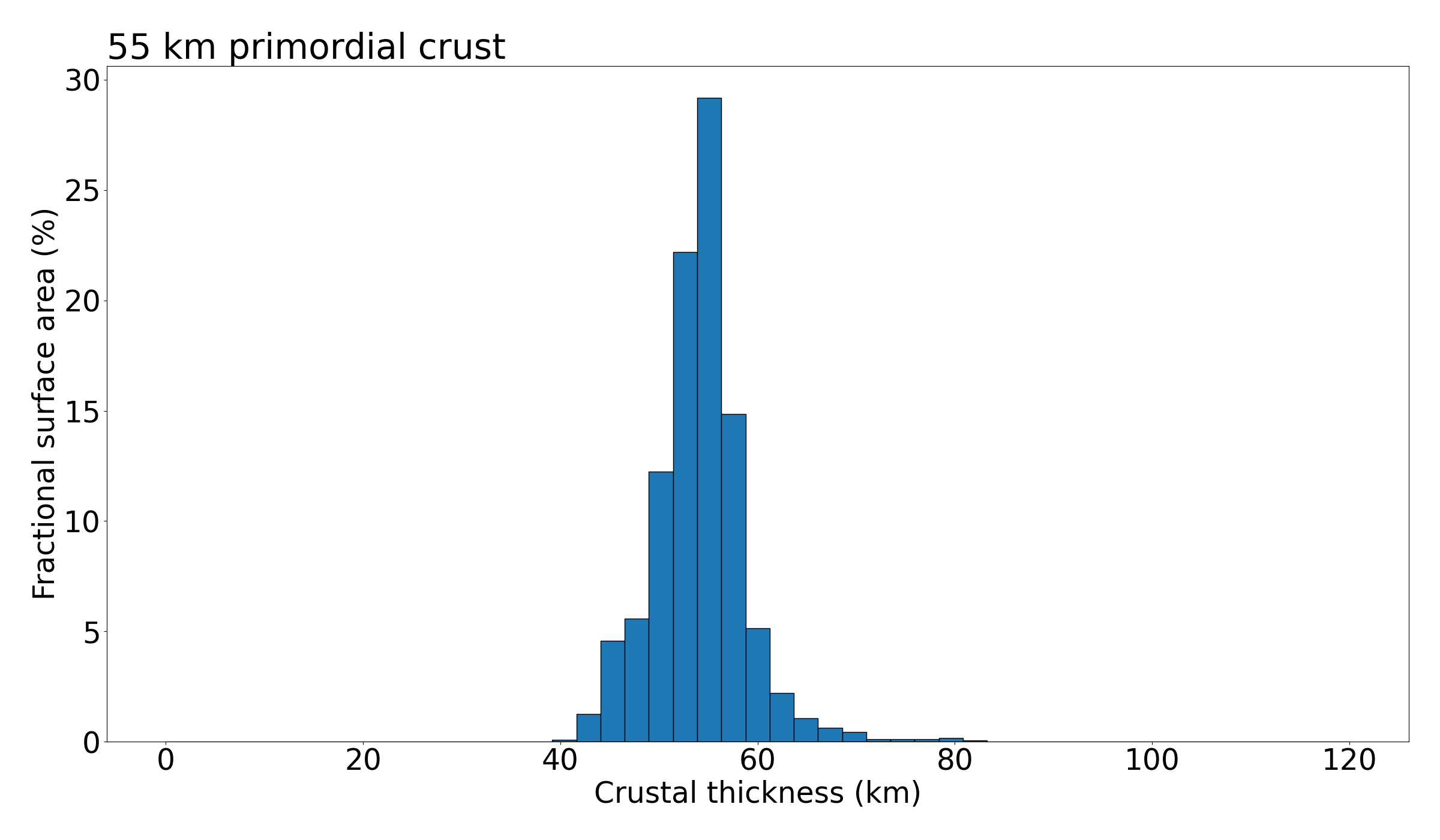}
    \end{subfigure}
    \begin{subfigure}{.49\textwidth}
        \adjincludegraphics[width=\linewidth]{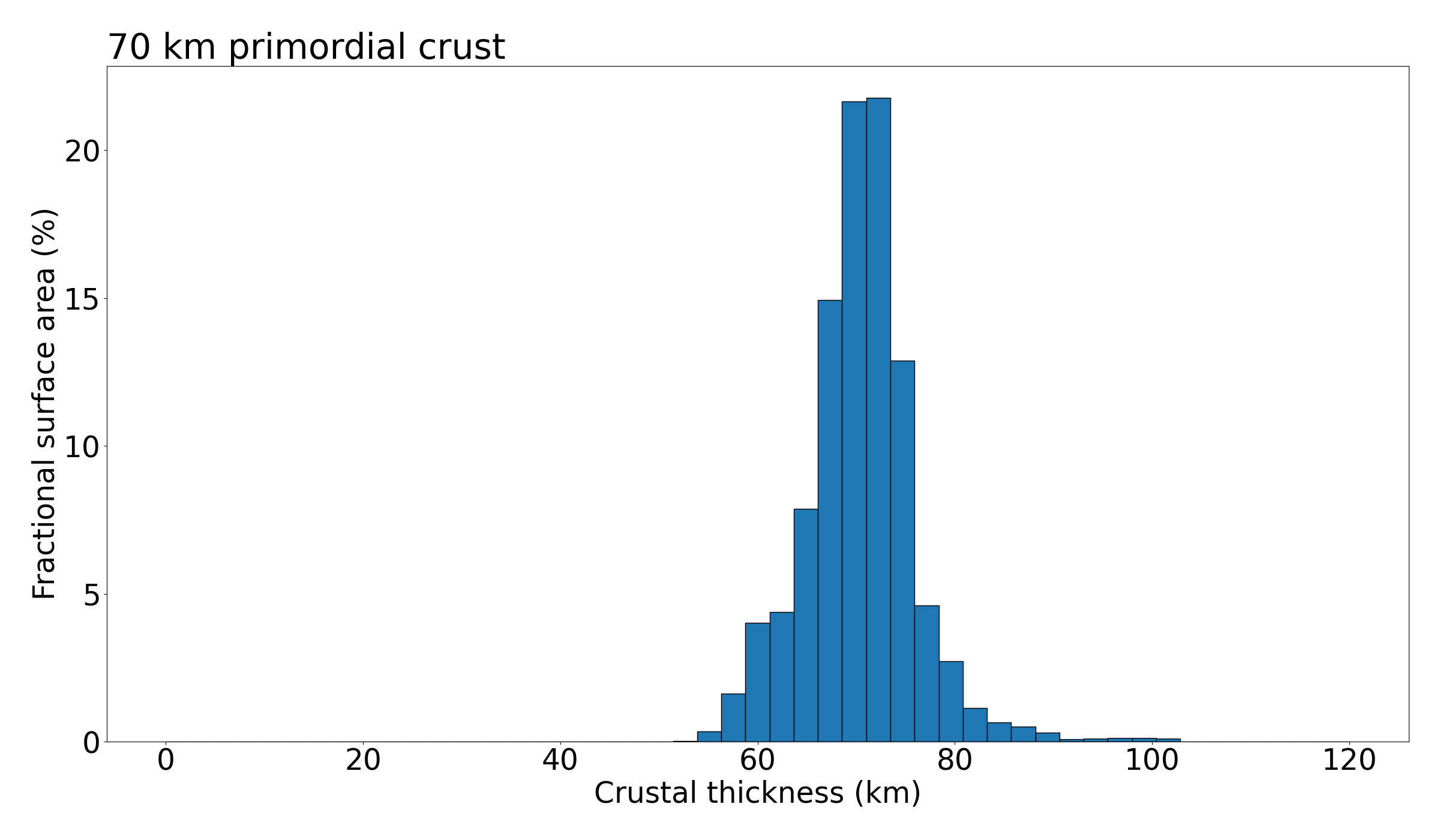}
    \end{subfigure}
    \caption{Histograms depicting the pre-impact distribution of crust on our Mars-like body, for each primordial thickness investigated in this study. Each bin displays the percentage of the Martian surface corresponding to the given crustal thickness range.}
    \label{fig:pre_impact}
\end{figure}

\section{Melt Extraction Threshold}
\begin{figure}[H]
    \centering
    \begin{subfigure}{.49\textwidth}
        \centering
        \begin{raggedright}
            \textbf{0.00\%}
        \end{raggedright}
        \adjincludegraphics[width=\linewidth]{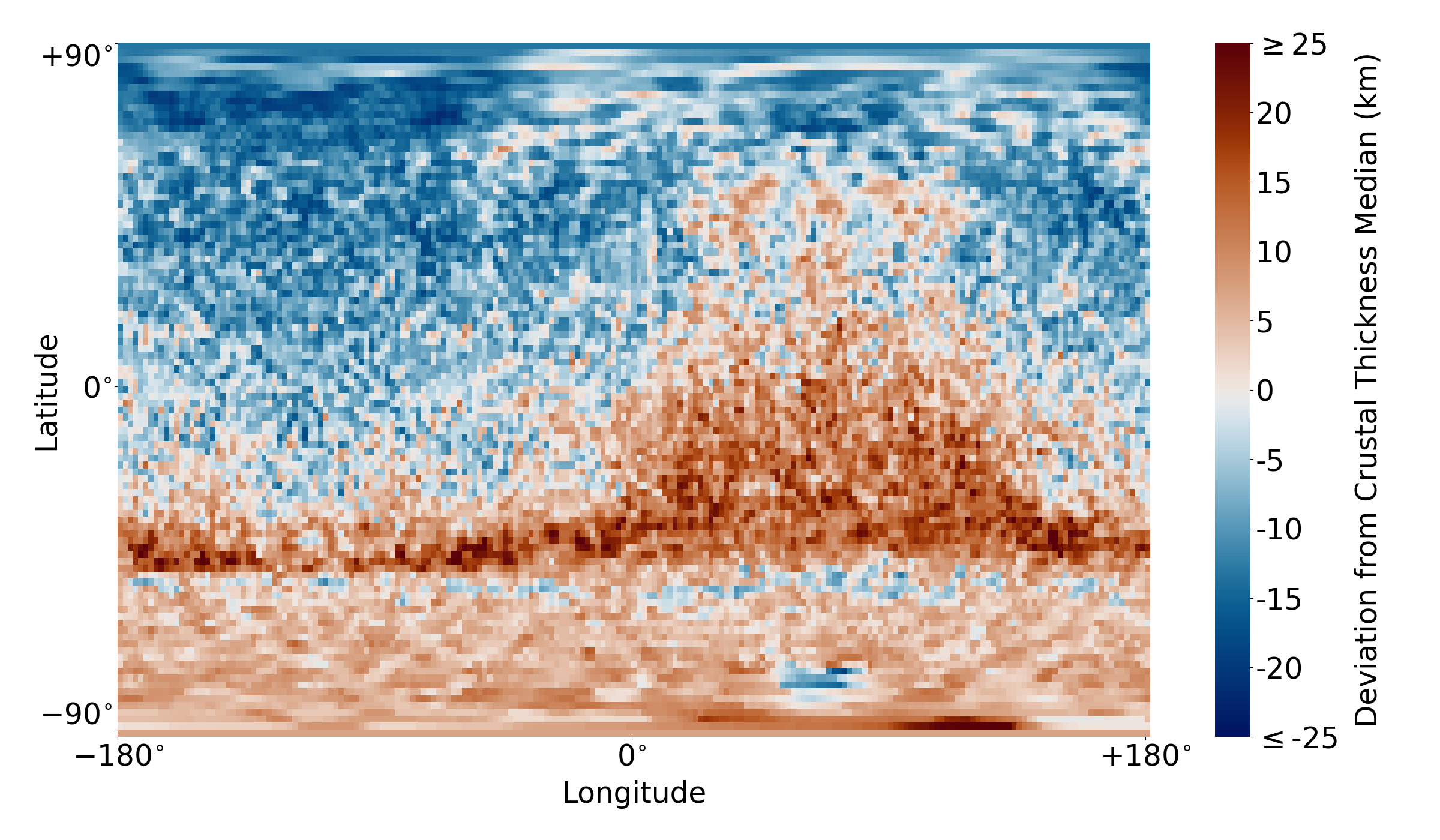}
    \end{subfigure}
    \begin{subfigure}{.49\textwidth}
        \centering
        \textbf{0.02\%}
        \adjincludegraphics[width=\linewidth]{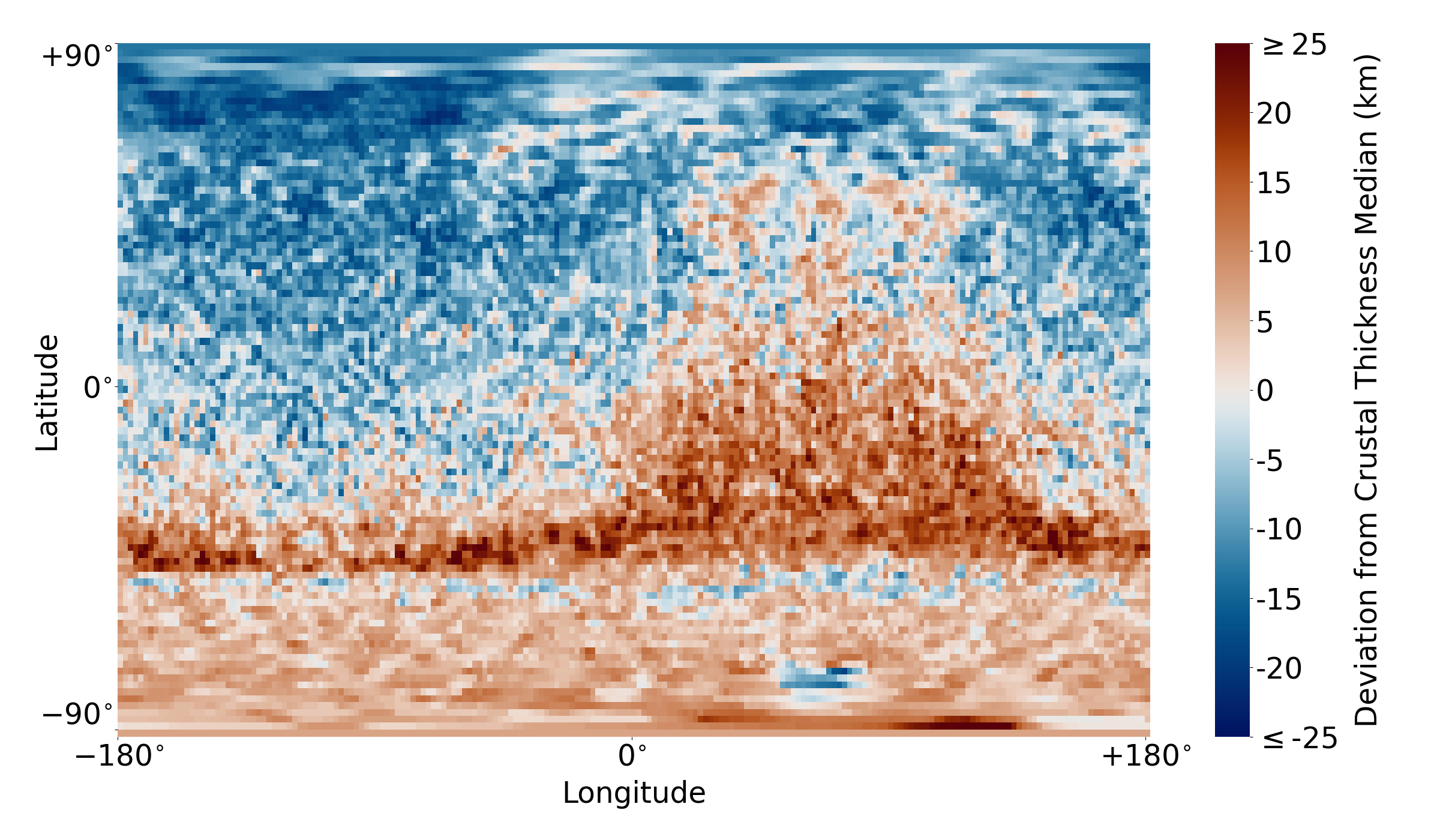}
    \end{subfigure}
    \begin{subfigure}{.49\textwidth}
        \centering
        \textbf{0.06\%}
        \adjincludegraphics[width=\linewidth]{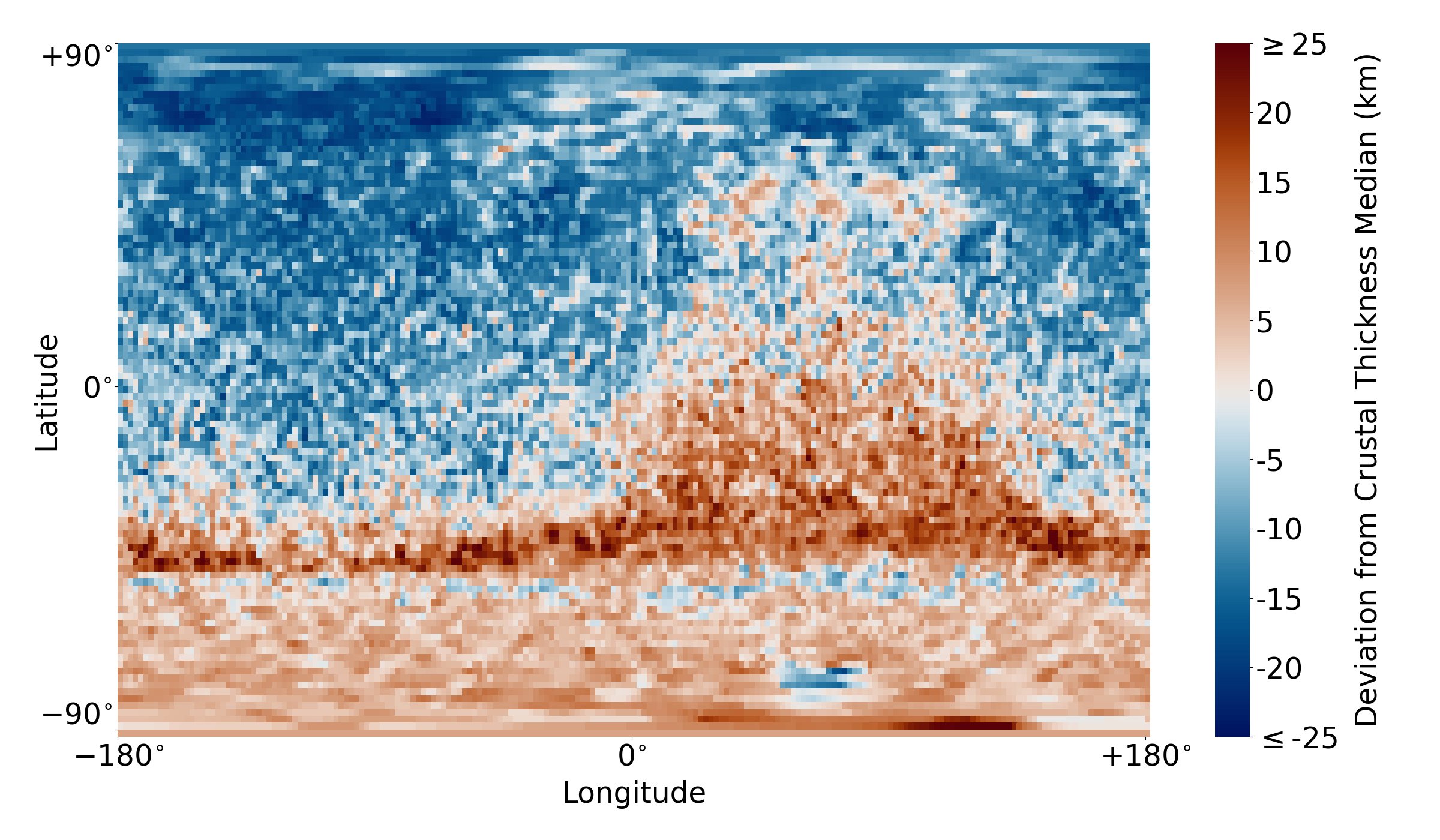}
    \end{subfigure}
    \begin{subfigure}{.49\textwidth}
        \centering
        \textbf{0.08\%}
        \adjincludegraphics[width=\linewidth]{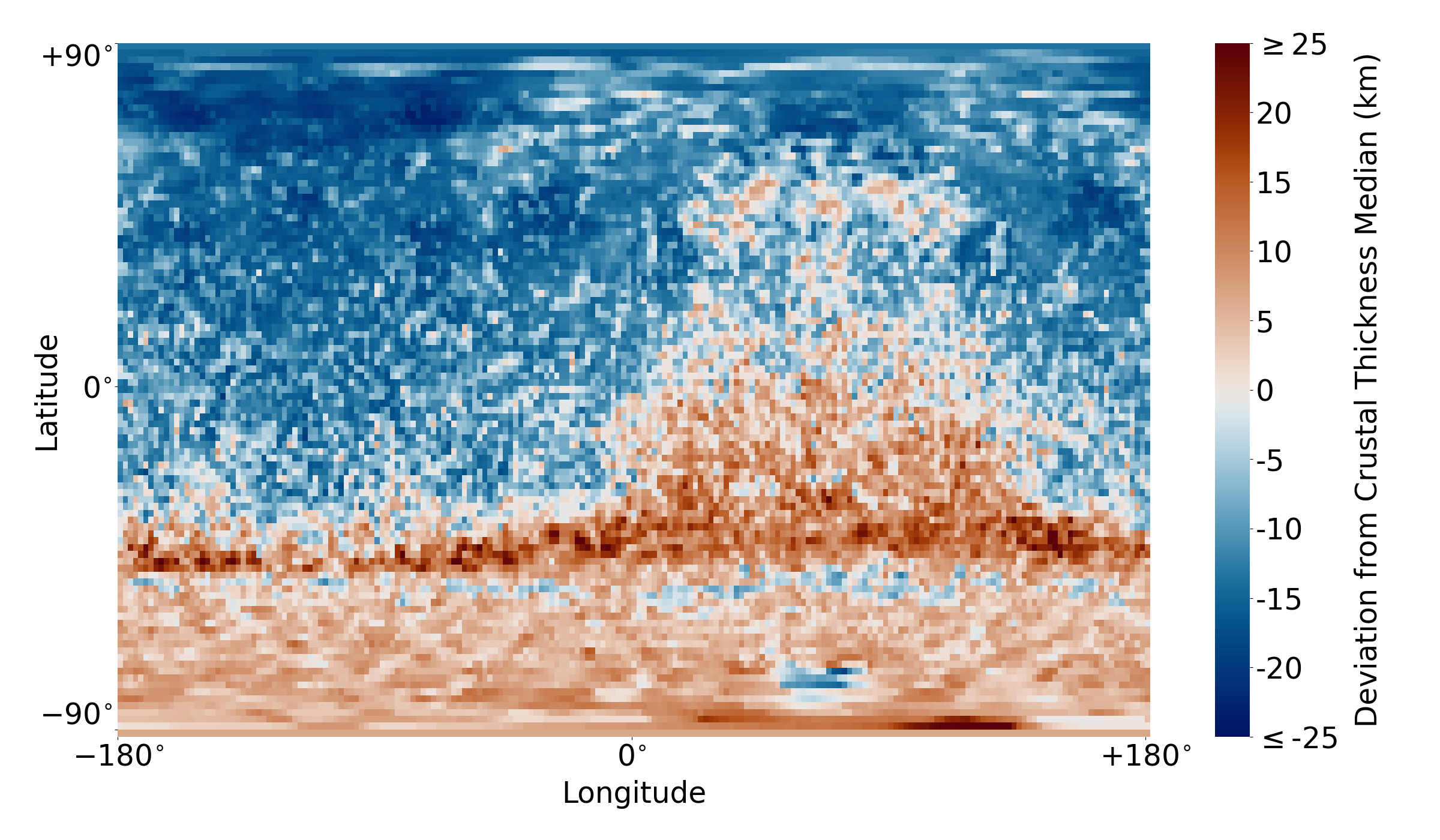}
    \end{subfigure}
    \caption{{Equirectangular projection of post-impact crustal thickness for the best-fitting case (same parameters as Figure~\mbox{\ref{fig:best_750 km}}) but with melt extraction thresholds of 0.00\% (top-left), 0.02\% (top-right), 0.06\% (bottom-left) and 0.08\% (bottom-right). At the main impact site, broadly corresponding with the region of crust excavation, melt contributing to the crust production scheme has very high melt fractions, and thus the predicted crustal thickness does not change for the different melt extraction thresholds. Beyond the main impact site, lower melt fractions are more significant, leading to different crustal thicknesses in these regions for the different extraction thresholds.}}
    \label{fig:melt_thresh}
\end{figure}

\section{Neutral Buoyancy Pressure}
\begin{figure}[H]
    \centering
    \begin{subfigure}{.49\textwidth}
        \adjincludegraphics[width=\linewidth]{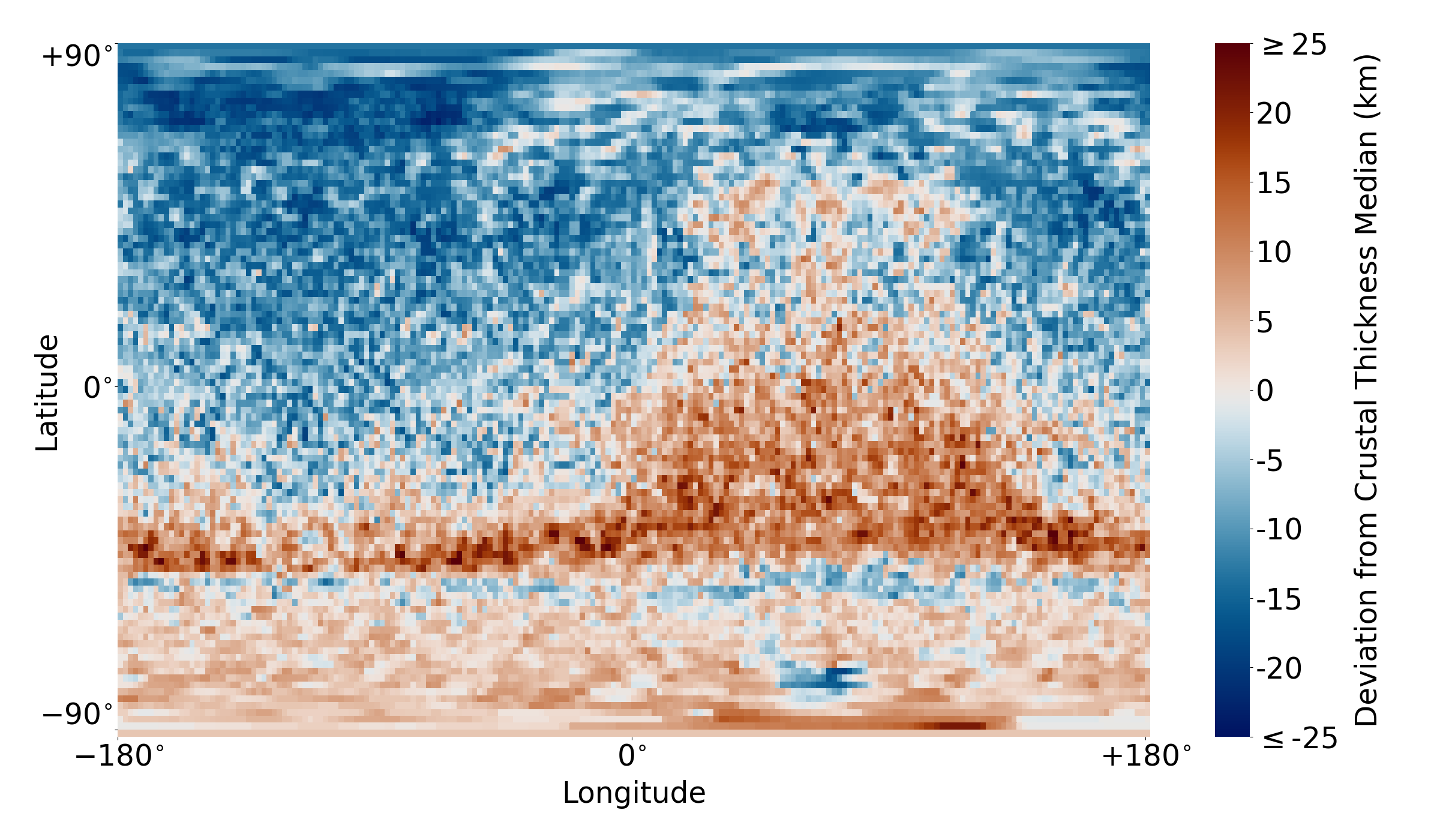}
    \end{subfigure}
    \begin{subfigure}{.49\textwidth}
        \adjincludegraphics[width=\linewidth]{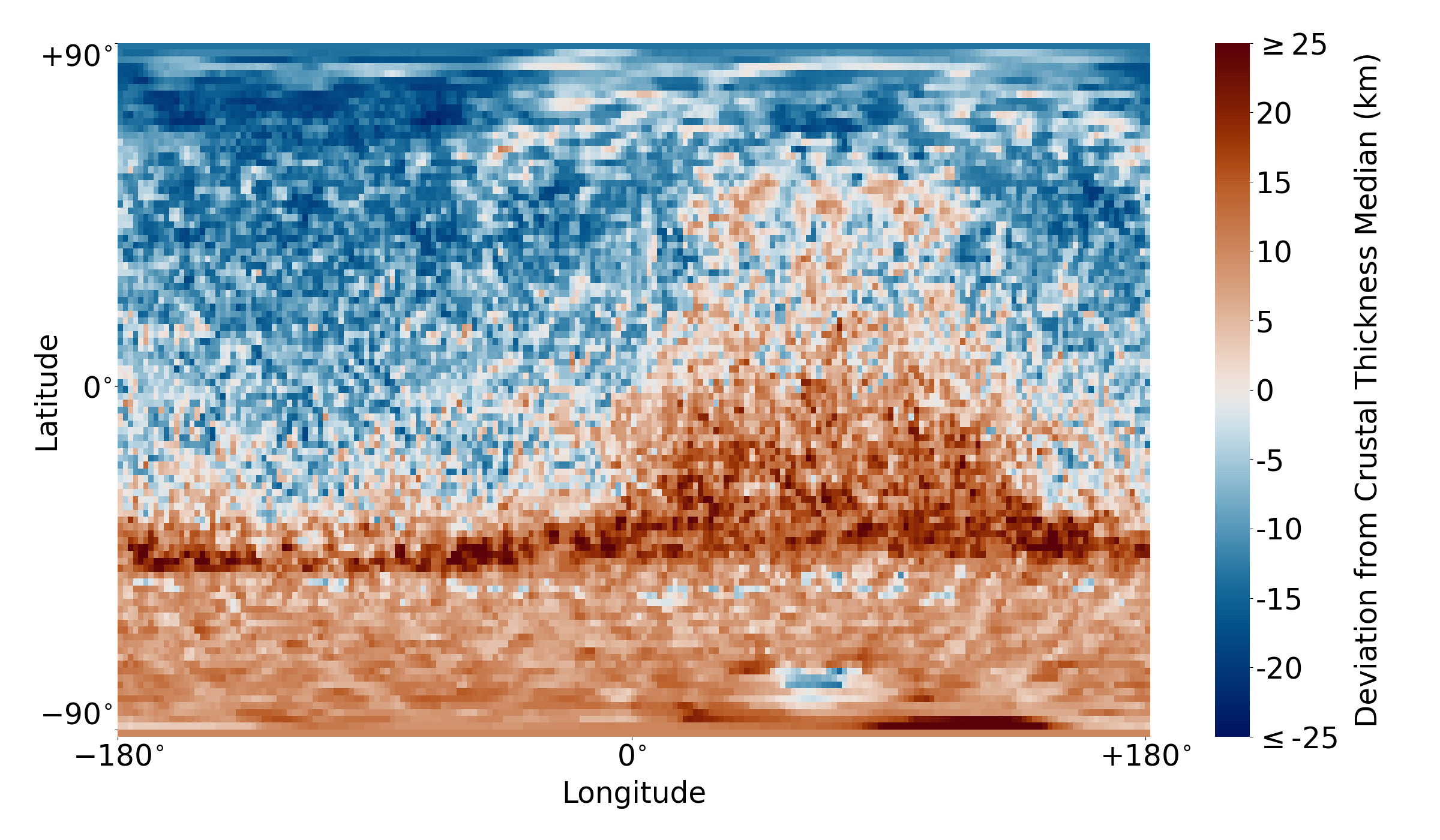}
    \end{subfigure}
    \caption{{Equirectangular projection of post-impact crustal thickness for the best-fitting case (same parameters as Figure~\mbox{\ref{fig:best_750 km}} but with neutral buoyancy depths of 7 GPa (left) and 8 GPa (right).}}
    \label{fig:buoyP}
\end{figure}

\section{Spherical Grid Accuracy}\label{grid_accuracy}
{In principle, smoothing the SPH data onto an arbitrary grid is fully consistent with the SPH method, as it is simply an extension of the fundamental smoothing equation of Equation~\mbox{\ref{sphEq}}. However, it can exaggerate some undesirable properties of the method, and for certain practical workarounds used in the SPH, lead to some inconsistencies.}

\begin{figure}[H]
    \centering
    \includegraphics[width=\textwidth]{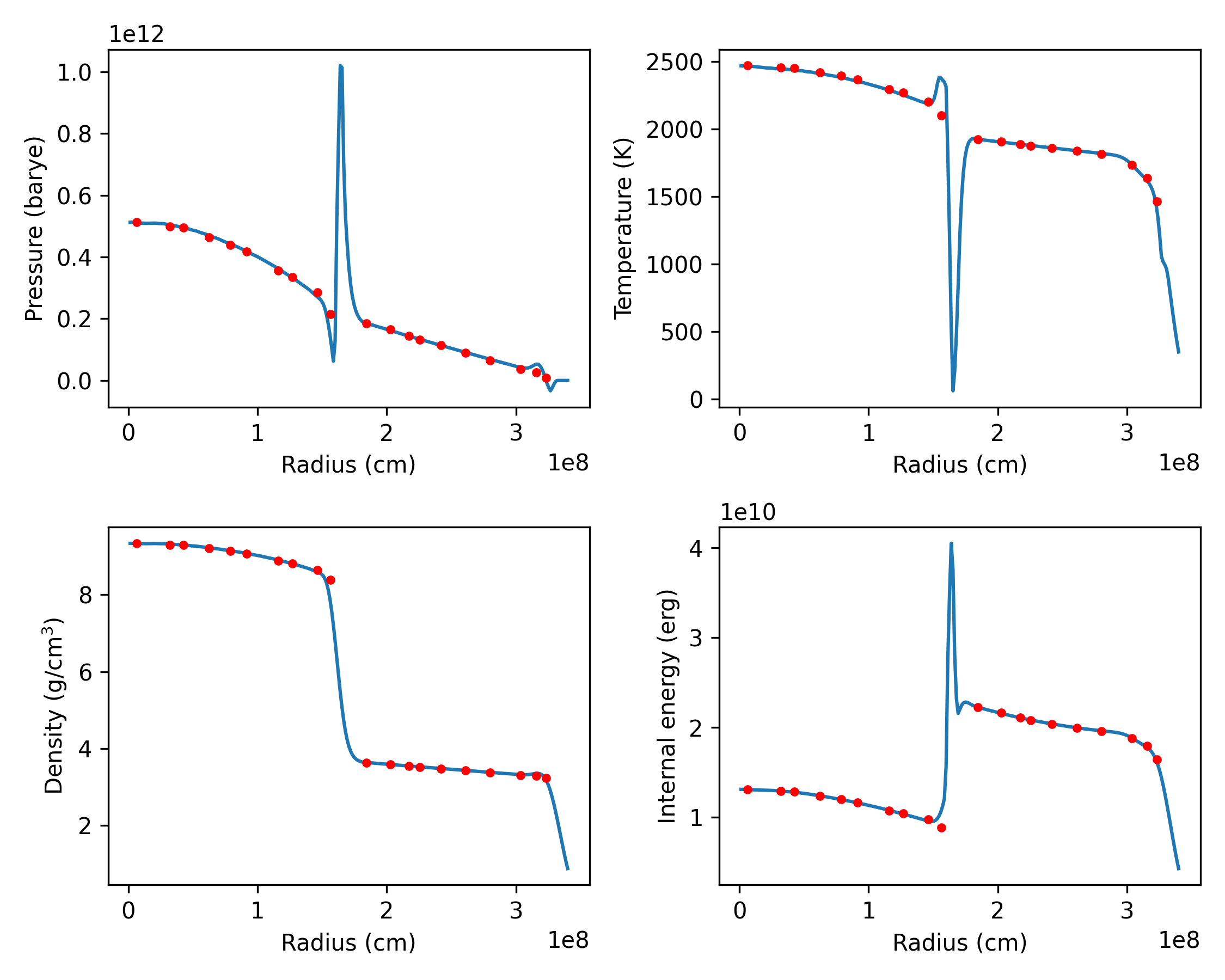}
    \caption{{Various quantities as a function of radius for a Mars-like body that has not undergone an impact (i.e. in its initial near-symmetric state directly after simulation setup). The red points represent individual SPH particles which were sampled by first dividing the body into 20 equal-length bins in radius and then randomly selecting a single particle from each bin. The blue line shows the average value for each radius on the spherical grid.}}
    \label{fig:profile_scatter}
\end{figure}

{Figure \mbox{\ref{fig:profile_scatter}} shows the most important thermodynamic quantities as a function of radius for a pre-impact Mars-like body used in this study, comparing the results taken directly from a sample of individual SPH particles to the average values of the corresponding spherical grid. For the majority of radii, the results agree very well and show a smooth distribution without any sharp gradients. At the core-mantle boundary, however, there is an abrupt spike in all quantities other than density. This effect is a known problem in SPH, where sharp contrasts in density lead to spurious results \mbox{\citep[e.g.][]{Woolfson2007,Emsenhuber2018,Ruiz-Bonilla2022}}. The amplitude of the peaks do appear larger than those seen in previous work; this may be a consequence of the increased resolution of the grid relative to the particles, as radii with the most dramatic values in the spherical grid may not actually contain any particles. 

The spurious results at the core-mantle boundary have no effect on the crustal thickness calculations as they correspond to depths much greater than the neutral buoyancy depth. The density contrast at the surface also leads to some oscillations in the variables, but these are far less pronounced and do not significantly affect the results of this work.}

\newpage
\bibliography{mycollection}

\end{document}